\pdfoutput=1
\documentclass[cernpreprint, atlasdraft=false, texlive=2023, UKenglish, texmf, orcidlogo]{atlasdoc}

\usepackage[subfigure]{atlaspackage}

\usepackage{atlasbiblatex}
 
\usepackage{atlasphysics}
\usepackage{atlasprocess}
\usepackage{PMGRefs-defs}
 
\usepackage{pdfpages}

\graphicspath{{logos/}{figures/}}
 
\usepackage{ANA-STDM-2020-14-PAPER-defs}

\AtlasTitle{
Measurements of $Z\gamma+$jets differential cross sections in $pp$ collisions at $\sqrt{s}=13$ TeV with the ATLAS detector}
 
\AtlasAbstract{
Differential cross-section measurements of $Z\gamma$ production in association with hadronic jets are presented, using the full 139~fb$^{-1}$ dataset of $\sqrt{s}=13$ TeV proton--proton collisions collected by the ATLAS detector during Run~2 of the LHC. Distributions are measured using events in which the $Z$ boson decays leptonically and the photon is usually radiated from an initial-state quark. Measurements are made in both one and two observables, including those sensitive to the hard scattering in the event and others which probe additional soft and collinear radiation. Different Standard Model predictions, from both parton-shower Monte Carlo simulation and fixed-order QCD calculations, are compared with the measurements. In general, good agreement is observed between data and predictions from MATRIX and MiNNLO$_\text{PS}$, as well as next-to-leading-order predictions from \textsc{MadGraph5\_aMC@NLO} and  \textsc{Sherpa}.}

\AtlasRefCode{STDM-2020-14} 
 
\PreprintIdNumber{CERN-EP-2022-240}

\AtlasJournal{JHEP}
\AtlasJournalRef{\JHEP 07 (2023) 72}
\AtlasDOI{10.1007/JHEP07(2023)072}

\hypersetup{pdftitle={ATLAS document},pdfauthor={The ATLAS Collaboration}}

\usepackage{multirow}
\usepackage{arydshln}
\usepackage{pgffor}

\begin{document}
 
\maketitle

\section{Introduction}
\label{sec:intro}

Precision measurements of cross sections for the production of a $Z$ boson and a photon (\Zg) at the Large Hadron Collider (LHC)~\cite{Evans:2008zzb} play a crucial role in the study of the Standard Model (SM) and are sensitive to physics beyond the SM. Differential cross sections for $Z\gamma$ in association with jet activity ($Z\gamma+$jets) can be used to test fixed-order perturbative QCD (pQCD) calculations and predictions with resummation of Sudakov logarithms~\cite{Wiesemann_2020}. This process is also sensitive to the parton distribution functions (PDFs) and can validate those PDFs extracted in global analyses~\cite{Ball2017}. In addition, the $Z\gamma+$jets differential cross sections can be used to constrain the Monte Carlo (MC) models, especially the parton-shower (PS) approximation~\cite{Krause_2018}.
 
In phase-space regions where the transverse momentum (\pT) of the system is much smaller than the mass ($m$) of the $Z$ boson or \Zg, fixed-order QCD calculations are dominated by Sudakov-logarithm terms, due to soft and collinear emission, of the order of $\alphas^n \ln^{n+1}(\pT/m)$, where $n$ is the fixed order considered. These terms are usually treated by resummation~\cite{Ebert_2017,Ebert_2021} and can give very precise predictions with next-to-leading logarithms (NLL) and up to next-to-next-to-next-to-leading logarithms (N3LL)~\cite{Wiesemann_2020}. These resummation models can be tested in phase-space regions where the logarithm terms dominate, i.e.\ in regions where the hard scale of the process is much larger than the value of the observable considered.
 
Such a phase-space region can be probed by simultaneously measuring two independent observables, providing a more complete description of the pattern of QCD emission~\cite{Lustermans_2019}. This is done with two-dimensional (2D) distributions, measuring an observable sensitive to the hard scale, called the \textit{hard variable}, as a function of another observable, called the \textit{resolution variable}, which probes the additional soft radiation. Thus, the \textit{hard variable} is an observable which is directly sensitive to the hard scale of the process and its value is non-zero at leading order (LO), e.g.\ $p_{\mathrm{T}}^Z$ , $p_{\mathrm{T}}^{\gamma}$, $m_{Z\gamma}$, or any linear combinations of these variables. On the other hand, a \textit{resolution variable} is an observable sensitive to the additional soft or collinear QCD radiation; the values of these observables, e.g.\  $\pT^{Z\gamma}$ or the number of jets ($N_\textrm{jet}$), are zero at LO and take non-zero values only beyond LO. An example of a 2D measurement is the differential cross section as a function of \DiffpTZ in different regions of \SumpTZ. In these measurements, \DiffpTZ is the \textit{resolution variable} that allows effects near the Jacobian peak to be studied, whereas \SumpTZ is the \textit{hard variable} that tests the different scales~\cite{Ebert_2021}.

Measurements of \Zg production have been performed by experiments at LEP~\cite{ZyLEP2004,LEPZZZ2007,OPAL2000}, the Tevatron~\cite{Abazov:2011qp,D0TeV2009}, and the LHC~\cite{STDM-2014-01,STDM-2018-04,CMS-SMP-13-014}. No new physics or deviations from the predictions of the SM have been observed so far. An example of physics beyond the SM is given by a model that includes axion-like particles (ALPs)~\cite{GEORGI198673}, which is particularly relevant for \Zg production; these particles were introduced to solve the strong CP problem and are also considered as a dark-matter candidate~\cite{PRESKILL1983127}. Measurements of \Zg production can help to constrain the ALP's couplings to the $Z$ boson and the photon~\cite{Carra:2021ycg}, which define the most general CP-conserving Lagrangian describing the ALP's bosonic interactions~\cite{Gavela_2020}. Another case where \Zg production can help is in the use of effective field theory~\cite{Brivio_2019}. These models describe different theories beyond the SM that introduce new-physics states at a mass scale $\Lambda$ that is large in comparison with the electroweak scale, using gauge-invariant combinations of SM fields. Previous measurements have not found any evidence of new physics in the \Zg final state. However, none of these measurements included any dedicated study of jet activity. Requiring the presence of jets in addition to the $Z\gamma$ pair, leads to configurations in the final state that enhance a region of the phase space different than that studied in the case of inclusive production. Therefore, measurements of differential cross sections for $Z\gamma+$jets production are expected to provide additional sensitivity to constrain ALPs and other models for physics beyond the SM.
 
This paper presents measurements of differential cross sections as functions of QCD-related observables associated with the $Z\gamma+$jets process. The measurements are performed differentially in either one or two observables. The analysis uses the full dataset of proton--proton ($pp$) collisions at a centre-of-mass energy of $\sqrt{s}=13~\TeV$ recorded by the ATLAS detector during \RunTwo (2015--2018) of the LHC. The results presented here build upon a previous analysis performed by ATLAS~\cite{STDM-2018-04}, which focused on more inclusive observables. The measurements extend the published results by including the hadronic activity associated with the \Zg system and by measuring double-differential cross sections.

\begin{figure}[h!]
\centering
\subfigure[]{\includegraphics[width=0.42\textwidth]{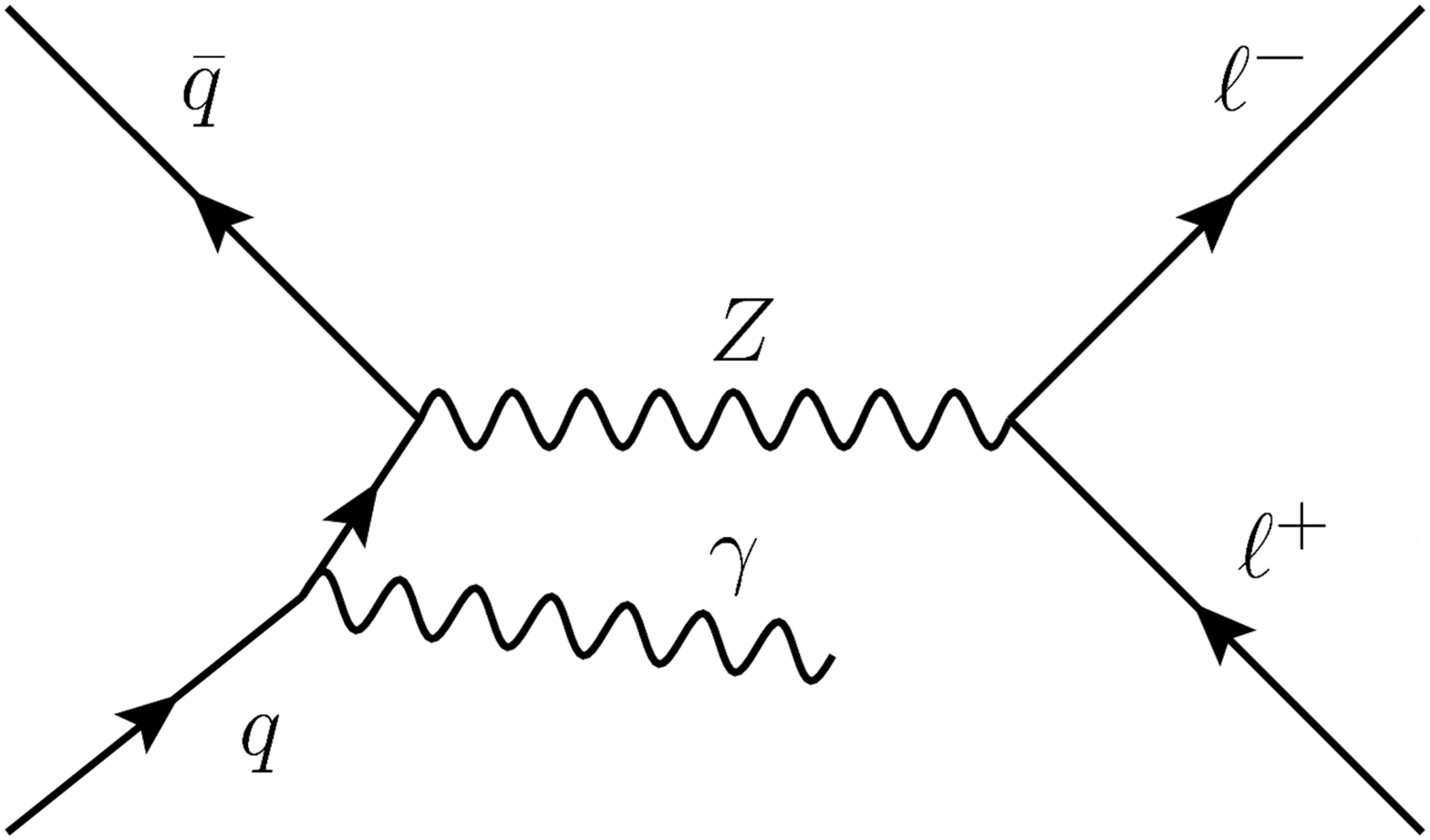}}
\hspace{0.07\textwidth}
\subfigure[]{\includegraphics[width=0.44\textwidth]{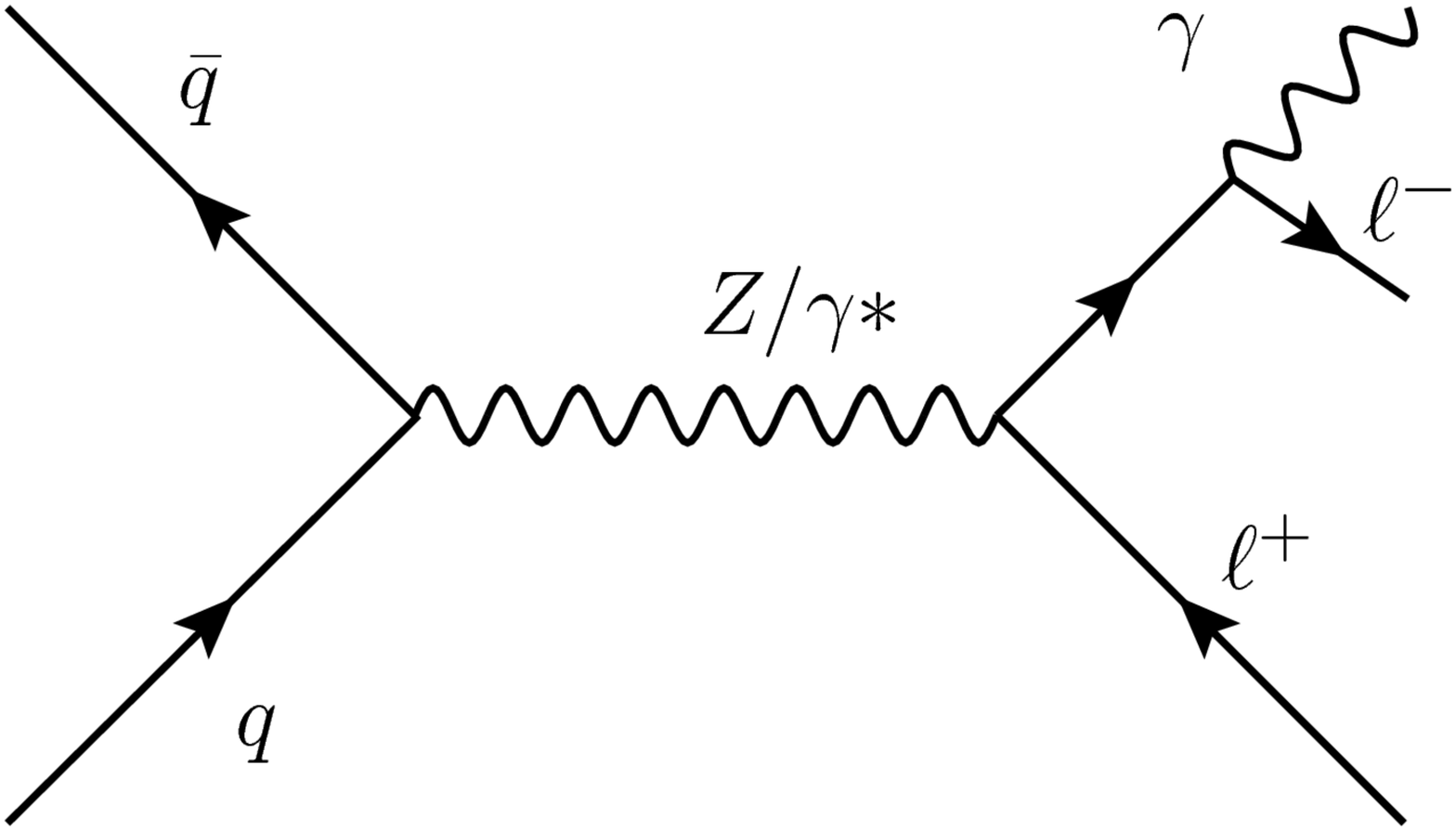}}
\caption{Diagrams for (a) $Z\gamma$ production via the ISR process and (b) $\ell\ell\gamma$ production via the FSR process.}
\label{fig:feynman}
\end{figure}

As in the previous analysis, only $Z$ bosons decaying into pairs of charged leptons ($\ell^+ \ell^-$, with $\ell =e,\mu $) are considered. This restriction makes it easier to fully reconstruct the final state with high resolution, and also provides a relatively large cross section with little background. Events are selected by requiring the invariant mass of the two leptons ($\mll$) to be greater than 40~\GeV, and the sum of the mass of the dilepton system and the mass of the  $\ell\ell\gamma$ system ($\mll+\mlly$) to be greater than 182~\GeV. These selections define a phase space that is enriched in photons from initial-state radiation (ISR), such as shown in Figure~\ref{fig:feynman}(a). In addition, these requirements reduce the contribution from final-state radiation (FSR), where the photons are radiated from the leptons as shown in Figure~\ref{fig:feynman}(b). In the \mll vs \mlly plane (see Figure~2 in Ref.~\cite{STDM-2018-04}) the second requirement forms a diagonal straight line that separates FSR events from ISR events; this is  because the FSR events are expected to lie in the region with  $\mlly $ around the nominal $Z$ boson mass, with $\mll $ at lower values.

The predictions of several MC models for \Zg production, which include multileg matrix elements  interfaced with parton-shower and hadronisation approximations, are compared with the measurements. Several models, which have different levels of precision, are considered: \Sherpa[2.2.4]~\cite{Bothmann:2019yzt} at LO and \Sherpa[2.2.11]~\cite{Bothmann:2019yzt} at next-to-leading order (NLO), \MADGRAPH at NLO~\cite{Alwall:2014hca} and MiNNLO$_\text{PS}$ at NNLO~\cite{MiNNLOPS2020,ZyNNLOPS2021}. The predictions of the fixed-order QCD calculations by MATRIX~\cite{Grazzini:2017mhc,MatrixZy_2015}  at NNLO are also compared with the data.


\section{The ATLAS detector}

The ATLAS experiment~\cite{PERF-2007-01} at the LHC is a multipurpose particle detector
with a forward--backward symmetric cylindrical geometry and a near \(4\pi\) coverage in
solid angle.\footnote{ATLAS uses a right-handed coordinate system with its origin at the nominal interaction point (IP)
in the centre of the detector and the \(z\)-axis along the beam pipe.
The \(x\)-axis points from the IP to the centre of the LHC ring,
and the \(y\)-axis points upwards.
Cylindrical coordinates \((r,\phi)\) are used in the transverse plane,
\(\phi\) being the azimuthal angle around the \(z\)-axis.
The pseudorapidity is defined in terms of the polar angle \(\theta\) as \(\eta = -\ln \tan(\theta/2)\).
Angular distance is measured in units of \(\Delta R \equiv \sqrt{(\Delta y)^{2} + (\Delta\phi)^{2}}\), where $y$ is the rapidity, defined as $y=(1/2)\ln[(E+p_z)/(E-p_z)]$.
}
It consists of an inner tracking detector (ID) surrounded by a thin superconducting solenoid
providing a \SI{2}{\tesla} axial magnetic field, electromagnetic and hadron calorimeters, and a muon spectrometer (MS).
The inner tracking detector covers the pseudorapidity range \(|\eta| < 2.5\).
It consists of silicon pixel, silicon microstrip, and transition radiation tracking detectors.
Lead/liquid-argon (LAr) sampling calorimeters provide electromagnetic (EM) energy measurements
with high granularity.
A steel/scintillator-tile hadron calorimeter covers the central pseudorapidity range (\(|\eta| < 1.7\)).
The endcap and forward regions are instrumented with LAr calorimeters
for both the EM and hadronic energy measurements up to \(|\eta| = 4.9\).
The muon spectrometer surrounds the calorimeters and is based on
three large superconducting air-core toroidal magnets with eight coils each.
The field integral of the toroids ranges between \num{2.0} and \SI{6.0}{\tesla\metre}
across most of the detector.
The muon spectrometer includes a system of precision tracking chambers and fast detectors for triggering.
A two-level trigger system is used to select events.
The first-level trigger is implemented in hardware and uses a subset of the detector information
to accept events at a rate below \SI{100}{\kHz}.
This is followed by a software-based trigger that
reduces the accepted event rate to \SI{1}{\kHz} on average
depending on the data-taking conditions.
An extensive software suite~\cite{ATL-SOFT-PUB-2021-001} is used in data simulation, in the reconstruction and analysis of real
and simulated data, in detector operations, and in the trigger and data acquisition systems of the experiment.


\section{Data and simulated samples}

The data used in this analysis were obtained from $pp$ collisions produced by the LHC in \RunTwo, and after applying the data quality criteria~\cite{DAPR-2018-01}, the total integrated luminosity recorded by the ATLAS detector is 139~\ifb. The uncertainty in the luminosity is $1.7\%$~\cite{ATLAS-CONF-2019-021}, obtained from measurements with the LUCID-2 detector~\cite{LUCID2}.
 
Three different MC samples are used to simulate the $Z\gamma+$jets process. The nominal sample was generated using the program \Sherpa[2.2.11] to calculate matrix elements with up to one additional parton at NLO and up to three additional partons at LO. The matrix element calculation includes all diagrams at order $\alpha_{\textrm{EW}}^2$, where $\alpha_{\textrm{EW}}$ is the electroweak coupling constant. The merging of the matrix element and parton shower (PS) was performed with  \MEPSatLO~\cite{Hoeche:2011fd,Hoeche:2012yf,Catani:2001cc,Hoeche:2009rj}.
The \NNPDF[3.0nnlo]~\cite{PDFNNLO2015} PDF set was used, with an additional set of tuned PS parameters developed by the \Sherpa authors~\cite{Bothmann:2019yzt}. Frixione isolation~\cite{Frixione:1998jh} was applied to the photon with the parameter choices $ \delta_0 = 0.1$, $\epsilon =0.1$ and $n=2$. This sample requires the transverse momentum of the photon ($\pT^{\gamma}$) to be greater than 7~\GeV. Throughout the paper the signal estimate refers to this sample, unless it is otherwise specified.

A second sample was produced using the program \Sherpa[2.2.4], with matrix elements at LO accuracy in QCD for up to three additional parton emissions matched and merged with the \SHERPA parton shower based on Catani--Seymour dipole factorisation~\cite{Gleisberg:2008fv,Schumann:2007mg} using the \MEPSatLO prescription~\cite{Hoeche:2011fd,Hoeche:2012yf,Catani:2001cc,Hoeche:2009rj}. The matrix element calculation includes all diagrams at order $\alpha_{\textrm{EW}}^2$. Samples were generated using the \NNPDF[3.0nnlo] PDF set~\cite{Ball:2014uwa}, along with the dedicated set of tuned parton-shower parameters developed by the \SHERPA authors. In the generation, \pTy is required to be larger than $7~\GeV$. Frixione isolation was also applied to this sample, with the same parameter values as in the \Sherpa[2.2.11] sample.
 
A third sample was generated using the program \MGNLO[2.3.3]~\cite{Alwall:2014hca} to calculate NLO matrix elements with up to one extra parton, using the  $\NNPDF[3.0nlo]\_$as\_0118 PDF set~\cite{Ball:2014uwa}. The matrix element calculation includes all diagrams at order $\alpha_{\textrm{EW}}^2$. This sample has the same Frixione isolation parameter values as in the \Sherpa samples.
 
Simulated samples of the purely electroweak production of \Zg in association with two jets are used at detector level. These samples were generated using the program \MGNLO[2.6.5]~\cite{Alwall:2014hca} at LO accuracy in $\alpha_{\textrm{EW}}^4$, using the  \NNPDF[3.0lo] PDF set~\cite{Ball:2014uwa}. It was interfaced with  \PYTHIA[8.240]~\cite{Sjostrand:2007gs} for parton showering, hadronisation,  and the underlying event.
 
The main backgrounds to the signal arise from $Z$ bosons produced in association with jets, from top-quark pairs and single top quarks produced in association with photons, from diboson and triboson events, and from multiple $pp$ interactions; the last of these are called pile-up events in the following.
 
The $Z$ boson in association with jets (\Zjets) background is estimated with a data-driven method  with signal and control regions (see Section \ref{ssec:Zjets}); still, MC samples of \Zjets events are used to account for signal leakage into the control regions. The \Zjets MC samples used in this analysis were generated using the \POWHEGBOX[v1] MC generator~\cite{Nason:2004rx,Frixione:2007vw,Alioli:2010xd,Alioli:2008gx}, with NLO accuracy for the hard-scattering processes. It was interfaced to \Pythia[8.186]~\cite{Sjostrand:2007gs} for the modelling of the PS, hadronisation, and underlying events, with parameters values set according to the \AZNLO tune~\cite{STDM-2012-23}. The \CT[10nlo] PDF set~\cite{Lai:2010vv} was used for the hard-scattering processes, whereas the \CTEQ[6L1] PDF set~\cite{Pumplin:2002vw} was used for the PS. This background is normalised to the cross section given by the generator.

The MC samples of the production of one or two top quarks and a photon (\ttg and $tW\gamma$) were generated using the program \MGNLO[2.3.3]~\cite{Alwall:2014hca} at LO with the \NNPDF[2.3lo]~\cite{Ball:2014uwa} PDF set. The events were interfaced with \Pythia[8.212]~\cite{Sjostrand:2014zea} using the ATLAS A14 tune~\cite{ATL-PHYS-PUB-2014-021} and the \NNPDF[2.3lo] PDF set. These samples are normalised to their NLO cross section~\cite{ttyCalc2011,deFlorian:2016spz}.

The MC samples of diboson processes, such as \ZZllll and \WZllln, were generated with \Sherpa[2.2.2] with matrix elements at NLO accuracy in QCD for up to one additional parton and at LO accuracy for up to three additional parton emissions. The matrix element calculations were matched and merged with the \SHERPA parton shower based on Catani--Seymour dipole factorisation~\cite{Gleisberg:2008fv,Schumann:2007mg} using the MEPS@NLO prescription~\cite{Hoeche:2011fd,Hoeche:2012yf,Catani:2001cc,Hoeche:2009rj}. The virtual QCD corrections were provided by the \OPENLOOPS library~\cite{Buccioni:2019sur,Cascioli:2011va,Denner:2016kdg}. The \NNPDF[3.0nnlo] set of PDFs was used~\cite{Ball:2014uwa}, along with the dedicated set of tuned parton-shower parameters developed by the \SHERPA authors.

The \WZg and \WWg processes constitute a small background and were simulated with \Sherpa[2.2.11] at NLO with zero jets, using the \NNPDF[3.0nnlo] PDF set. The multiboson background samples are normalised to the cross section given by the generator.

For all these MC samples, pile-up from additional $pp$ collisions in the same and neighbouring bunch crossings was simulated by overlaying each MC event with a variable number of simulated inelastic $pp$ collisions generated using \PYTHIA[8.186] with the ATLAS set of tuned parameters for minimum-bias events (the A3 tune)~\cite{ATL-PHYS-PUB-2016-017}. The MC events are weighted (\enquote{pile-up reweighting}) to reproduce the distribution of the average number of interactions per bunch crossing observed in the data.
 
All the samples of generated events were passed through the $\GEANT$-based~\cite{Agostinelli:2002hh} ATLAS detector- and trigger-simulation programs~\cite{SOFT-2010-01}. They are reconstructed and analysed by the same program chain as the data. Table~\ref{tab:SamplesUsed} gives an overview of the generators used in this analysis, their precision in QCD, and the PDF set used.
 
\begin{table}[tbp]
\centering
\caption{Summary of MC samples used in the analysis.}
\label{tab:SamplesUsed}
{\scriptsize
\begin{tabular}{ lcccc }
\toprule
Process & Generator & Order & PDF set &   PS/UE/MPI \\
\midrule
\midrule
$\Zg+$jets  & \Sherpa[2.2.11] & 0,1j@NLO + 2,3,4j@LO & \NNPDF[3.0nnlo] &  \Sherpa[2.2.11] \\
$\Zg+$jets  & \Sherpa[2.2.4]  & 0,1,2,3j@LO & \NNPDF[3.0nnlo] &  \Sherpa[2.2.4] \\
$\Zg+$jets & \MGNLO      & 0,1j@NLO & $\NNPDF[3.0nlo]\_$as\_0118  & \Pythia[8.212] \\
Purely EW $\Zg jj$ &  \MGNLO  & LO & \NNPDF[3.0lo] & \PYTHIA[8.240] \\
\midrule
\Zjets & \POWHEGBOX & 0j@NLO  &  \CT[10nlo] & \Pythia[8.186] \\
\ttg, $tW\gamma$   & \MGNLO & LO & \NNPDF[2.3lo] & \Pythia[8.212] \\
\ZZllll, \WZllln & \Sherpa[2.2.2] & 0,1j@NLO + 2,3j@LO & \NNPDF[3.0nnlo] & \Sherpa[2.2.2]\\
\WZg, \WWg & \Sherpa[2.2.11] &  0j@NLO + 1,2j@LO & \NNPDF[3.0nnlo] & \Sherpa[2.2.11] \\
\bottomrule
\end{tabular}
}
\end{table}

Additional signal theory predictions at NNLO matched to the parton shower~\cite{ZyNNLOPS2021} were calculated using \POWHEGBOX~\cite{PowhegBox2010}. They provide a prediction consistently matched to the parton shower, including spin correlation, intereference and off-shell effects, using the MiNNLO$_\text{PS}$~\cite{MiNNLOPS2020}  approach. These predictions are compared with the data in Section~\ref{sec:result}. Photon infrared-safe predictions were obtained by imposing the same Frixione isolation parameter values as for the \Sherpa[2.2.11] sample. The central renormalisation and factorisation scales were set to $\mZy$. Events were generated using the \NNPDF[3.0nnlo] PDF set with the strong coupling constant taken as $\alphas(m_Z)=0.118$.

Fixed-order QCD calculations are also considered by using the program MATRIX~\cite{Grazzini:2017mhc,MatrixZy_2015}, which relies on \OPENLOOPS~\cite{Cascioli:2011va} for all amplitudes up to the one-loop level, and on other dedicated calculations~\cite{Gehrmann:2011ab,Denner:2016kdg,Catani:2012qa,Catani:2007vq}. The MATRIX  predictions are obtained with \CT[14nnlo] PDFs~\cite{Dulat:2015mca}. Frixione isolation parameter values are the same as for the \Sherpa[2.2.11] sample.
 
The measurements presented in this paper are obtained at particle level, where leptons are corrected for collinear photon radiation by adding to the four-momentum of the lepton the four-momenta of the photons within $\Delta R=0.1$ of given lepton. This procedure is called dressing. The MATRIX predictions are obtained instead at parton level, since no QED radiation is included in these calculations. For this reason, corrections are applied to the fixed-order calculations. These corrections are calculated by comparing the predictions from \Sherpa[2.2.11] MC before and after applying the dressing. In addition, MATRIX does not include non-perturbative effects, corrections are applied to allow comparisons with the measurements. These corrections are obtained by calculating the ratio of the MC cross sections with and without hadronisation. The MC sample used for these corrections was generated with  \MGNLO with up to one extra parton at NLO in the matrix element, and then interfaced with \PYTHIA[8]~\cite{Sjostrand:2014zea}, using the A14 tune~\cite{ATL-PHYS-PUB-2014-021}.


\section{Event selection}
\label{sec:evsel}

Selected events must have at least one reconstructed vertex with at least two associated tracks with $\pT>500~\MeV$. In events with multiple vertices, the one with the highest $\sum \pT^2$ of associated tracks is selected as the primary vertex. Candidate events must also pass at least one unprescaled single-muon or single-electron trigger~\cite{TRIG-2018-05,TRIG-2018-01}. For data recorded in 2015, the lowest \pT threshold was 24~\GeV\ for the electron trigger, and 20~\GeV\ for the muon trigger. For data recorded during the period 2016$-$2018, these thresholds were both raised to 26~\GeV\ and tighter isolation criteria were applied, to compensate for the increase in instantaneous luminosity. Triggers with a higher \pT threshold, but looser isolation, are also used because they increase the total trigger efficiency. The trigger efficiency for events satisfying all the selection criteria is about 99\%. This is estimated using signal simulated samples.

\subsection{Lepton, photon, and jet selections}
Photons and electrons are reconstructed from energy clusters in the electromagnetic calorimeter (ECAL). Electron candidates are required to have a matching track in the ID. Photon candidates must have  $|\eta|<2.37$ and $\pT>30~\GeV$, while electron candidates must have $|\eta|<2.47$ and $\pT>25~\GeV$. Both electron and photon candidates are rejected if they lie in the transition region between the barrel and endcaps of the ECAL ($1.37<|\eta|<1.52$). Electrons are identified using a likelihood function based on shower shape variables in the ECAL, track variables, and the quality of the track--cluster matching. Electrons are required to satisfy the \textit{Medium} criteria, as described in Ref.~\cite{EGAM-2018-01}. Photons are identified using shower shape variables in the ECAL and are required to satisfy the \textit{Tight} criteria~\cite{EGAM-2018-01}. Photons are classified as \textit{converted} to electron--positron pairs if the ECAL cluster is matched to a conversion vertex formed by the tracks of oppositely charged particles, or by a single track consistent with having originated from a photon conversion. Photon candidates are classified as \textit{unconverted} if it is not possible to match clusters to tracks. Both types of photons are used in this analysis, and the distinction between converted and unconverted photons has no impact on the result. The photon and electron energy scale is calibrated using $Z\rightarrow ee$ events, as described in Ref.~\cite{EGAM-2018-01}.

Muons are reconstructed by matching the tracks in the MS with tracks in the ID. The momentum is obtained by combining the MS measurement, corrected for the energy deposited in the calorimeter, and the measurement in the ID. Muon candidates are also required to satisfy the \textit{Medium} identification criterion, as described in Ref.~\cite{MUON-2018-03}. This criterion is based on the number of hits matched to the muon's tracks reconstructed in the ID and the MS, and on the compatibility of the ID and MS measurements of the muon's transverse momentum. Muon candidates are required to have $|\eta|<2.5$ and $\pT>25~\GeV$.
 
Electrons and muons must be compatible with originating from the primary vertex. This requirement is fulfilled by requiring that the transverse impact parameter ($d_0$) relative to the beam-spot divided by its uncertainty ($\sigma(d_0)$), i.e.\ the significance,  satisfy $|d_0/\sigma(d_0)| < 5$ for electrons and $|d_0/\sigma(d_0)| < 3$ for muons. Additionally, for both electrons and muons, the longitudinal impact parameter ($z_0$), i.e.\ the $z$-distance from the primary vertex to the point where $d_0$ is measured, must satisfy $|z_0 \sin \theta| < 0.5$~mm.
 
Leptons and photons are required to be isolated, i.e.\ without additional activity in their proximity. Isolation requirements are based on tracking information and calorimeter energy clusters. The isolation variable $\pT^{\textrm{iso}}$ is computed as the $\sum \pT$ of nearby tracks with $\pT>1~\GeV$, excluding tracks associated with the lepton or photon candidate. The variable $E_{\textrm{T}}^{\textrm{iso}}$ is obtained as the scalar sum of the transverse energies of nearby topological clusters ~\cite{TopoClust2017}, corrected for the energy deposited by the photon or lepton candidate itself and the contribution from the underlying event and pile-up~\cite{STDM-2014-09,PERF-2017-02}.
 
Photons must satisfy an isolation criterion, as described in Ref.~\cite{EGAM-2018-01}, with $\pT^{\textrm{iso}}/E_{\textrm{T}} < 0.05$ and $E_{\textrm{T}}^{\textrm{iso}}/E_{\textrm{T}} < 0.065$ in a cone of size $\Delta R=0.2$ around the photon candidate, where $E_{\textrm{T}}$ is the transverse energy of the photon. Electrons must satisfy $\pT^{\textrm{iso}}/\pT < 0.15$ in a cone of $\pT$-dependent size up to $\Delta R=0.2$ around the electron candidate, and $E_{\textrm{T}}^{\textrm{iso}}/\pT < 0.2$ in a cone of size $\Delta R=0.2$.
Muon isolation~\cite{MUON-2018-03} requires $\pT^{\textrm{iso}}/\pT < 0.15$ in a cone of $\pT$-dependent size up to $\Delta R=0.3$ ($\Delta R=0.2$) for muons with \pT less (greater) than 50~\GeV, and $E_{\textrm{T}}^{\textrm{iso}}/\pT < 0.3$ in a cone of fixed size $\Delta R=0.2$.
 
Jets are reconstructed with the anti-$k_t$ algorithm~\cite{Cacciari:2008gp,Fastjet} with a radius parameter of $R=0.4$, using a particle-flow~\cite{PERF-2015-09} procedure, with clusters of energy deposited in the calorimeter as inputs. Jets are calibrated and their energy is corrected to account for detector effects, using methods based on MC and in-situ techniques~\cite{JETM-2018-05}. Jets with $ \pT < 60~\GeV$ and $|\eta| < 2.4$ are removed if they are identified as pile-up jets by the jet vertex tagger (JVT)~\cite{PERF-2014-03}. Jets are required to have $ \pT > 30~\GeV$ if $|\eta| < 2.5$, or $ \pT > 50~\GeV$ if $|\eta| > 2.5$, to further suppress pile-up. Distributions with jets require at least one jet, unless it is otherwise explicitly stated.
 
Ambiguities in the identity of reconstructed leptons, jets, and photons are resolved with an overlap-removal procedure. First, jets are removed if they are within $\Delta R = 0.4$ of a photon, or within $\Delta R = 0.2$ of an electron. Then leptons are removed if they are within  $\Delta R = 0.4$ of a jet, while photons are removed if they are within  $\Delta R = 0.4$ of a lepton. Finally, electrons are removed if they are within  $\Delta R = 0.2$ of a muon.

\subsection{Signal region and control region definitions}
\label{ssec:SRCRdef}
The signal region (SR) is defined by events with at least two opposite-sign (OS) same-flavour (SF) leptons and a photon. The leading lepton (with the highest transverse momentum) is required to have $\pT>30~\GeV$. Events must also have at least one photon with $\pTy>30~\GeV$. Events are further selected by requiring $\mll> 40~\GeV$, to avoid low-mass resonances. As mentioned in Section~\ref{sec:intro}, FSR events are suppressed by requiring that the sum of the invariant mass of the leptons and the invariant mass of the leptons and the photon is greater than twice the mass of the $Z$ boson, i.e.\ $\mll+\mlly>182~\GeV$. In the \mll vs \mlly plane, this requirement is a diagonal straight line that separates FSR and ISR events since FSR events are expected to lie in the region with $\mlly \sim 90$~\GeV\ and $\mll < 90$~\GeV\ (see Figure~2 in Ref.~\cite{STDM-2018-04}).
 
The \ttg background modelling is checked in a dedicated control region ($\ttg$-CR) obtained by applying all of the SR requirements except the SF-lepton requirement, which is replaced by a requirement of different-flavour (DF) leptons. The signature in the $\ttg$-CR is then $e\mu\gamma$. Table~\ref{tab:EveSel} shows a summary of these selection criteria.

\begin{table}[tbph]
\centering
\caption{Summary of the selection criteria used in this analysis.}
\label{tab:EveSel}
\begin{tabular}{l cc}
\toprule
Observable & Signal Region & \ttg Control Region \\
\midrule
Number of signal leptons & $\geq$ 2 opposite sign, same flavour & $\geq$ 2 opposite sign, different flavour  \\
Lepton &  \multicolumn{2}{c}{$\pT(\ell_1)>30$~\GeV, $\pT(\ell_2)>25$~\GeV} \\
Photon &  \multicolumn{2}{c}{$\geq$ 1 photon with $\pTy > 30~\GeV$} \\
\mll &  \multicolumn{2}{c}{$> 40~\GeV$} \\
$\mll+\mlly$ &  \multicolumn{2}{c}{$> 182~\GeV$} \\
\bottomrule
\end{tabular}
\end{table}


\section{Measured observables}

Differential cross sections are measured for the following one-dimensional observables:
\begin{itemize}
\item[•] \njet, the number of jets
\item[•] \pTjetOne (\pTjetTwo), the transverse momentum of the leading jet (subleading jet)
\item[•] \Rptj, the ratio of the \pT of the subleading jet and leading jet
\item[•] \mllyj, the invariant mass of the lepton-pair--photon--leading-jet system
\item[•] \mjj, the invariant mass of the two leading jets
\item[•] \HT, the scalar sum of the \pT of all jets, leptons, and photons
\item[•] \ptysqrtHT, the ratio of the \pT of the photon to the square root of \HT
\item[•] \DPhiJetY, the azimuthal angle between the leading jet and the leading photon
\item[•] \DRll, the angular distance  $\Delta R$ between the two leptons, measured in units of \(\Delta R \equiv \sqrt{(\Delta\eta)^{2} + (\Delta\phi)^{2}}\).
\item[•] \pTZ, the transverse momentum of the two-lepton system
\item[•] \DiffpT, the difference between the transverse momenta of the $\ell\ell$ system (\pTZ) and the photon (\pTy)
\item[•] \SumpT, the scalar sum of the transverse momenta of the $\ell\ell$ system and the photon
\item[•] \pTZyj, the transverse momentum of the $\ell\ell\gamma j$ system.
\end{itemize}

The QCD-sensitive 2D observables measured in this paper are:
\begin{itemize}
\item[•]  The \textit{resolution  variable} \RZy, the ratio of the transverse momentum of the $\ell\ell\gamma$ system to its mass, is measured in bins of the \textit{hard  variable} \mZy
\item[•] The \textit{resolution  variable} \DiffpT is measured in three different bins of the \textit{hard  variable} \SumpT
\item[•] The \textit{resolution  variable} \pTZyj  is measured in bins of the \textit{hard variable} \pTZy
\end{itemize}
 
Additionally, 2D observables sensitive to polarisation effects of the $Z$ boson are considered~\cite{Stirling_2012}:
\begin{itemize}
\item[•] $\cos\theta_\mathrm{CS}$, the cosine of the angle between the negatively charged lepton and the lepton pair in the  Collins--Soper (CS) frame~\cite{Collins:1977iv} in bins of \pTZ
\item[•] $\phi_\mathrm{CS}$, the azimuthal angle between the negatively charged lepton and the lepton pair in the CS frame in bins of \pTZ.
\end{itemize}
The polarisation-sensitive observables are estimated in the Collins--Soper frame, where the $Z$~boson is at rest; the CS frame is commonly used when extracting angular coefficients of the $Z$~boson~\cite{CMS-SMP-13-010,STDM-2018-03}.
 
For the 2D distributions, computational complications in the unfolding of a 2D distribution are avoided by unfolding the resolution observable in wide bins of the hard observable. These wide bins are chosen such that the migration effects in the hard observable are negligible.


\section{Background estimation}
\label{sec:bkgest}

The main background to the $Z\gamma+$jets signal arises from \Zjets events, in which one of the jets is misidentified as a photon. This background is estimated using a data-driven method. Pile-up events, in which the leptons and the photon originate from two different $pp$ interactions during the same bunch crossing, also constitute a background and are estimated using a data-driven method. Another large background, especially at high jet multiplicity,  is \ttg production, where the top-quark decays can also produce same-flavour leptons. The \ttg background is estimated using MC samples normalised to data in the dedicated $\ttg$-CR defined in Section~\ref{ssec:emuyCR}. Other small backgrounds that can also produce the same signature as the signal, such as triboson events from \WWg, \WZg, and $ZZ\gamma$ production, are estimated using MC samples. The background from diboson events, such as $WZ(\rightarrow \ell\nu\ell\ell)$ and $ZZ(\rightarrow \ell\ell\ell\ell)$, where one electron is misidentified as a photon is also taken into account using MC samples.

\subsection{$\Zjets$ background}
\label{ssec:Zjets}
A two-dimensional sideband method~\cite{STDM-2014-09}, similar to the one in Ref.~\cite{STDM-2018-04}, is used to estimate the background in each bin of each distribution. In addition to the SR and the $\ttg$-CR, three $\Zjets$-CRs are created to estimate this background by inverting the isolation and/or identification criteria for the photon. Photons that fail to satisfy the \textit{Tight} identification criteria must still satisfy a loose identification criterion, where the requirements on four of the EM calorimeter shower shape variables are removed, as described in Ref.~\cite{STDM-2010-08}.
The photon isolation is modified such that only the calorimeter-based component is considered, while the track-based isolation is applied in all regions. Photon candidates fail to satisfy the isolation criteria when $E_{\textrm{T}}^{\textrm{iso}} >0.065\times E_{\textrm{T}}^{\gamma} + E_{\textrm{T}}^{\textrm{gap}}$, where $E_{\textrm{T}}^{\textrm{gap}}$ is an energy gap set to $E_{\textrm{T}}^{\textrm{gap}}=2~\GeV$ and helps to reduce the number of $\Zg +$jets signal events leaking into the $\Zjets$-CRs (signal leakage).
 
The $\Zjets$-CRs described above are dominated by \Zjets events. The leakage of signal events into the $\Zjets$-CRs is removed via the signal leakage fractions estimated using the MC simulation. These factors are inclusively about 6\% (1.4\%) for the control region with modified identification (isolation) criteria, and less than 0.2\% for the control region where both the identification and isolation criteria are modified. Backgrounds from other processes are subtracted using the MC simulation of each process. The purity in the CR is $0.90 \pm 0.02$, with values varying from 0.86 to 0.92, depending on the exact bin. The yields of \Zjets events in the SR can then be derived from the number of events in the SR and in the three $\Zjets$-CRs, using the formulas described in Ref.~\cite{STDM-2014-09}.

Possible correlations between the isolation and identification variables are estimated with \Zjets MC samples with the use of a correlation factor $R$, which is the ratio of the fraction of \Zjets events satisfying the photon isolation requirement $E_{\textrm{T}}^{\textrm{iso}} < 0.065\times E_{\textrm{T}}^{\gamma}$ in events satisfying the identification criteria, to those not satisfying the identification criteria. In the absence of correlation, this ratio is equal to unity. To preserve the correlation and reduce the statistical uncertainties, $R$ is computed in larger bins than those used in the sideband method or integrated, depending on the observable. Results  of the \Zjets estimation with larger intervals for the correlation computation are compatible within uncertainties with the results with finer binning, but with reduced systematic and statistical uncertainties.

The uncertainty in the correlation factor $R$ is obtained by varying the definition of the $\Zjets$-CRs in both data and MC simulation. The $E_{\textrm{T}}^{\textrm{gap}}$ requirement is varied by $\pm 1$~\GeV\ and different loose identification criteria are used, for which three or five of the EM calorimeter shower shape variables are removed from the \textit{Tight}  criteria instead of four. In the inclusive phase space, the correlation factor is $R = 1.30 \pm 0.04 \text{ (stat.)} \pm 0.23  \text{ (syst.)} $, estimated using the MC samples, as mentioned above. A cross-check of this estimate is performed by computing $R$ in a $\Zjets$-CR where photons also fail the track isolation, and in this CR the estimate is  $R = 1.29 \pm 0.02 \text{ (stat.)}$, in agreement with the nominal estimate. Another source of uncertainty arises from the estimation of the $\Zg+$jets signal leakage into the $\Zjets$-CRs. The signal leakage factors are computed using \Sherpa[2.2.11] and are found to be small; the uncertainty is estimated by using \MADGRAPH instead of \Sherpa. Additional uncertainties in $R$ arise from the subtraction of other backgrounds (such as diboson events, or \ttg). For these, the uncertainties in the cross sections are propagated to the final \Zjets estimate. The total uncertainty (including statistical uncertainties) in the integrated \Zjets estimate is $22\%$ and is dominated by the uncertainty of the data-driven method.

\subsection{Pile-up background}
The selected photons may originate from different $pp$ collisions in the same bunch crossing because photons do not have requirements on the longitudinal position of their origin ($z_{\gamma}$) with respect to the primary vertex, since it is not a well-measured quantity. The reconstructed photon $z_{\gamma}$ is determined by using a weighted mean of the intersections of the directions obtained from the electromagnetic clusters by taking into account the longitudinal segmentation of the calorimeter, with a constraint from the beam-spot position, and has a typical resolution of 15~mm.

This background is estimated, using a method similar to the one described in Ref.~\cite{STDM-2018-04}, by evaluating the fraction of pile-up events in data (\fpu) after the $\ell\ell\gamma$ selection, and it is briefly described here. To select photons with a better position resolution $\sigma (z_{\gamma})$, only photons that converted to electron--positron pairs with two tracks in the pixel detector are considered. Additionally, the radial conversion position of the photons must be between 5~mm from the beam-spot (outside the beam pipe) and 125~mm (before the end of the pixel detector). The \fpu of converted photons is assumed to be the same as for unconverted ones. This assumption is checked using a sample of signal MC events. In this sample, the fraction of events with \enquote{MC truth}-matched photons is the same for events with and without photon conversion. This is expected since the two effects (i.e.\ pile-up fraction and conversion fraction) should not be correlated.

The primary vertex position $z_{\textrm{vtx}}$ has a Gaussian distribution with a measured width of $\sigma(z_{\textrm{vtx}})\sim 35$~mm~\cite{STDM-2018-04}, corresponding to the width of the luminous region. The fraction \fpu can then be written as:

\begin{equation*}
\fpu = \frac{1}{N_{\textrm{data}}} \cdot \frac{N_{\textrm{data}}^{\textrm{PU}} -N_{ \textrm{MC} }^{\textrm{PU}}}{ P_{\textrm{PU}}},
\end{equation*}
where  $ N_{\textrm{data}(\textrm{MC})}^{\textrm{PU}}$ is the number of data (MC) events in a region dominated by pile-up, defined as the region with $|\Dz|=|z_{\textrm{vtx}}-z_\gamma| >50$~mm. Since the pile-up events are Gaussian-distributed with a width $\sigma(z_{\textrm{vtx}}-z_\gamma)=\sqrt{2} \sigma(z_{\textrm{vtx}})\sim 50$~mm, the probability of observing events with $|\Dz|>50\,\textrm{mm}$ is estimated to be $P_{\textrm{PU}}=0.32$. The term $N_{\textrm{MC} }$ describes the MC events where the $Z$ boson and the photon come from the same $pp$ collision, and is taken from signal MC simulation. The MC sample is normalised to the data with $|\Delta z|<5$~mm. The $|z_{\textrm{vtx}}-z_\gamma|$  distribution is shown in Figure~3 of Ref.~\cite{STDM-2018-04}. To have a better description of the pile-up events in the differential observables, \fpu is  computed  as a function of \njet and \pTy. The estimated \fpu varies from 0.02 to 0.08.

The procedure described above gives the total fraction of pile-up events in bins of  \pTy and \njet, while the shape for the other distributions is taken from the MC samples at particle level, as described in the following. A sample is built by adding together a generated single-photon sample and a generated \Zjets sample. Only jets from the \Zjets sample are considered, since in data and MC events the jets are required to be matched to the vertex with the highest $\sum\pT^{2}$ of associated tracks through the JVT requirement, which is likely to reject the jets produced in association with the photon.
 
The difference between the nominal particle-level sample and a pile-up enriched sample is assigned as an uncertainty. This additional pile-up enriched sample is selected from the data, by selecting only events where $z_{\gamma}$ is closer to the vertex with the second highest $\sum\pT^{2}$ of associated tracks than to the primary vertex. By definition, these events will be pile-up-like events. Additionally, only in observables that depend on jets, the difference between this particle-level distribution and the one obtained by considering all the jets is added as a further uncertainty.

\subsection{Other backgrounds}
\label{ssec:emuyCR}

Background contributions from \ttg, triboson, and diboson events are estimated with simulated samples. Since the \ttg process is about four times larger than all other backgrounds, the modelling is estimated using the dedicated \ttg-CR defined in Section~\ref{ssec:SRCRdef}. The \ttg MC sample is scaled by a normalisation factor of 1.44, and a relative uncertainty of $15\%$ is assigned to this normalisation~\cite{TOPQ-2017-14}. 
 
Figure~\ref{fig:emuyDataMC1} shows a comparison between data and MC events in the \ttg-CR as functions of \njet and \ptysqrtHT. The \Zjets  estimate is obtained using the same method as previously described in Section~\ref{ssec:Zjets}, but using $e\mu\gamma$ events instead of $ee\gamma/\mu\mu\gamma$ events. The correlation factor in the \Zjets background estimation is fixed to $R=1.30\pm 0.04 (\textrm{stat.})$, as explained in Section~\ref{ssec:Zjets}. Another background also present in this region is from diboson events where one lepton is misidentified as a photon ($WZ\rightarrow \ell\nu\ell\ell$). A $30\%$ uncertainty is assigned to this background, which accounts for uncertainties in the inclusive cross-sections due to possible higher-order contributions. Good agreement is seen between data and MC events in the \ttg-CR. The largest discrepancy can be seen in the 0-jet bin. Such mismodelling has negligible impact on the analysis since the contribution of \ttg and diboson processes for events with no jets in the SR is more than one order of magnitude smaller than the signal.

\begin{figure}[h!]
\centering
 
\subfigure[]{\includegraphics[width=.4\columnwidth]{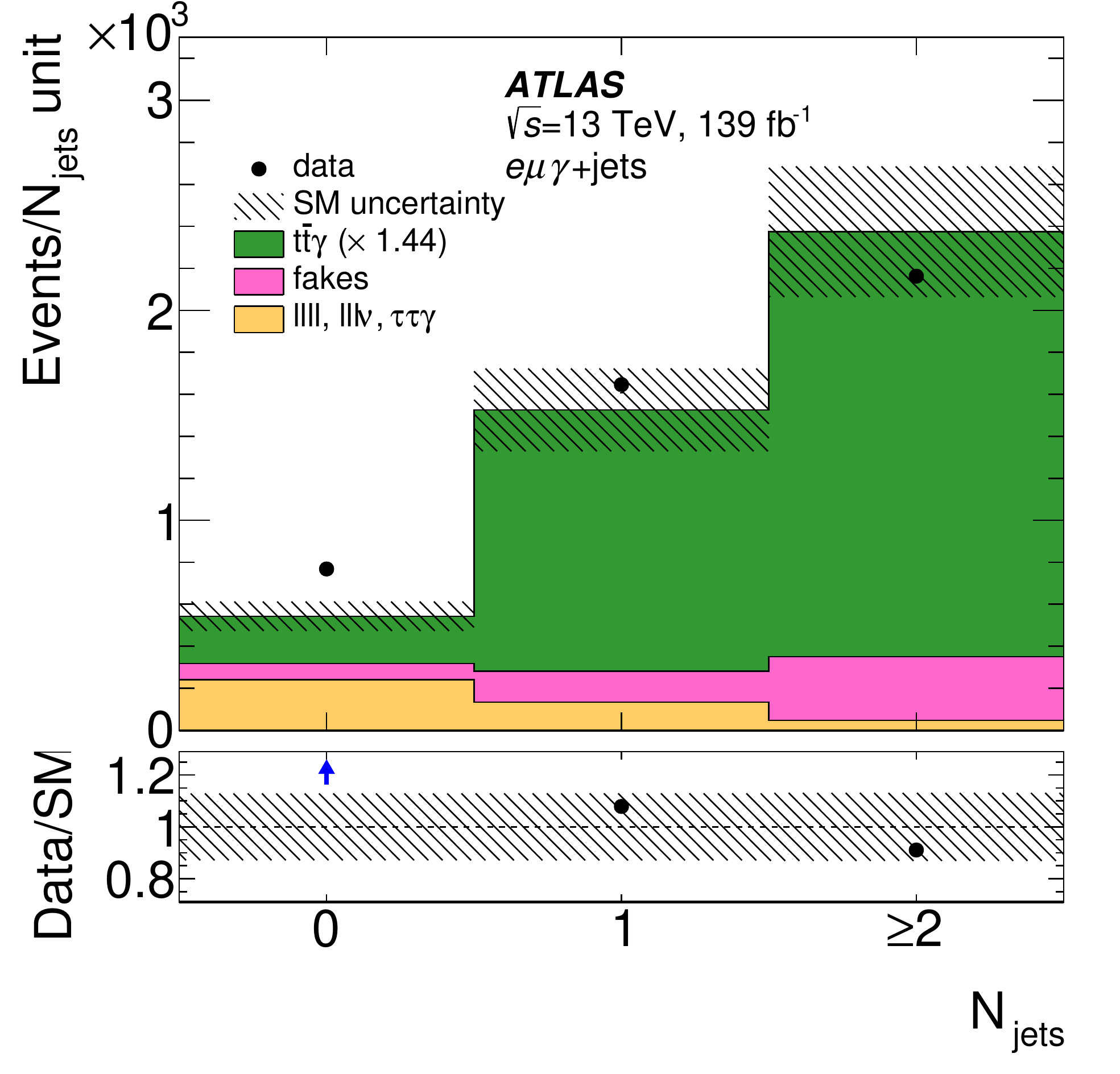}}
\subfigure[]{\includegraphics[width=.4\columnwidth]{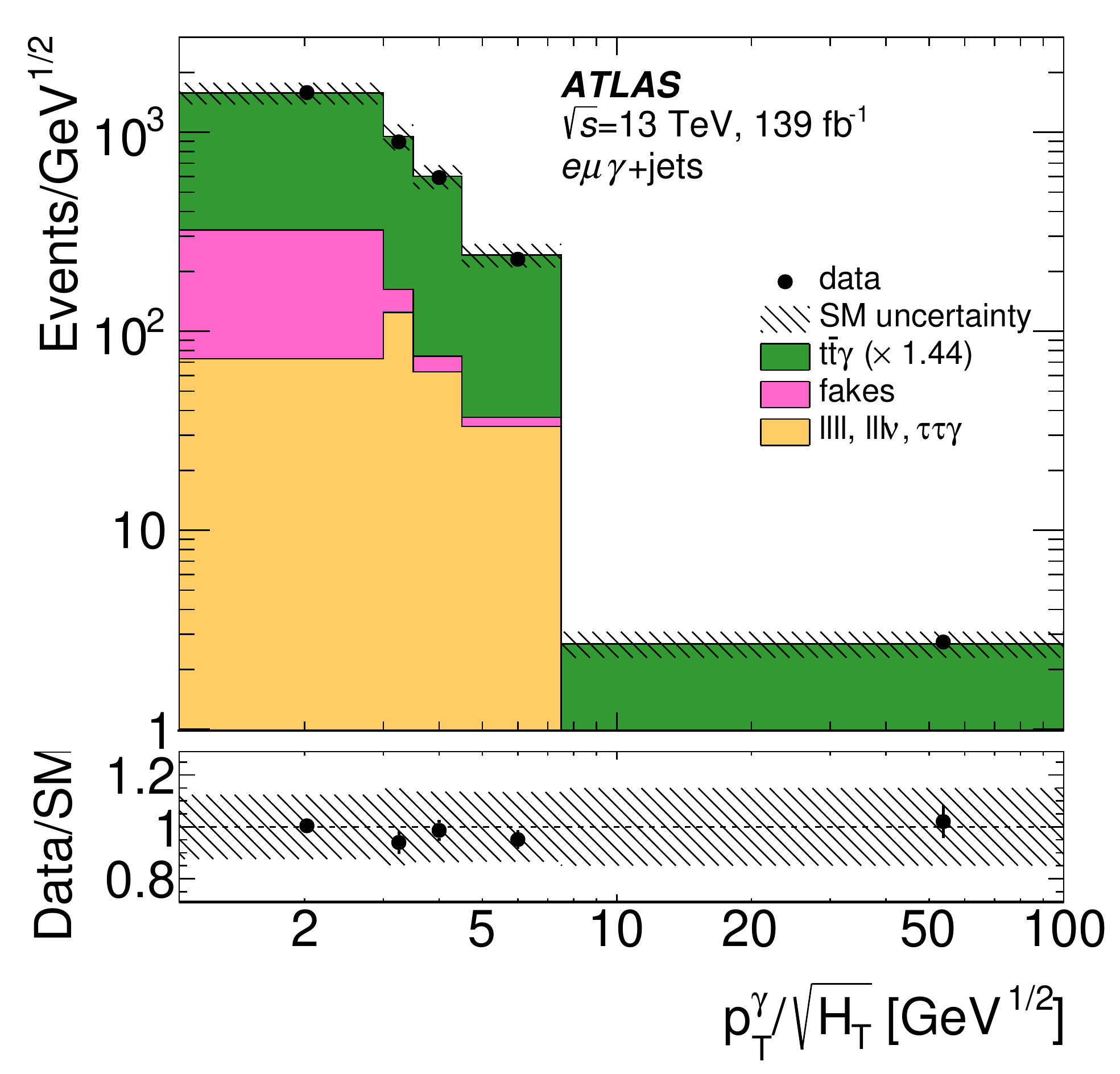}}
\caption{Data (black dots) in the \ttg-CR compared with scaled simulated \ttg events, simulated diboson events and fake photons estimated with a sideband method. The bottom panel shows the ratio of observed data events to the sum of the estimates. The simulated signal and background distributions are stacked to produce the figures.  The hatched band represents the statistical and systematic uncertainties of the SM background yields added in quadrature.}
\label{fig:emuyDataMC1}
\end{figure}
The other backgrounds (tribosons and dibosons) contribute around $1\%$ of the total expected yield in the SR. For this reason, they are estimated directly from MC simulation. Other even smaller backgrounds (such as $H\rightarrow Z\gamma$) are neglected, since they contribute less than $0.03\%$ of the events in total.

\subsection{Data event yield, signal, and background estimate comparisons in the signal region}
Table~\ref{tab:total_syst} shows the data event yield and the signal and background estimates in the SR. The \Sherpa[2.2.11] MC sample is used for the $\Zg+$jets process, together with the purely electroweak production of $\Zg jj$. Table~\ref{tab:total_syst}  includes the statistical uncertainties, experimental uncertainties (see Section~\ref{sec:systunc}), and background systematic uncertainties (as described in this section). The $ee\gamma$ and  $\mu\mu\gamma$ event yields are compatible with each other, once differences in efficiency are accounted for.

\begin{table}[h!]
\centering
\caption{Data yield and the signal and background estimates in the SR. The systematic uncertainty includes  experimental uncertainties and background uncertainties.}
\label{tab:total_syst}
\begin{tabular}{l | c}
\hline\hline
Source & $ee$ + $\mu\mu$ \\
\hline
$\Zg+$jets signal 		& $73{\,}500			\pm \phantom{0}50 \text{ (stat.)} 	\pm 2{\,}600 \text{ (syst.)}$ \\
\Zjets 				& \phantom{0}$9{\,}800 	\pm 460 \text{ (stat.)} 			\pm 2{\,}100 \text{ (syst.)}$ \\
\ttg 				& \phantom{0}$3{\,}600 	\pm \phantom{0}10 \text{ (stat.)} 	\pm \phantom{0\,}540 \text{ (syst.)}$ \\
Pile-up 				& \phantom{0}$2{\,}500	\pm \phantom{0}70 \text{ (stat.)} 	\pm \phantom{0\,}700 \text{ (syst.)}$ \\
Multiboson 			& \phantom{00\,}$950	\pm \phantom{00}5 \text{ (stat.)} 	\pm \phantom{0\,}280 \text{ (syst.)}$ \\
\tWg 				& \phantom{00\,}$150	\pm \phantom{00}1 \text{ (stat.)} 	\pm \phantom{00\,}45 \text{ (syst.)}$ \\
\hline
Total prediction 	& $90{\,}500			\pm 500 \text{ (stat.)} 			\pm 3{\,}500 \text{ (syst.)}$ \\
\hline
Data 				& $96{\,}410 			\phantom{~\pm 000 \text{ (stat.)} 			~\pm 0{\,}000 \text{ (syst.)}}$  \\
\hline\hline
\end{tabular}
\end{table}

Figures~\ref{fig:SRDataMC1} and \ref{fig:SRDataMC2} show a comparison between the data and the expected SM events in a subset of distributions. The \Sherpa[2.2.11] signal sample is scaled by a normalisation factor of 1.08 to match the rate in the data. The normalisation factor is obtained from the ratio of the measured yields to the predicted yields from \Sherpa[2.2.11], as shown in Table~\ref{tab:total_syst}. The hatched band in the figures shows the impact of the systematic uncertainties, as also shown in Table~\ref{tab:total_syst}. After the normalisation of the backgrounds, good agreement is observed between the data and the SM estimates. Observables  inclusive in the number of jets are well modelled; in some bins of some observables, small differences are observed which, when comparing the measured differential cross sections with the predictions, are covered by the theoretical uncertainties (see Section~\ref{sec:result}).

\begin{figure}
\centering
\subfigure[]{\includegraphics[width=.4\columnwidth]{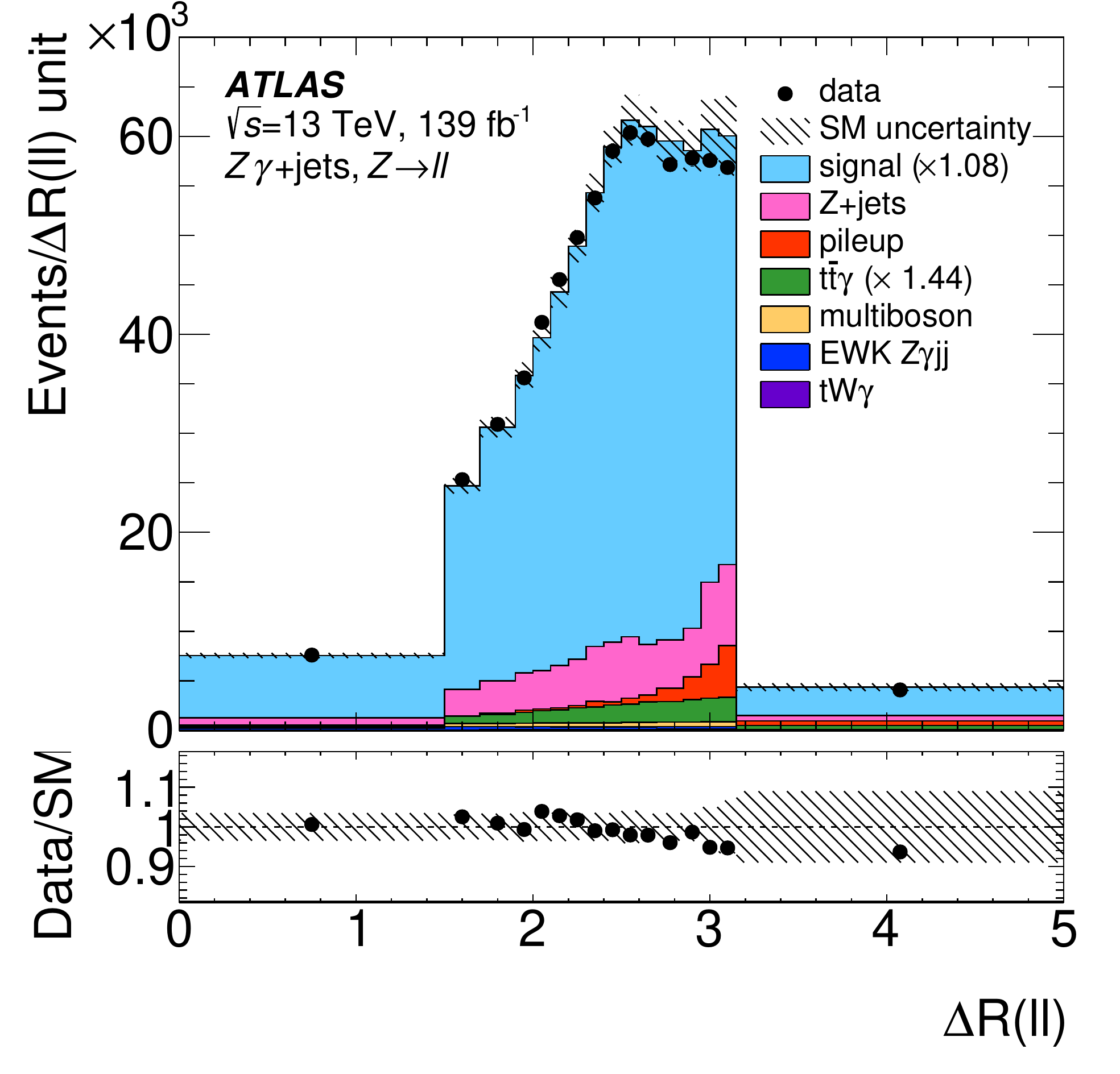}}
\subfigure[]{\includegraphics[width=.4\columnwidth]{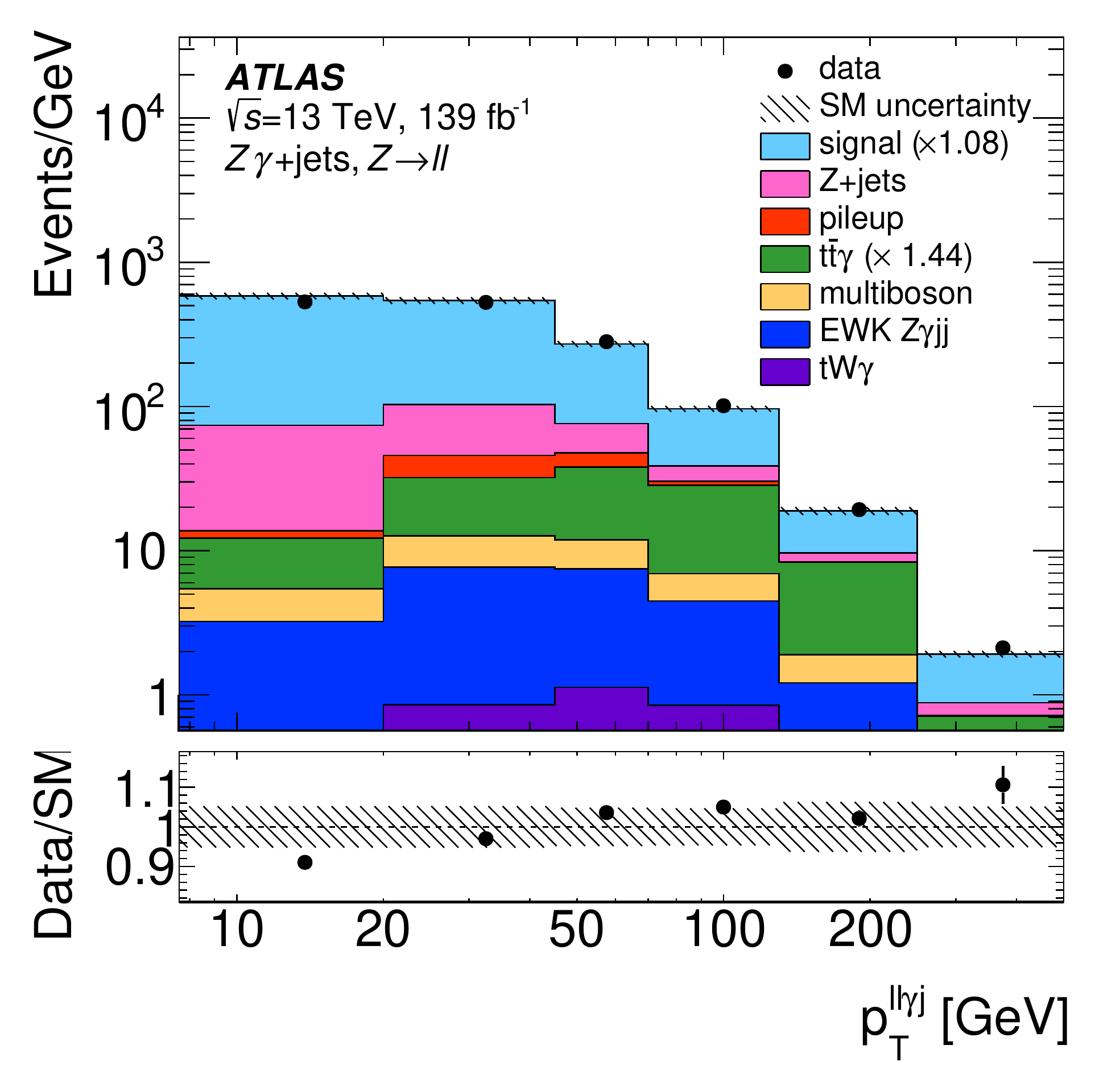}}\\
\subfigure[]{\includegraphics[width=.4\columnwidth]{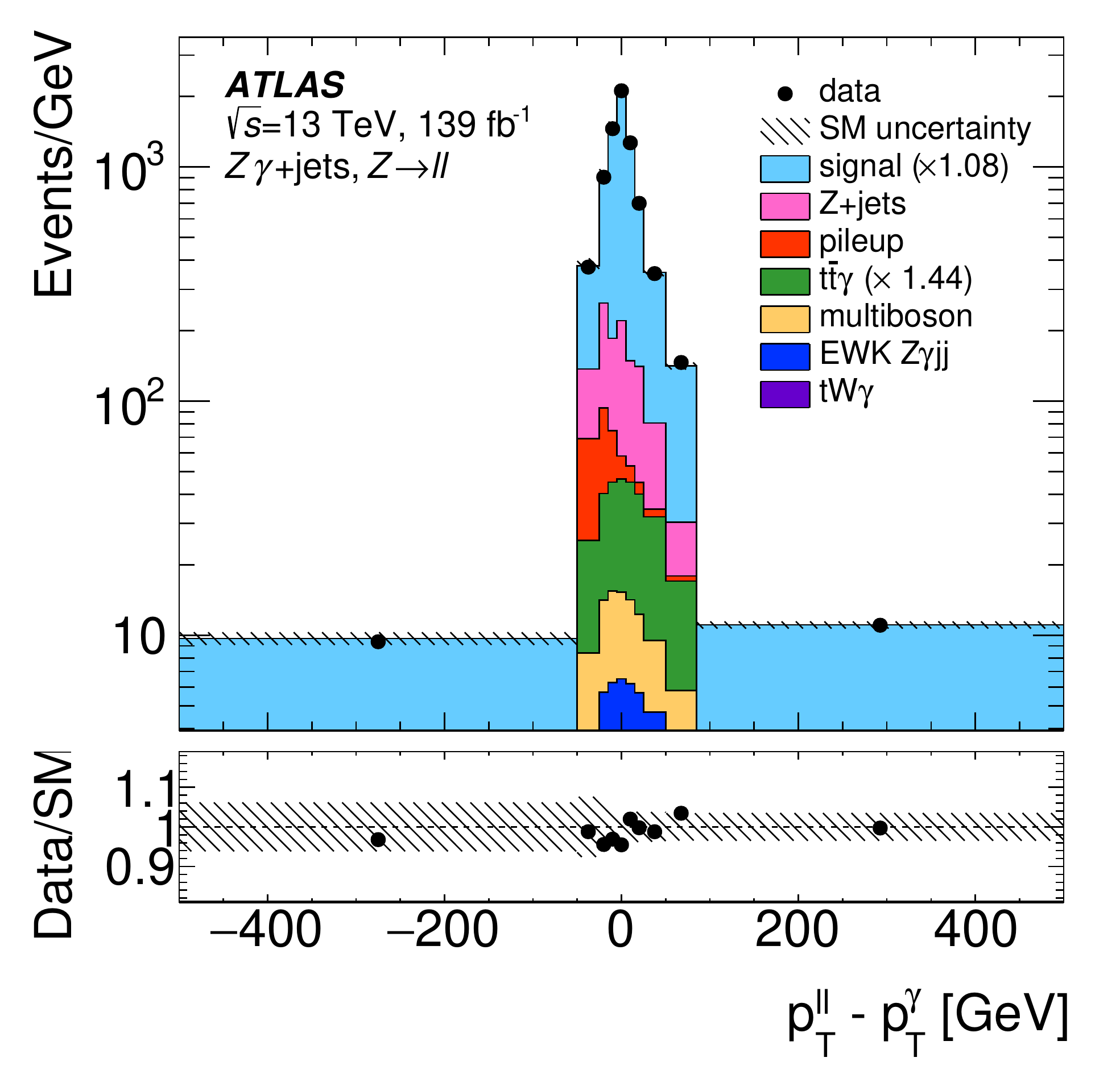}}
\subfigure[]{\includegraphics[width=.4\columnwidth]{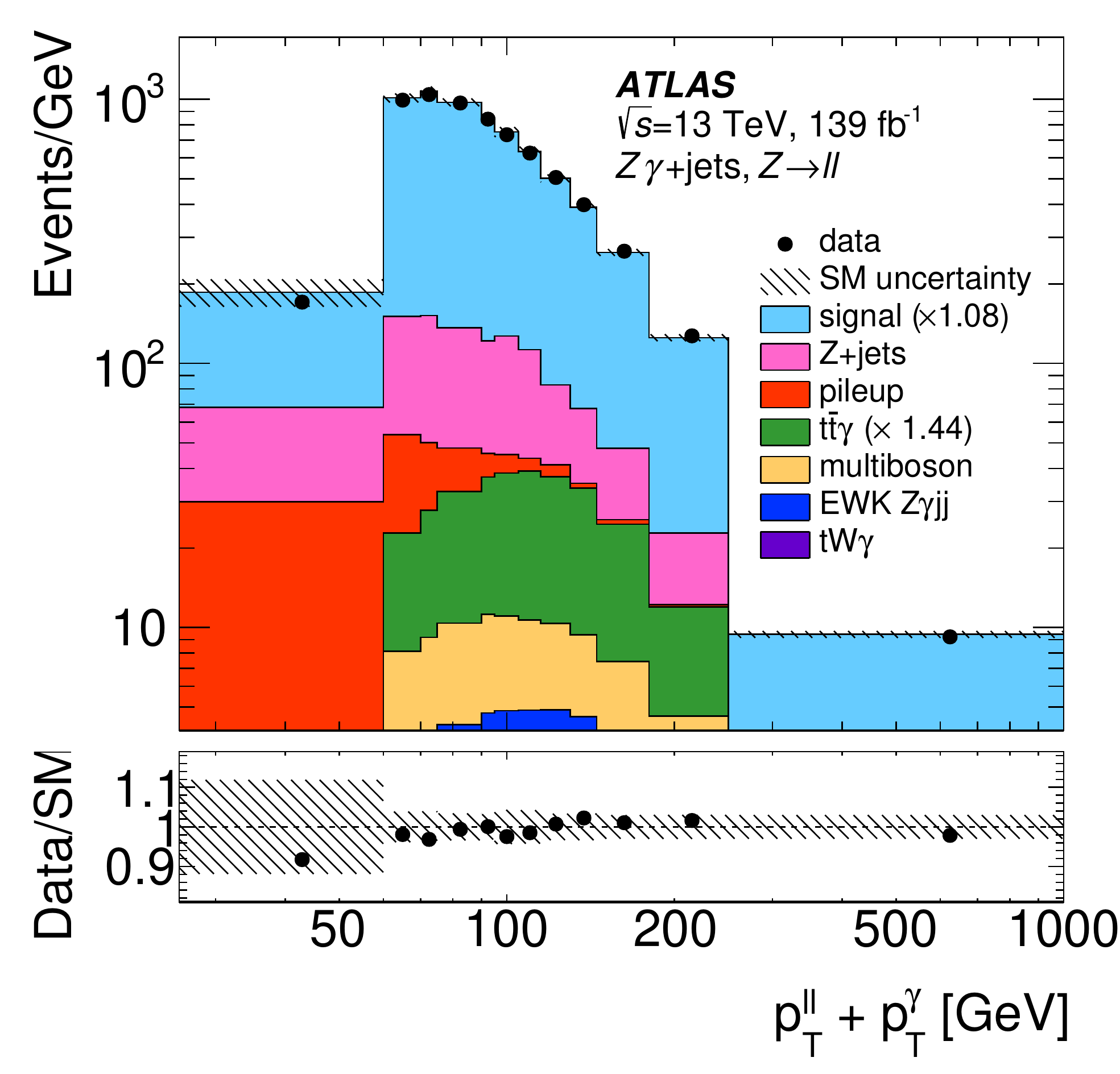}}
\caption{The measured  (a) \DRll ,  (b) \pTZyj , (c) \DiffpT , and (d) \SumpT distribution (dots) in the signal region. The error bars represent the data statistical uncertainty; for most of the points, the error bars are smaller than the marker size and, thus, not visible. The MC simulation of the signal from \Sherpa[2.2.11] (blue histograms)   and various backgrounds are also included. The signal and background distributions are stacked to produce the figures. The variable bin width is taken into account in the vertical scale. The lower part of each figure shows the ratio of the data to the expected total SM distribution. The hatched band represents the statistical and systematic uncertainties of the SM background yields added in quadrature, excluding theory uncertainties of the signal.}
\label{fig:SRDataMC1}
\end{figure}
 
\begin{figure}
\centering
\subfigure[]{\includegraphics[width=.4\columnwidth]{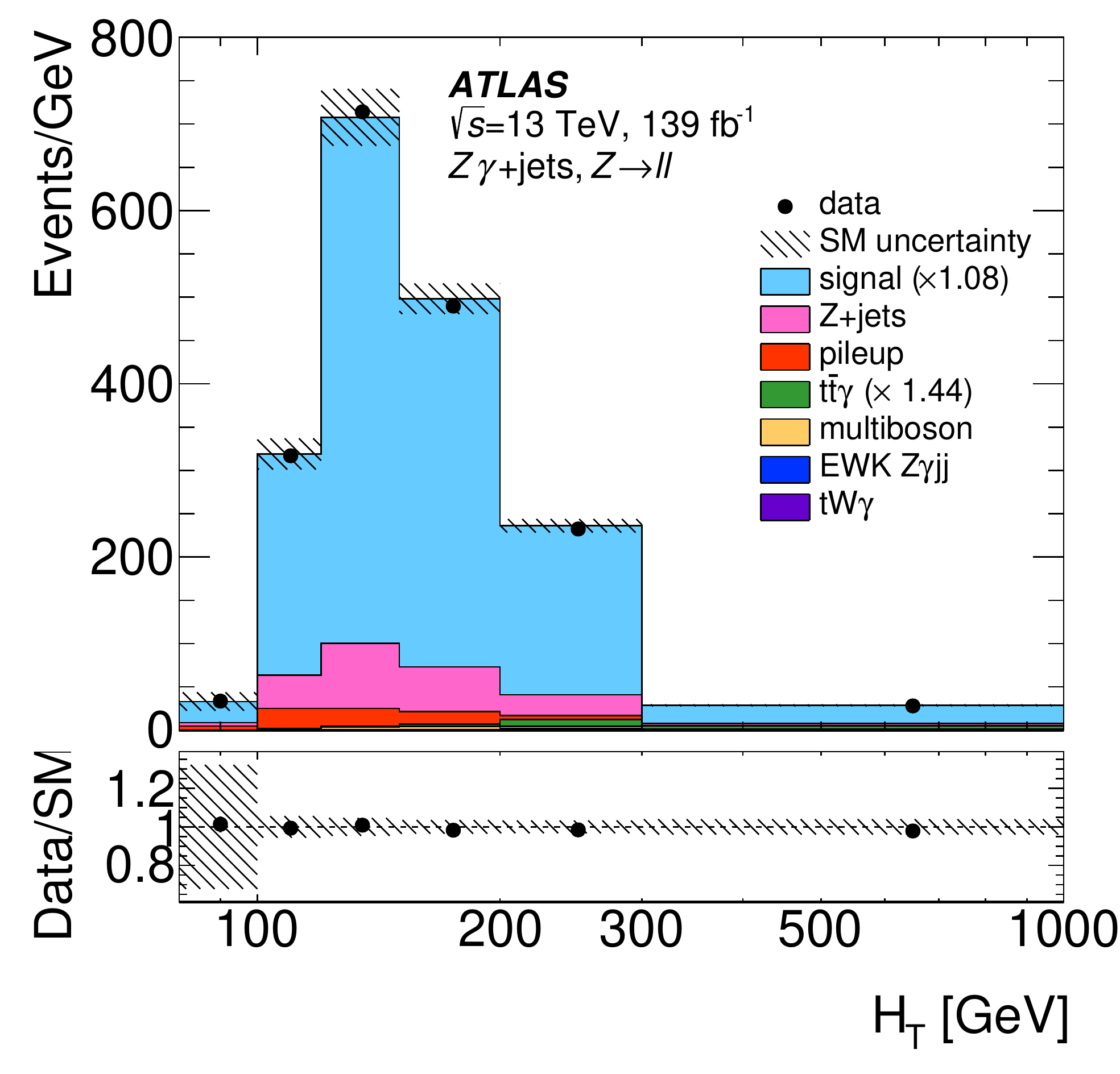}}
\subfigure[]{\includegraphics[width=.4\columnwidth]{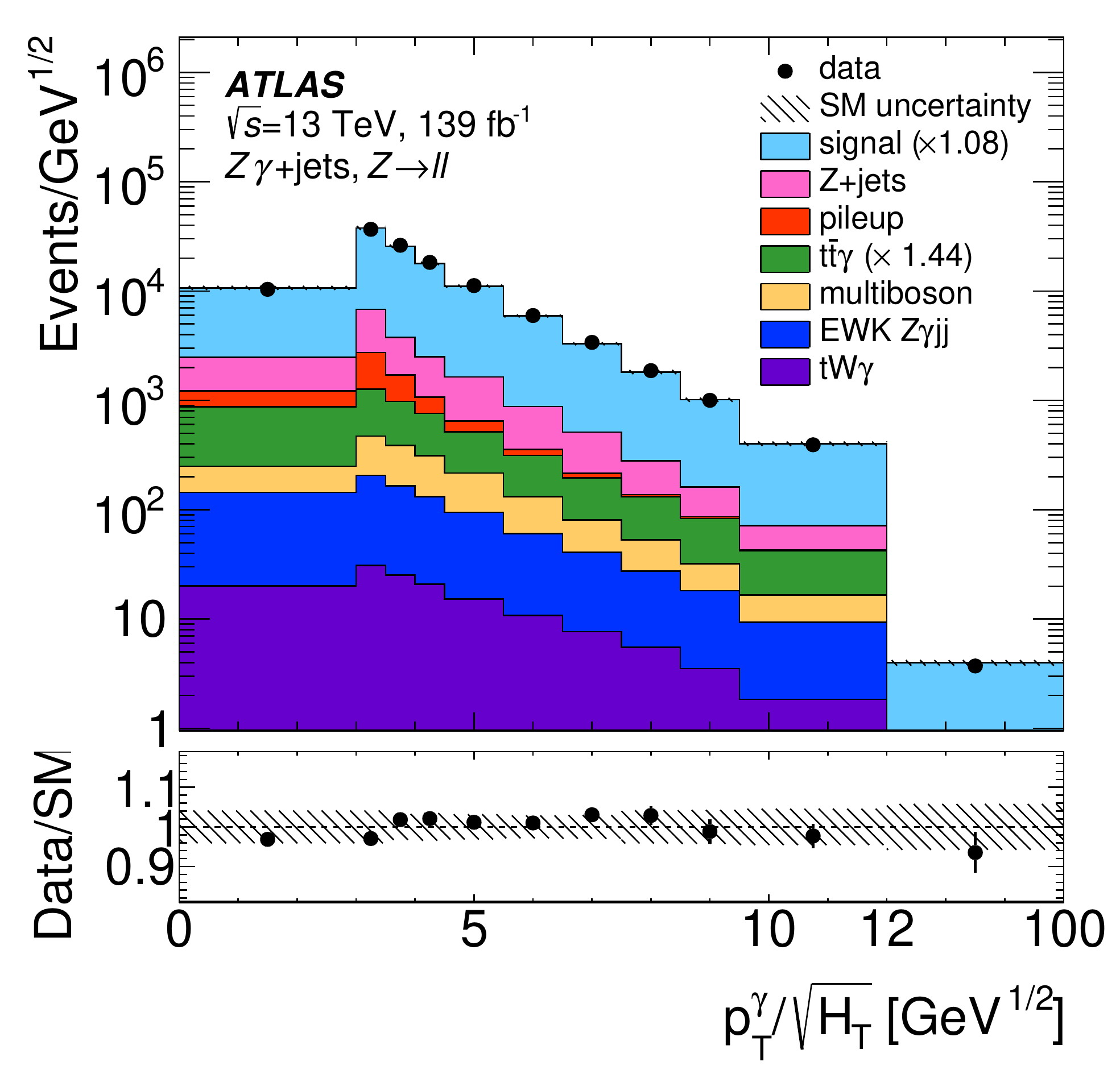}}\\
\subfigure[]{\includegraphics[width=.4\columnwidth]{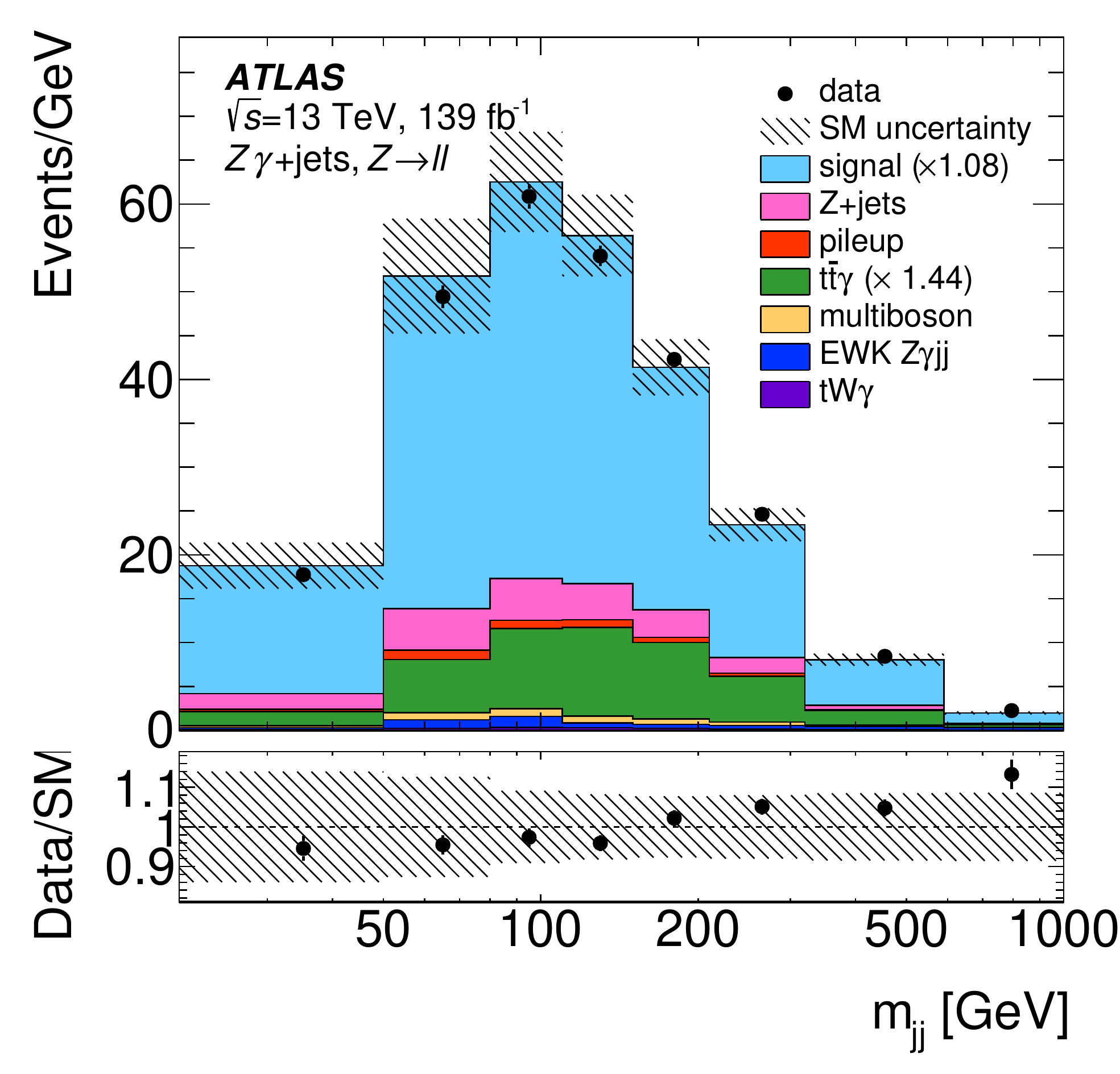}}
\subfigure[]{\includegraphics[width=.4\columnwidth]{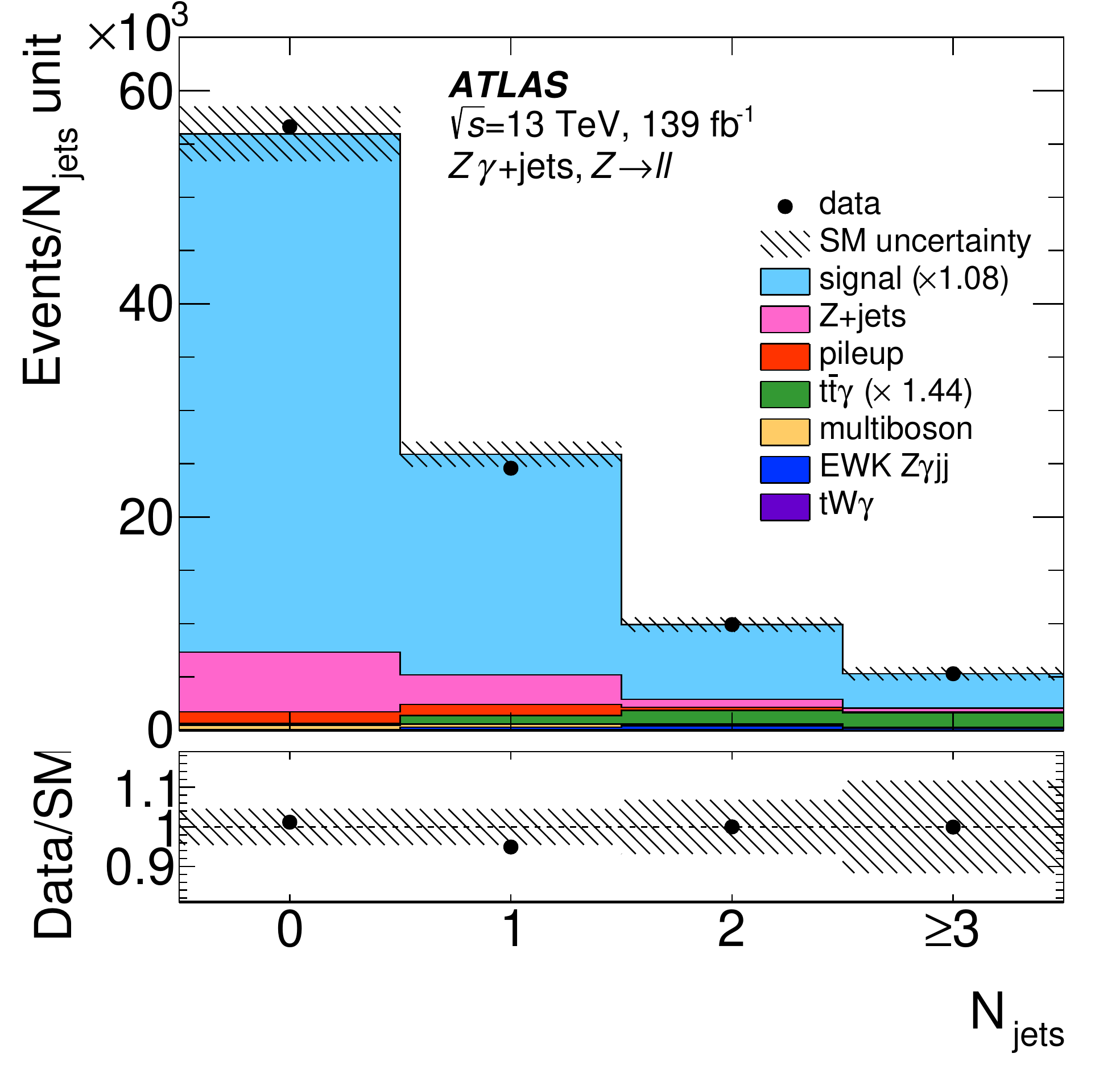}}
\caption{
The measured (a) \HT , (b) \ptysqrtHT , (c) \mjj , and (d) \njet distribution (dots) in the signal region. The error bars represent the data statistical uncertainty; for most of the points, the error bars are smaller than the marker size and, thus, not visible. The MC simulation of the signal from \Sherpa[2.2.11] (blue histograms)  and various backgrounds are also included. The signal and background distributions are stacked to produce the figures. The variable bin width is taken into account in the vertical scale. The lower part of each figure shows the ratio of the data to the expected total SM distribution. The hatched band represents the statistical and systematic uncertainties on the SM background yields added in quadrature, excluding theory uncertainties of the signal.}
\label{fig:SRDataMC2}
\end{figure}


\section{Cross-section determination}

\subsection{Fiducial region at particle level}
The measurements are unfolded to a fiducial phase space defined by particle-level quantities. The fiducial phase space in this analysis is built to be as close as possible to the detector-level selection discussed in Section~\ref{sec:evsel}, with selection criteria that minimise the extrapolation and allow comparisons with theoretical predictions. The phase space is selected for $Z\gamma\rightarrow\ell^+\ell^-\gamma$ events, with $\ell$ being either an electron or muon. Only stable particles (with a mean lifetime $c\tau>10$~mm) are used in the definition of the fiducial region. Additionally, only \enquote{prompt}  leptons (dressed) and photons (only those that do not originate from hadron decays) are considered.

Leptons are required to pass the same \pT requirements as in the SR: $\pT(\ell_1)>30$ GeV, $\pT(\ell_2)>25$ GeV. However, the $\eta$ requirements are different: for both electrons and muons $|\eta(\ell)|<2.47$ is required, since at particle level the discontinuities in the detector are not present. A particle-level isolation requirement is applied to photons: the scalar sum of the $E_{\textrm{T}}$ of all particles, except muons and neutrinos, within a cone of size $\Delta R = 0.2$ around the photon must be less than $7\%$ of the transverse energy of the photon, $E_{\textrm{T}}^{\gamma}$.  This selection is the same as in Ref.~\cite{STDM-2018-04} and is optimised to achieve the same level of acceptance in both the detector-level and particle-level selections. Photons are rejected if they are within $\Delta R=0.4$ of any lepton. Jets are obtained by clustering stable particles, excluding prompt leptons and using the anti-$k_t$ algorithm with a radius parameter of $R=0.4$. Photons within a cone of size $\Delta R = 0.1$ around prompt leptons are also excluded. Jets are defined in the same way as for the SR, by requiring $\pT>30~\GeV$ for $ |\eta|<2.5$ and $\pT>50~\GeV$ for $2.5< |\eta|<4.5 $. Jets are rejected if they are within $\Delta R=0.4$ of any photon. As in the SR, $\mll > 40~\GeV$ and $\mll+\mlly>182~\GeV$ requirements are applied. A pair of opposite-sign, same-flavour leptons is selected, and no additional veto on the number of leptons is applied. Table \ref{tab:FiducialDef} summarises the fiducial selection used in the analysis.\\
 
\begin{table}[h!]
\begin{center}
\caption{Definition of the fiducial region at particle level.}
\label{tab:FiducialDef}
\begin{tabular}{l|r} \hline\hline
\rule{0pt}{1.0em}
Quantity & Selection criteria \\[2pt] \hline\hline
\rule{0pt}{1.0em}
Lepton kinematics & $\pT(\ell_1)>30$~\GeV, $\pT(\ell_2)>25$~\GeV, $|\eta|<2.47$ \\
Photon kinematics & $\pT>30$~\GeV, $|\eta|<2.37$, $\Delta R (\gamma, \ell) >0.4$ \\
Photon isolation & $\et^{\text{iso}}/\et^{\gamma}<0.07$ \\
Jet kinematics &  $ (\pT > 30 ~\GeV$ if $|\eta| < 2.5$)  or ($ \pT > 50 ~\GeV$ if $2.5< |\eta|<4.5 $), $\Delta R (\gamma, \textrm{jet}) >0.4$  \\
Invariant mass  & $\mll>40$~\GeV, $\mll+\mlly>182$~\GeV \\
\hline\hline
\end{tabular}
\end{center}
\end{table}
 
\subsection{Fiducial and differential cross section}
The fiducial cross section is evaluated in the fiducial region described in the previous subsection. It is obtained with the following formula:
$$
\sigma^{\text{fid}} =\frac{N_{\text{obs}} - N_{\text{bkg}} }{C\times \mathcal{L}},
$$
where $N_{\text{obs}}$ and  $N_{\text{bkg}} $ are the observed number of events and the expected number of background events, respectively, $ \mathcal{L}$ is the integrated luminosity, and $C$ is the  correction factor which accounts for detector inefficiency and resolution effects. Events from inside (outside) the fiducial region at particle level that migrate outside (inside) the SR are also accounted for by this correction factor. The factor $C$ is calculated as the number of simulated $\Zg+$jets events entering the detector-level SR divided by the number of simulated $\Zg+$jets events entering the fiducial volume. The inclusive $C$ factor is obtained with \Sherpa[2.2.11] MC samples, combining the $ee$ and $\mu\mu$ channels; its value is found to be $C=0.543 \pm 0.001\,(\textrm{stat}) \pm 0.020\,(\textrm{syst})$, with the systematic uncertainties estimated as described in Section ~\ref{sec:systunc}.
 
Differential cross sections are evaluated in the fiducial signal region for several observables. The event yields in the $e^+e^-$ and $\mu^+\mu^-$  decay channels are added together and unfolded in a single step. The distributions are unfolded using an iterative Bayesian method~\cite{DAGOSTINI1995487}, with two iterations as the nominal number. The $\Zg+$jets events simulated with \Sherpa[2.2.11] are used to produce the response matrices needed to correct for the migration between bins in the detector- and particle-level distributions. These migrations are mainly due to the jet reconstruction. Additionally, the unfolding corrects for fiducial and reconstruction efficiencies. These are respectively the probability of having particle-level events satisfy the detector-level SR criteria, and the probability that detector-level events originate from outside the fiducial region.


\section{Systematic uncertainties}
\label{sec:systunc}

 
Systematic uncertainties from several different sources affect this measurement: experimental uncertainties due to detector reconstruction, uncertainties in the background estimate (from simulated samples as well as data-driven methods, as described in Section~\ref{sec:bkgest}), systematic uncertainties in the unfolding, and theoretical uncertainties in the signal prediction. The individual sources of uncertainty are varied by $\pm 1\sigma$ in the MC simulations and propagated through the analysis separately. The uncertainties are propagated to the cross sections by modifying the migration matrix and computing the resulting deviation from the nominal cross section. This deviation is taken as the systematic uncertainty.

Experimental uncertainties account for the finite resolution of the objects reconstructed by the ATLAS detector, their calibration, and the modelling of the reconstruction in the simulation. Uncertainties affecting the electrons and photons include the uncertainties in the energy scale and resolution~\cite{EGAM-2018-01}, while for muons, uncertainties are considered for the momentum resolution~\cite{MUON-2018-03}. Both leptons and photons  have uncertainties in the efficiency of the identification and the isolation~\cite{EGAM-2018-01,MUON-2018-03}. Uncertainties in the lepton trigger efficiencies are also considered~\cite{TRIG-2018-05,TRIG-2018-01}. Jet uncertainties account for both the energy scale (JES) and the resolution (JER)~\cite{ATLAS:JETM-2018-05}. The JES uncertainties take into account detector modelling, statistical effects, flavour composition, and the description of pile-up jets.  The JVT efficiency uncertainties are also considered~\cite{PERF-2014-03}. Additional uncertainties are added to take into account the modelling of the number of $pp$ collisions. An uncertainty of  $1.7\%$ in the total integrated luminosity is considered in this analysis.

The statistical uncertainty in the measured cross sections is evaluated using `toy experiments' (bootstrap technique~\cite{as:7:1}). Statistically-independent replicas of the data distributions are used and each one of them is then unfolded; the root mean square (RMS) of the replicas distribution is used as the uncertainty. For the MC samples, the limited number of simulated events mainly affects the estimation of the migration matrices. This statistical uncertainty is also calculated using toy experiments and found to be small.

The unfolding procedure is based on an assumption, namely our choice of a simulated signal sample. This choice can bias the results, and a systematic uncertainty to account for this effect is obtained through a data-driven closure test. The simulated signal distributions are  reweighted with a smooth function obtained by requiring that the detector-level distribution matches the data (after background subtraction). The reweighted distribution is then unfolded, treating this sample as pseudo-data, and using the migration matrix from the reweighted distributions. The uncertainty is obtained by comparing this result with the nominal unfolded result.

Systematic uncertainties in the cross sections due to the theoretical modelling are obtained by unfolding the data with a migration matrix calculated using alternative signal simulations. Uncertainties in the signal predictions are due to missing higher-order contributions in the cross-section calculation, the uncertainties from the PDF choice, and the uncertainties in \alphas. The effect of QCD scale uncertainties is estimated by halving and doubling the renormalisation and factorisation scales in the signal simulation relative to their nominal values.  Uncertainties are obtained  by taking an envelope: in each bin the largest resulting change is used as the uncertainty. Additional uncertainties are added to account for the choice of a specific PDF in the cross-section calculation. Following the PDF4LHC recommendation~\cite{Butterworth_2016}, the $\NNPDF[3.0nnlo]\_$as\_0118 PDF set is used as the nominal set, and is compared with results obtained with weights stored in the \Sherpa samples. An envelope is then taken of all the variations. A similar approach is used for the \alphas variations, where the $\NNPDF[3.0nnlo]\_$as\_0117 and $\NNPDF[3.0nnlo]\_$as\_0119 PDF sets are used. For the theory predictions, uncertainties are obtained by taking an envelope of the difference between the nominal unfolded results and their variations.

Uncertainties for the background estimates are taken to be 30\% in the diboson cross section and 15\% in the \ttg cross section (which corresponds to the uncertainty in the normalisation factor from the LO cross section to the NLO cross section~\cite{ttyCalc2011}). The uncertainty of $30\%$ in the diboson cross section covers both the nominal uncertainty \cite{STDM-2018-34} and the typical size of the mismodelling of non-prompt objects. The systematic uncertainties for the \Zjets background are estimated as described in Section~\ref{ssec:Zjets} and propagated through the unfolding framework.
 
Table~\ref{Tab:SystBreackdownNjet} shows the breakdown of the systematic uncertainties in the cross section as a function of \njet. The last row in the table is the total relative uncertainty obtained as the sum in quadrature of each systematic uncertainty and statistical uncertainty. The increase in jet systematic uncertainties with the number of jets, is due to the modelling of pile-up jets, forward jet modelling, and statistical fluctuations. The combination of the pileup-jet tagger, and the jets definition as described in Section \ref{sec:evsel} allows to reduce the uncertainty in the bins with at least 1 jet.

\begin{table}[tbp]
\centering
\caption{Impact of the different systematic uncertainties on the measured $\Zg +$jets cross section as a function of \njet in each bin of the distribution. }
\label{Tab:SystBreackdownNjet}
\begin{tabular}{lcccc}
\hline\hline
$\njet$ & 0 & 1 & 2 & $>2$ \\
\hline
Source & \multicolumn{4}{c}{Uncertainty [\%]} \\
\hline
Electrons &  1.0 &  0.9 &  0.8 &  0.8 \\
Muons &  0.3 &  0.3 &  0.3 &  0.4\\
Jets &  1.7 &  1.7 &  4.5 &  8.8 \\
Photons & 1.4 &  1.3 &  1.3 &  1.2 \\
Pile-up &  2.1 &  0.8 &  0.2 &  0.3 \\
Background &  1.8 &  1.8 &  3.0 &  4.4 \\
MC statistical &  0.1 &  0.2 &  0.3 &  0.4 \\
Data statistical &  0.8 &  1.5 &  1.8 &  1.9 \\
Luminosity &  1.7 &  1.7 &  1.7 &  1.7 \\
Theory &  0.6 &  0.2 &  1.4 &  1.0\\
\hline
Total &  4.2 & 3.8 & 6.3 & 10.3\phantom{0} \\
\hline\hline
\end{tabular}
\end{table}


\section{Results}
\label{sec:result}


The measured fiducial cross section for $\Zg$ production is
${\sigma = 533.7 \pm 2.1 (\textrm{stat}) \pm 12.4 (\textrm{syst}) \pm 9.1 (\textrm{lumi})}$~fb, as presented in Ref.~\cite{STDM-2018-04}. The predicted fiducial cross sections for the $\Zg+$jets are ${479.5 \pm 0.3 (\textrm{stat})}$~fb from \Sherpa[2.2.11] interfaced with  \MEPSatLO and ${493.0 \pm 3.0 (\textrm{stat})}$~fb from MiNNLO$_\text{PS}$; both predictions do not include the contribution from purely electroweak $\Zg jj$ production.

The measured differential cross sections as functions of the different observables are shown in Figures~\ref{fig:DiffTheoryPredData1} to~\ref{fig:DiffTheoryPredData8}. To obtain these results the unfolding uses as signal the $\Zg+$jets MC samples added together with the MC sample for purely electroweak production of $\Zg jj$. The theoretical predictions from \Sherpa[2.2.4], \Sherpa[2.2.11] (both interfaced with  \MEPSatLO and using the \NNPDF[3.0nnlo] PDF set) and \MGNLO (interfaced with \Pythia[8.212] and using the \NNPDF[3.0nlo] PDF set), and the NNLO predictions from MiNNLO$_\text{PS}$ (using the \NNPDF[3.0nnlo] PDF set) and MATRIX (using the \CT[14nnlo] PDF set) are compared with the measurements in these figures. The purely electroweak production of $\Zg jj$ has not been added to the theoretical predictions shown in these figures; this contribution is estimated to be around 1\%. In general, \Sherpa samples underestimate the total cross section, with \Sherpa[2.2.11] (NLO) being higher than \Sherpa[2.2.4], while \MGNLO, MiNNLO$_\text{PS}$ and MATRIX show a better agreement for the total cross section. Between the two \Sherpa samples, \Sherpa[2.2.11] also shows generally better agreement in the distribution shapes, especially for the number of jets.
 
The $Z$ boson \pT is a fundamental observable, correlated with the jet activity, and its difference from \pTy is an observable that probes pQCD over a wide range of scales, while \SumpT describes the hard scale of the process. Figure~\ref{fig:DiffTheoryPredData1} shows the differential cross sections as functions of the observables \pTZ  , \DiffpT,  \SumpT, and \DRll. All the predictions show good agreement with the measurements, although  \Sherpa generally underestimates the data. The MATRIX calculations are also in good agreement with the measurements.
 
Jet multiplicity is a fundamental observable to probe QCD and additional soft radiation~\cite{ZyNNLOPS2021}. The ratio \Rptj in particular is an observable that tests the limits of PS effects and resummation of Sudakov logarithms.  Differential cross sections for jet observables are shown in Figure~\ref{fig:DiffTheoryPredData2}. The differential cross section is dominated by events with zero jets, and rapidly falls off with increasing QCD emission. The leading and subleading jets are mostly produced with similar \pT; however, the cross section is not zero at $\Rptj = 0.1$, which means that the subleading jet  has only 10\% of the \pT of the leading jet. In general, the MC samples are in good agreement with the data; however, at high jet multiplicity and high jet \pT \Sherpa[2.2.4] predicts higher yields than \Sherpa[2.2.11], while \MGNLO has lower yields in \njet but it is comparable to the \Sherpa samples in the leading (\pTjetOne) and subleading (\pTjetTwo) jet momenta. The MiNNLO$_\text{PS}$ calculation instead predicts softer jets and lower jet multiplicity, while the MATRIX prediction models the jet \pT spectrum better, but predicts higher jet multiplicity. It is worth noting that MATRIX produces no more than two jets, so the last bin is empty for this calculation.  The ratio \Rptj   is equally  well  described by both \Sherpa models and \MGNLO. The MATRIX prediction is also in good agreement, while MiNNLO$_\text{PS}$  underestimates the data.
 
The invariant mass of the two leading jets is an important observable that describes the hard scale of the process, and its precise modelling is fundamental for the $\Zg jj$ QCD background in measurements of purely electroweak $\Zg jj$ production or searches for new physics~\cite{STDM-2015-21}. The \mjj distribution in Figure~\ref{fig:DiffTheoryPredData3} is  well modelled in general, except for MiNNLO$_\text{PS}$, which underestimates the highest bins. The MATRIX prediction shows good agreement, except for a few bins with some overestimations between 60 and 100~\GeV. The invariant mass \mllyj is  a variable sensitive to the hard scale of the process, and is also well modelled by the predictions, except for the last bins in the case of MiNNLO$_\text{PS}$, where an underestimation is observed. The MATRIX prediction also shows good agreement, with a small overestimation, although within systematic uncertainties, in the last bins.

Figure~\ref{fig:DiffTheoryPredData4} shows additional jet observables \HT , \ptysqrtHT , \DPhiJetY , and \pTZyj. The \HT and \ptysqrtHT observables are both well described by \Sherpa[2.2.4] and \Sherpa[2.2.11], while \MGNLO, MATRIX, and MiNNLO$_\text{PS}$ underestimate the last bin of \HT.  The \DPhiJetY  distribution is important for PS corrections in QCD predictions in phase-space regions with soft QCD emission~\cite{ZyNNLOPS2021}. The measurement shows a preference for events in which the photon and the leading jet are back-to-back. The \DPhiJetY  observable is well modelled by all the predictions, except in the very first bin of the MATRIX prediction. The \pTZyj observable is well modelled within uncertainties by both \Sherpa samples and \MGNLO, whereas the MiNNLO$_\text{PS}$ and MATRIX predictions show some deviation from the data distributions.

Figure~\ref{fig:DiffTheoryPredData5} shows the cross sections as functions of $\cos\theta_\text{CS}$ and $\phi_\text{CS}$, which are sensitive to the polarisation of the $Z$ boson. The variation of $\cos\theta_\text{CS}$ with \pTZ is mainly due to the lepton selection, which translates into differences in acceptance in each \pTZ region. For these variables, the \Sherpa, \MGNLO and MiNNLO$_\text{PS}$ predictions are in general in good agreement for the shapes, although, as already mentioned, \Sherpa[2.2.4] and 2.2.11 generally underestimate the data. The MATRIX predictions show some small disagreement in $\cos\theta_\text{CS}$ in the very first \pTZ bin, but otherwise show very good agreement.
 
The ratio \RZy is an important resolution variable for probing the effects of Sudakov-logarithm terms in different regimes of the hard scale of the process. The ratio \RZy in all the \mZy slices  is presented in Figure~\ref{fig:DiffTheoryPredData6}. These observables are  better modelled by \Sherpa[2.2.11] than by \Sherpa[2.2.4]. The MiNNLO$_\text{PS}$ prediction shows good agreement in all the \mZy slices, while MATRIX overestimates the data in the lowest bin in \RZy.
 
The  \DiffpT and \pTZyj distributions offer an important probe of additional soft QCD emissions, in increasing hard scale of the process, described by \SumpT and \pTZy, respectively. The \DiffpT distributions become more asymmetric in regimes with a harder scale of the process. All the distributions as function of \DiffpT in different regions of \SumpT (Figure~\ref{fig:DiffTheoryPredData7}) are well modelled by all the MC models and the MATRIX predictions, except for the last two bins in the region with $\SumpT>300~\GeV$, where \MGNLO slightly overestimates the data. The \pTZyj distributions in different regions of \pTZy (Figure~\ref{fig:DiffTheoryPredData8}) are well modelled by \Sherpa and \MGNLO, while the MiNNLO$_\text{PS}$ predictions underestimate the data for $\pTZy>75~\GeV$. The MATRIX prediction exhibits a softer spectrum than the other predictions.
 
In summary, the \Sherpa and \MGNLO predictions describe the data well, especially for observables involving jets, although \Sherpa underestimates the measured total cross section. The MiNNLO$_\text{PS}$ and MATRIX predictions give an adequate description of the measurements, but some deviations from the data are observed at high jet multiplicity.

\begin{figure}[h!]
\centering
\subfigure[]{\includegraphics[width=.49\columnwidth]{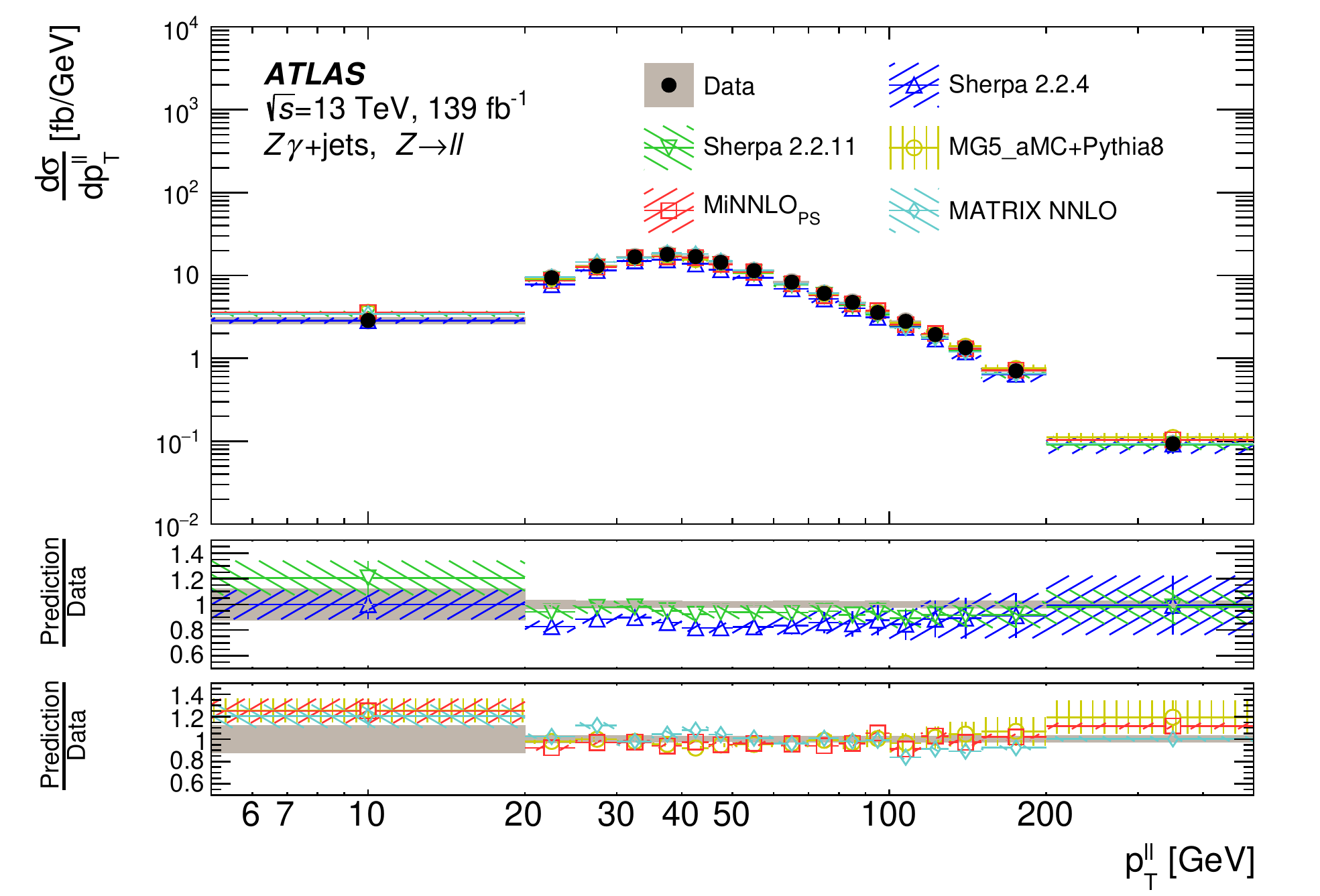}}
\subfigure[]{\includegraphics[width=.49\columnwidth]{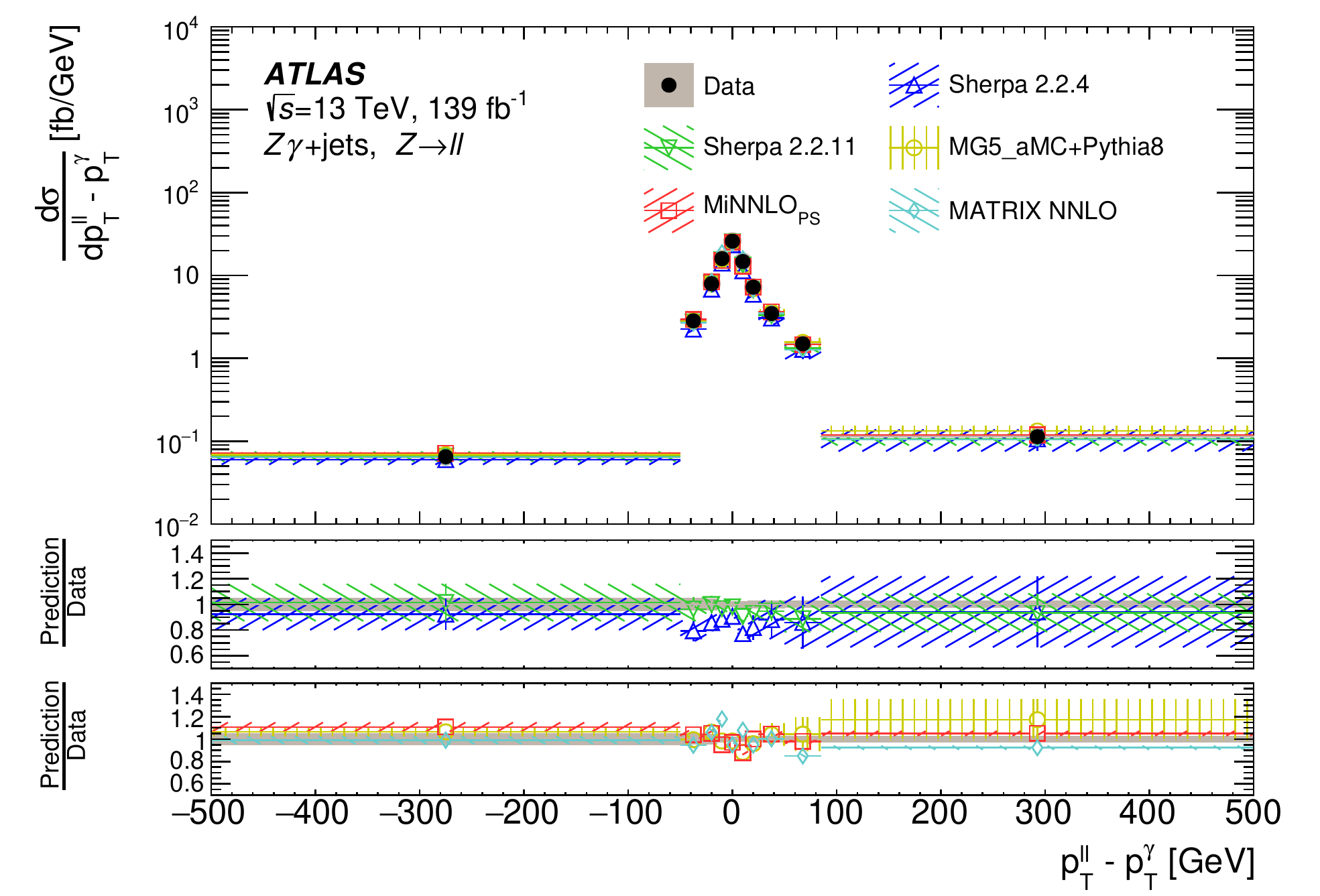}}\\
\subfigure[]{\includegraphics[width=.49\columnwidth]{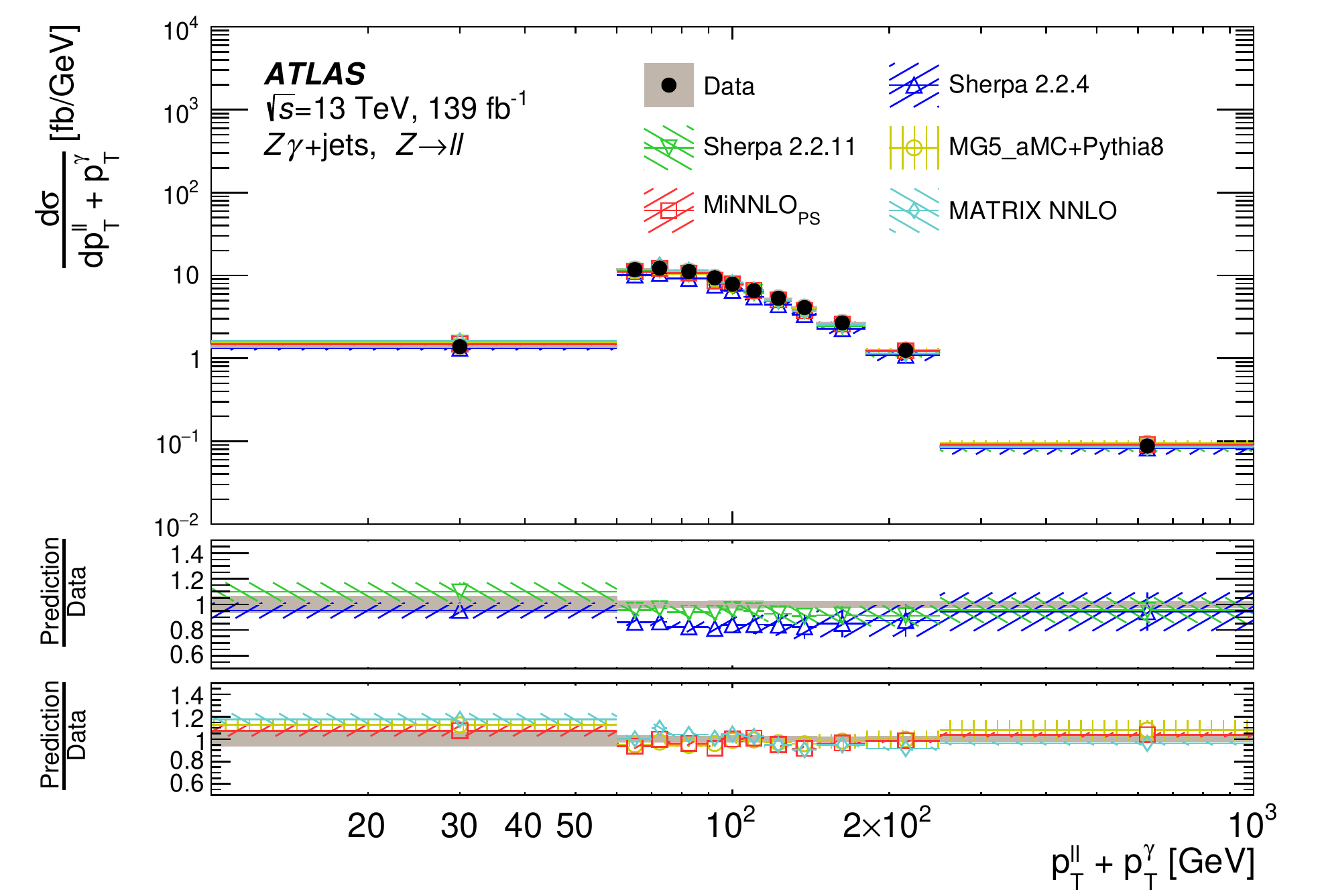}}
\subfigure[]{\includegraphics[width=.49\columnwidth]{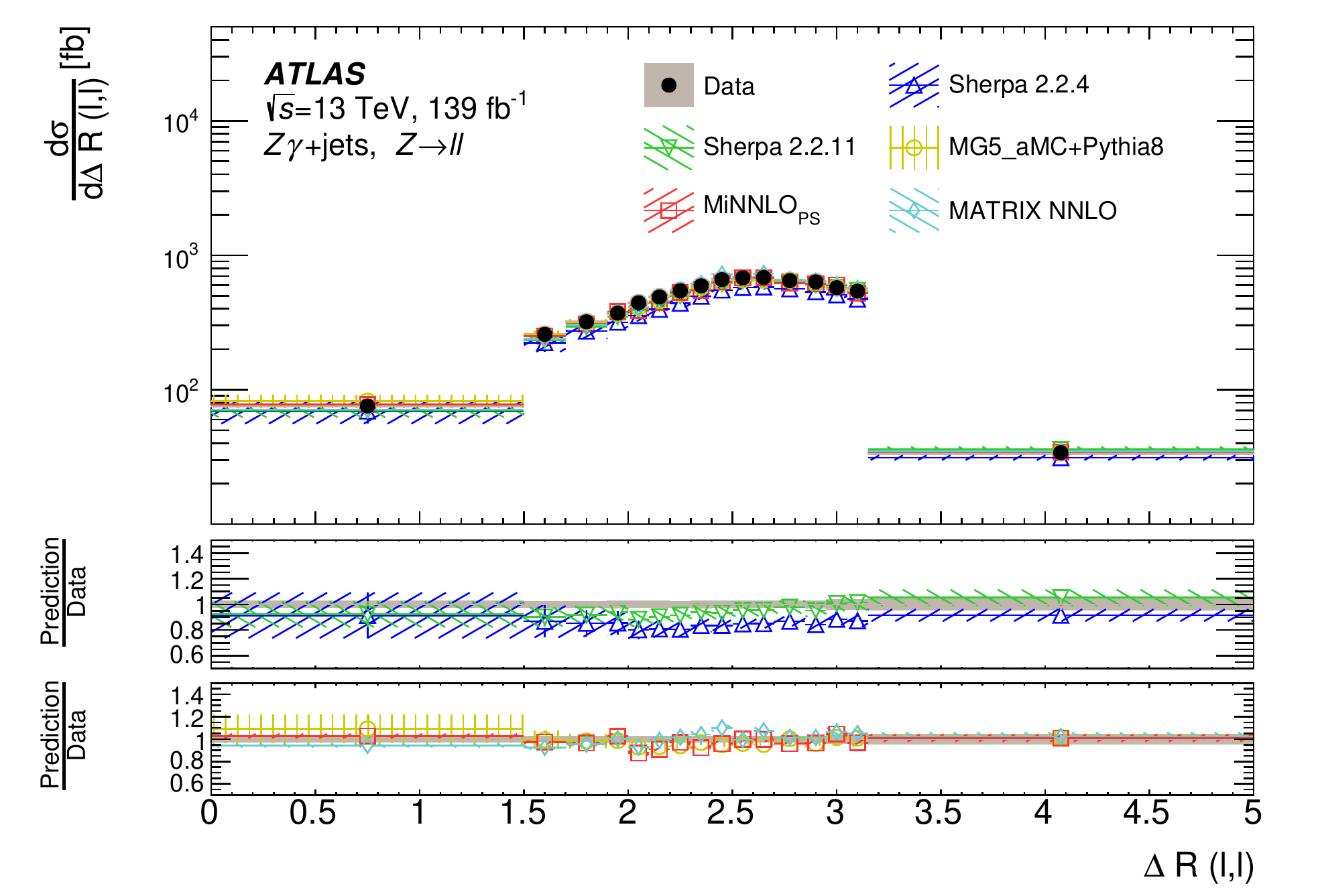}}
\caption{Measured differential cross section (black data points) as a function of the observables
(a) \pTZ , (b) \DiffpT, (c) \SumpT, and (d) \DRll. Error bands on the data points show the statistical uncertainty, while the grey area shows the total uncertainty in the unfolded result. Measured cross sections are compared with SM predictions from event generators at particle level: \Sherpa[2.2.4], \Sherpa[2.2.11], \MGNLOPY[8] (MG5\_aMC+Pythia8 in the legend), and MiNNLO$_\text{PS}$. Fixed-order calculation results from MATRIX NNLO are also shown. Dashed bands represent the statistical uncertainty and theoretical uncertainty (PDF and scale variations). The bottom panels show the ratio of the SM prediction to the measured cross section.}
\label{fig:DiffTheoryPredData1}
\end{figure}
 
\begin{figure}[h!]
\centering
\subfigure[]{\includegraphics[width=.49\columnwidth]{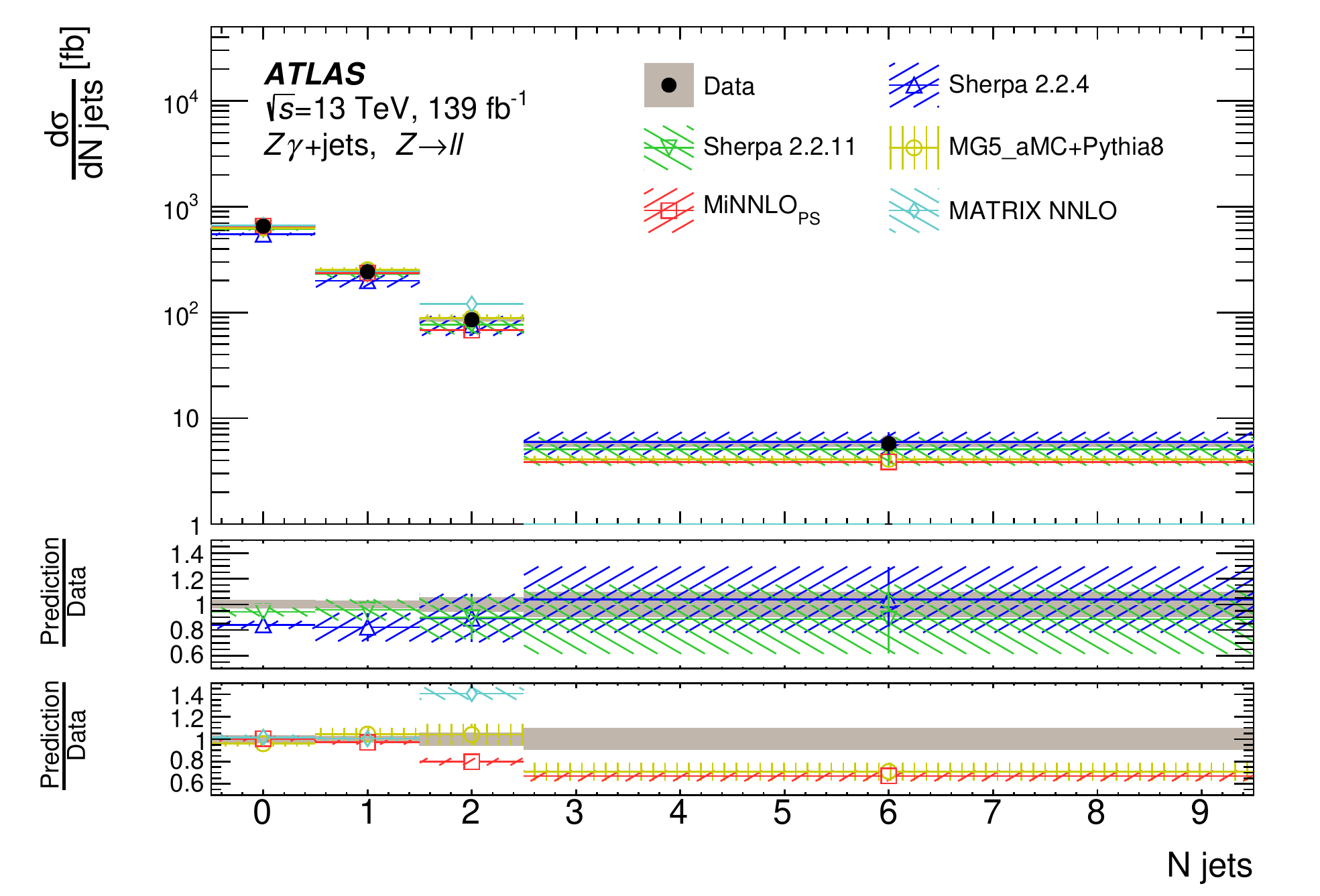}}
\subfigure[]{\includegraphics[width=.49\columnwidth]{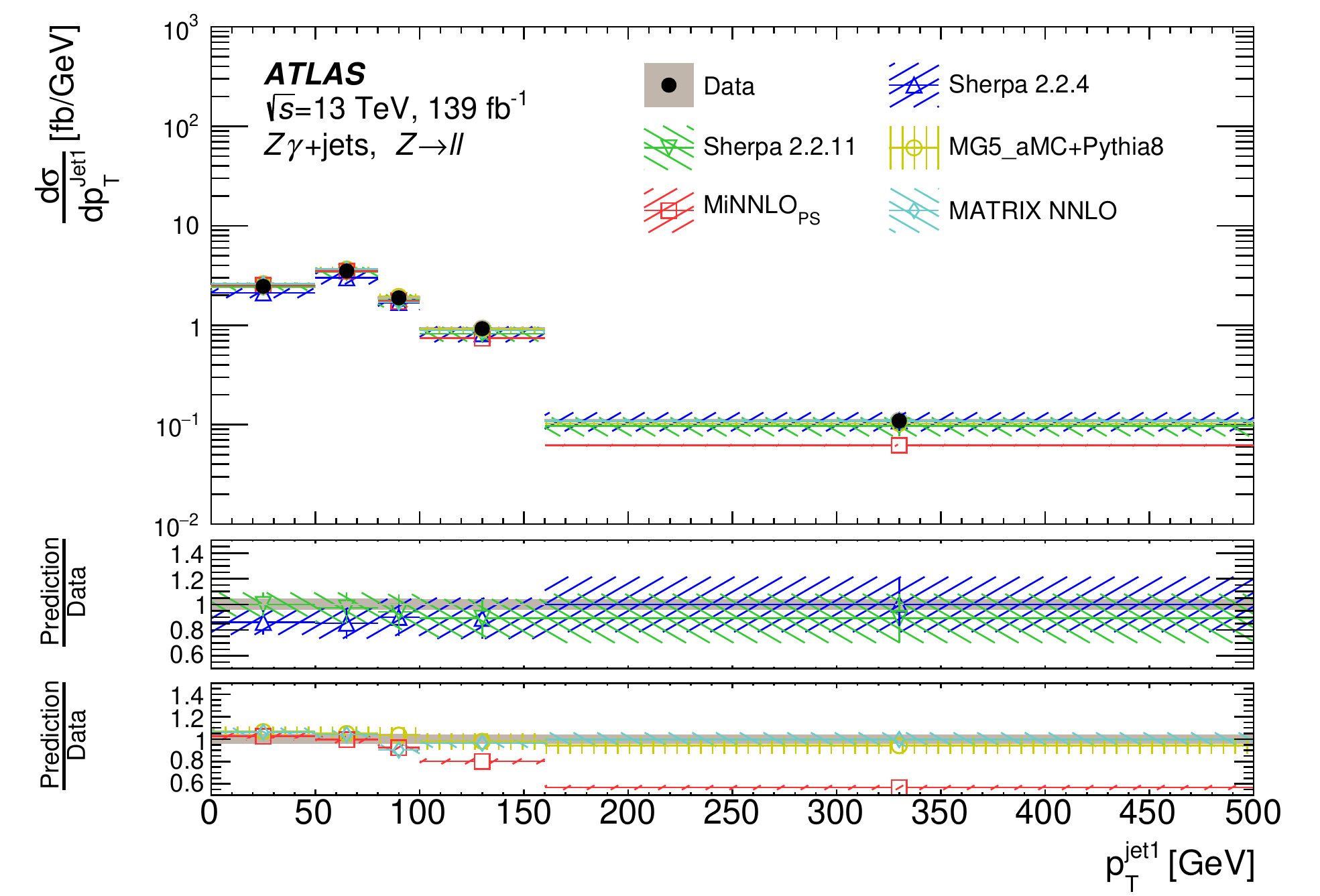}}\\
\subfigure[]{\includegraphics[width=.49\columnwidth]{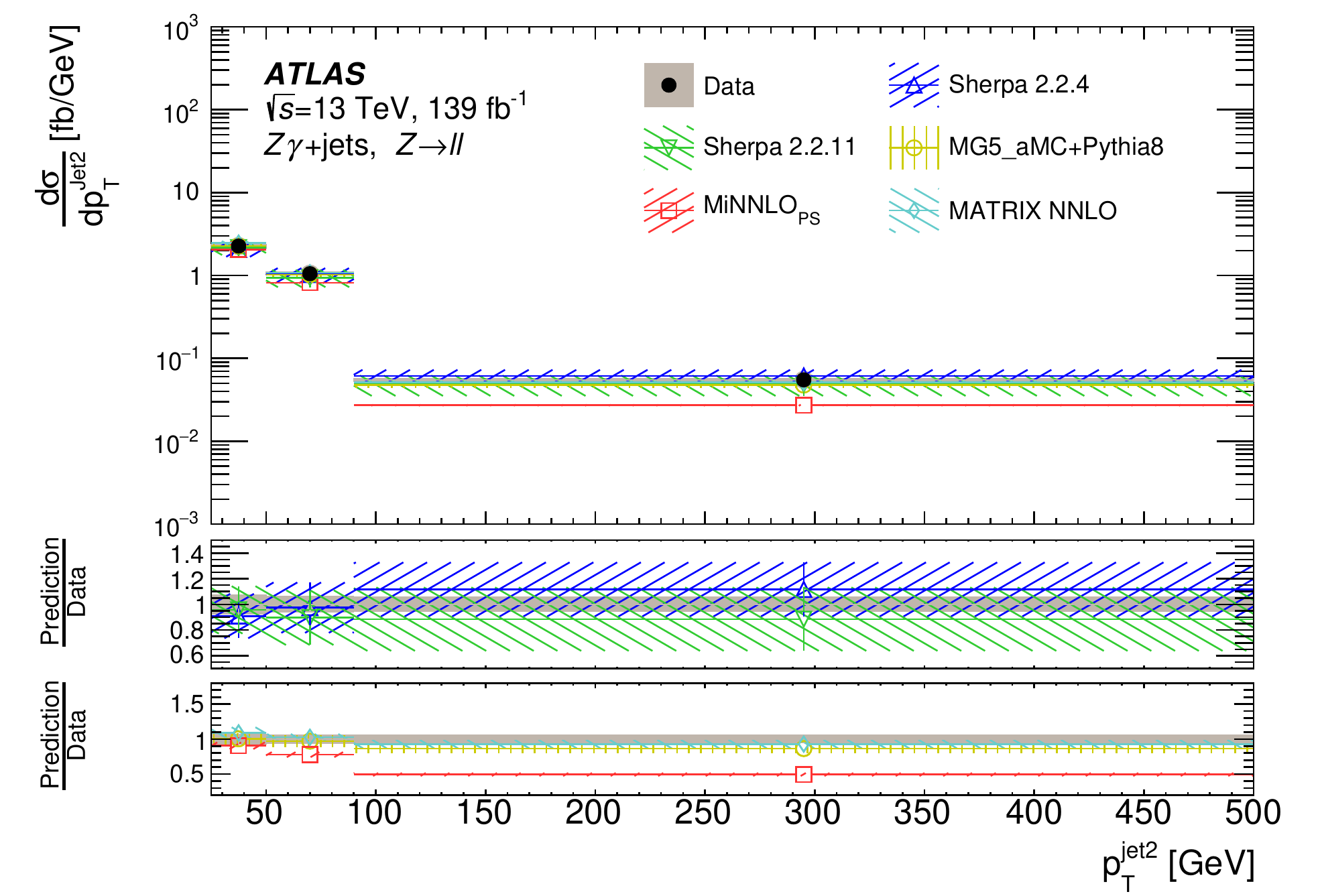}}
\subfigure[]{\includegraphics[width=.49\columnwidth]{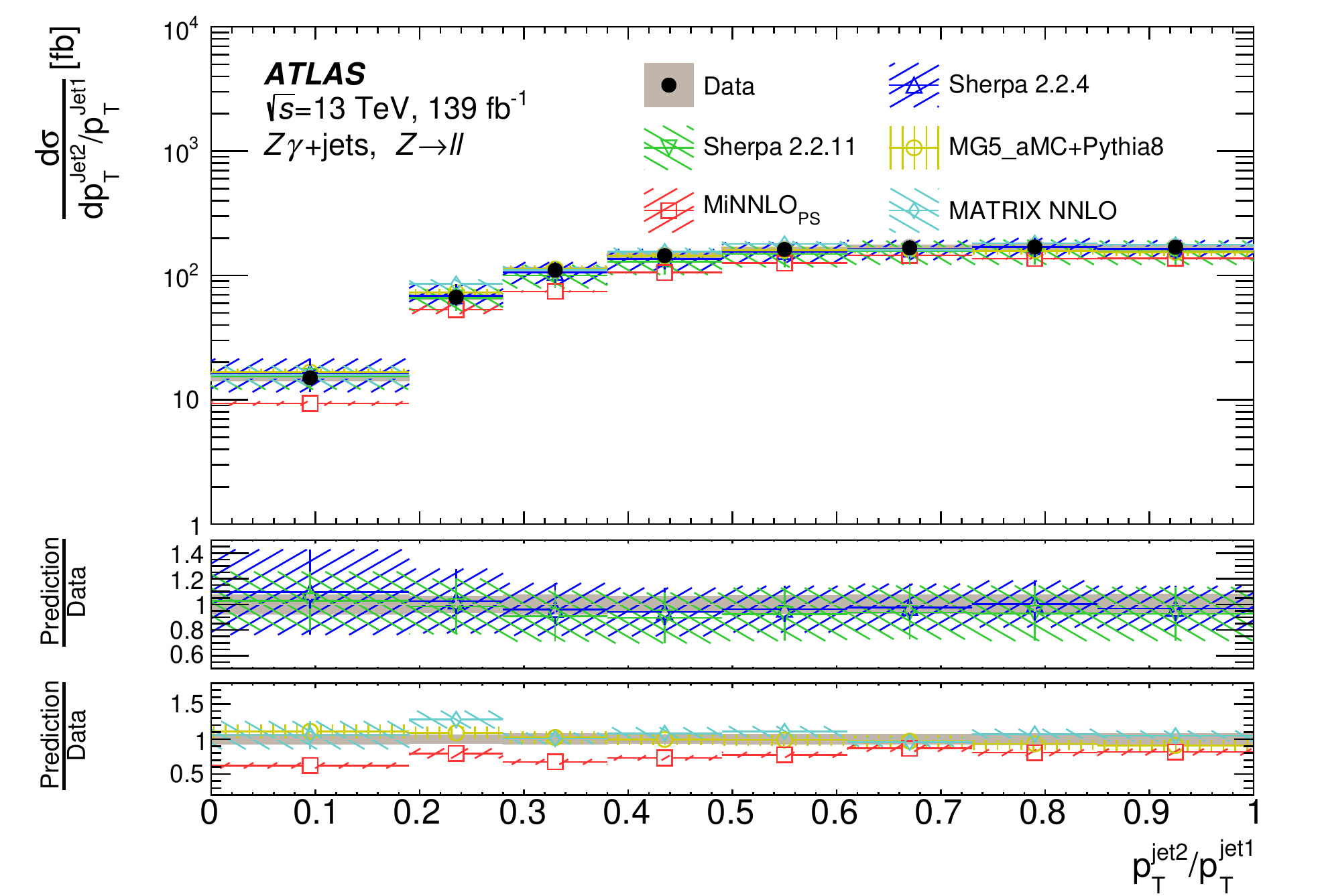}}
\caption{Measured differential cross section (black data points) as a function of the observables
(a) \njet , (b) \pTjetOne, (c) \pTjetTwo, and (d) \Rptj. Error bands on the data points show the statistical uncertainty, while the grey area shows the total uncertainty in the unfolded result. Measured cross sections are compared with SM predictions from event generators at particle level: \Sherpa[2.2.4], \Sherpa[2.2.11], \MGNLOPY[8] (MG5\_aMC+Pythia8 in the legend), and MiNNLO$_\text{PS}$. Fixed-order calculation results from MATRIX NNLO are also shown. Dashed bands represent the statistical uncertainty and theoretical uncertainty (PDF and scale variations). The bottom panels show the ratio of the SM prediction to the measured cross section.}
\label{fig:DiffTheoryPredData2}
\end{figure}
 
\begin{figure}[h!]
\centering
\subfigure[]{\includegraphics[width=.49\columnwidth]{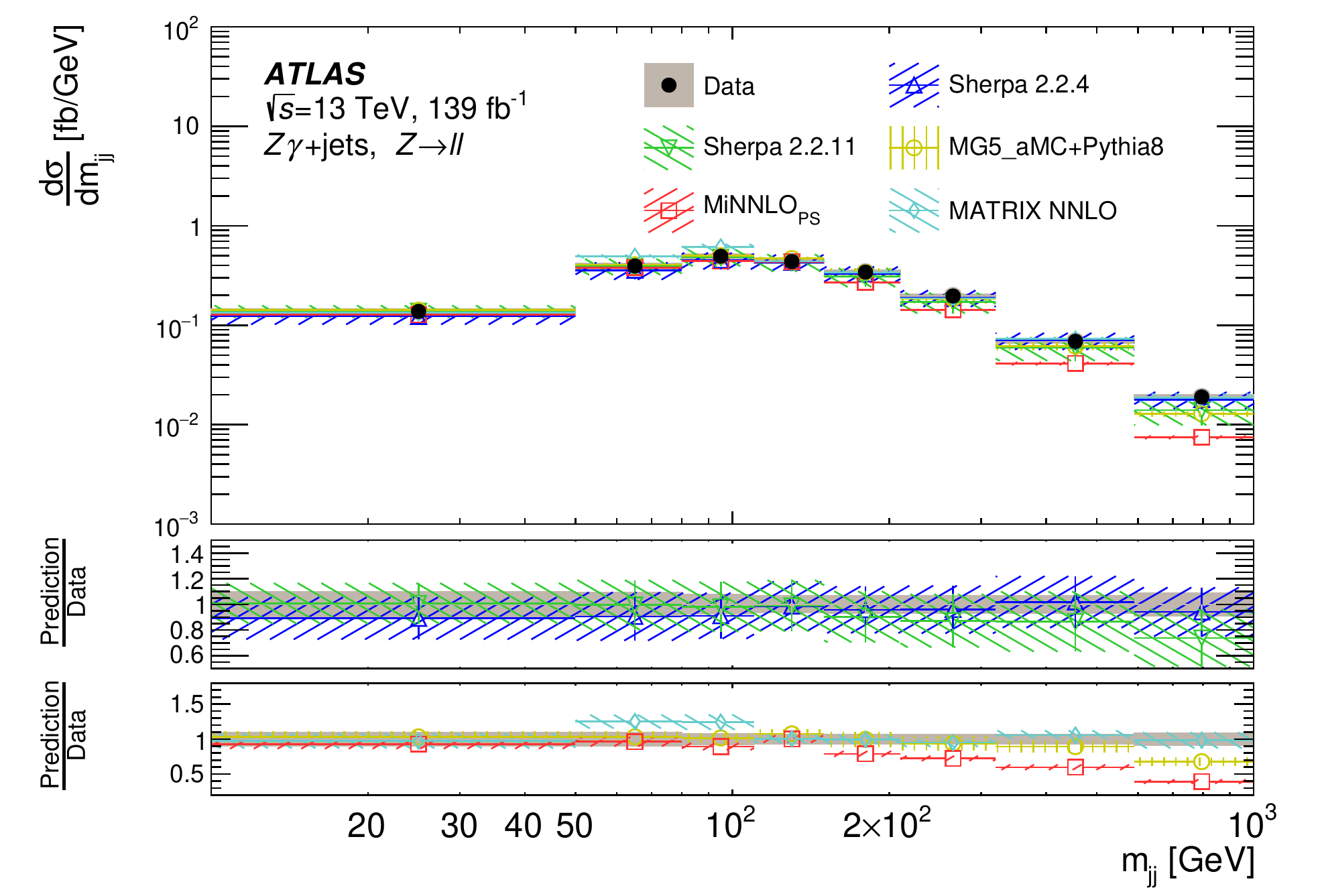}}
\subfigure[]{\includegraphics[width=.49\columnwidth]{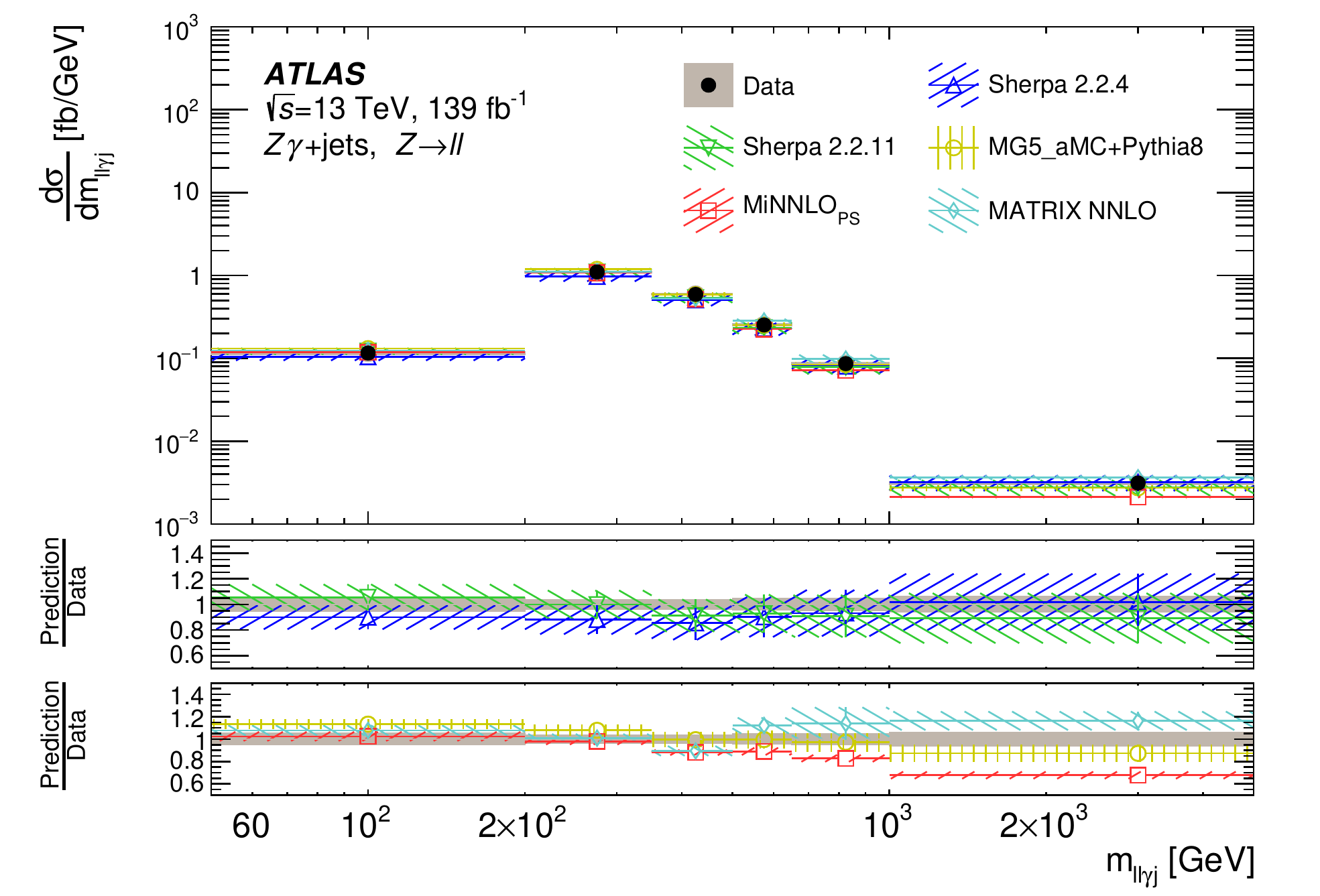}}
\caption{Measured differential cross section (black data points) as a function of the observables
(a) \mjj\ and (b) \mllyj. Error bands on the data points show the statistical uncertainty, while the grey area shows the total uncertainty in the unfolded result. Measured cross sections are compared with SM predictions from event generators at particle level: \Sherpa[2.2.4], \Sherpa[2.2.11], \MGNLOPY[8] (MG5\_aMC+Pythia8 in the legend), and MiNNLO$_\text{PS}$. Fixed-order calculation results from MATRIX NNLO are also shown. Dashed bands represent the statistical uncertainty and theoretical uncertainty (PDF and scale variations). The bottom panels show the ratio of the SM prediction to the measured cross section.}
\label{fig:DiffTheoryPredData3}
\end{figure}

\begin{figure}[h!]
\centering
\subfigure[]{\includegraphics[width=.49\columnwidth]{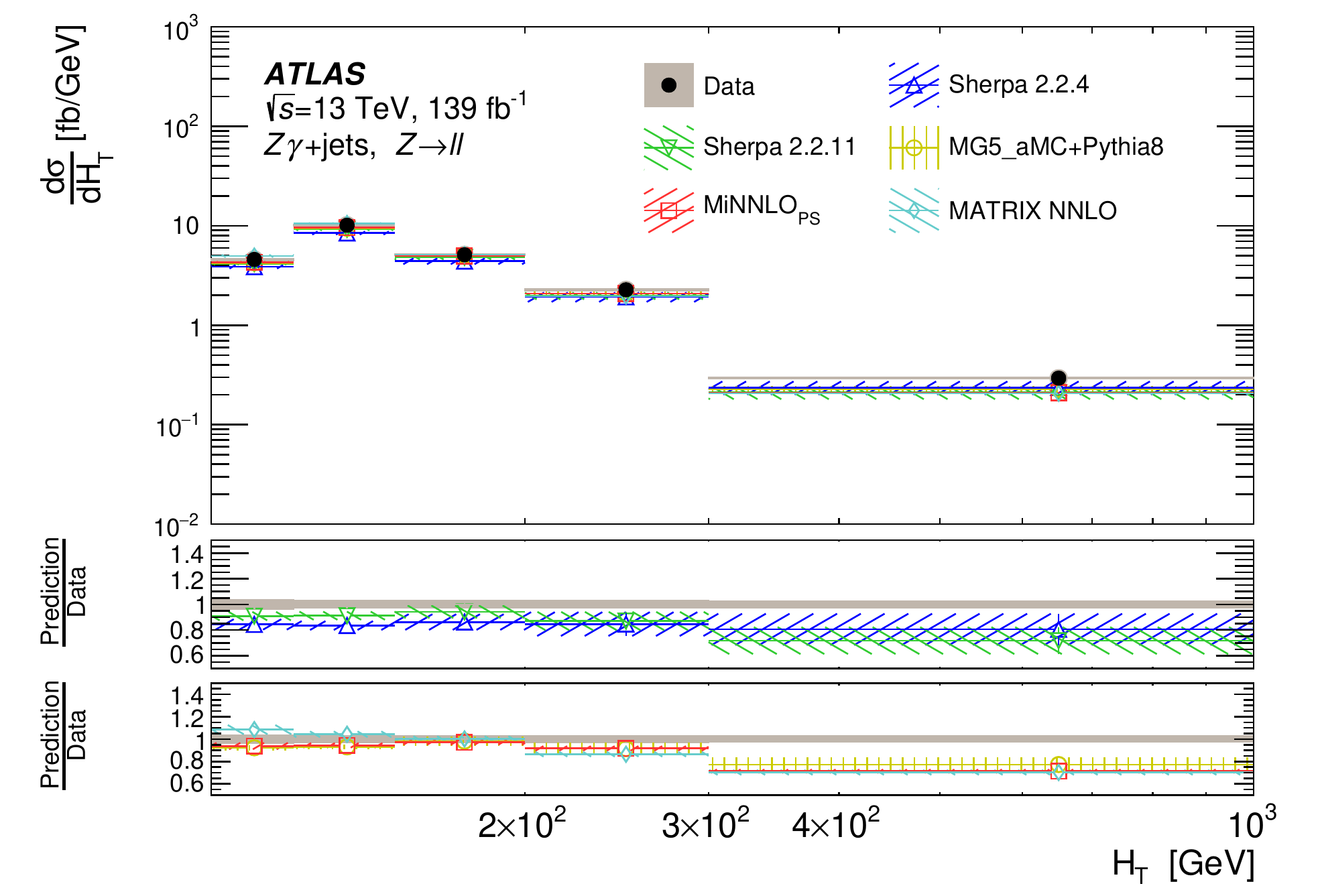}}
\subfigure[]{\includegraphics[width=.49\columnwidth]{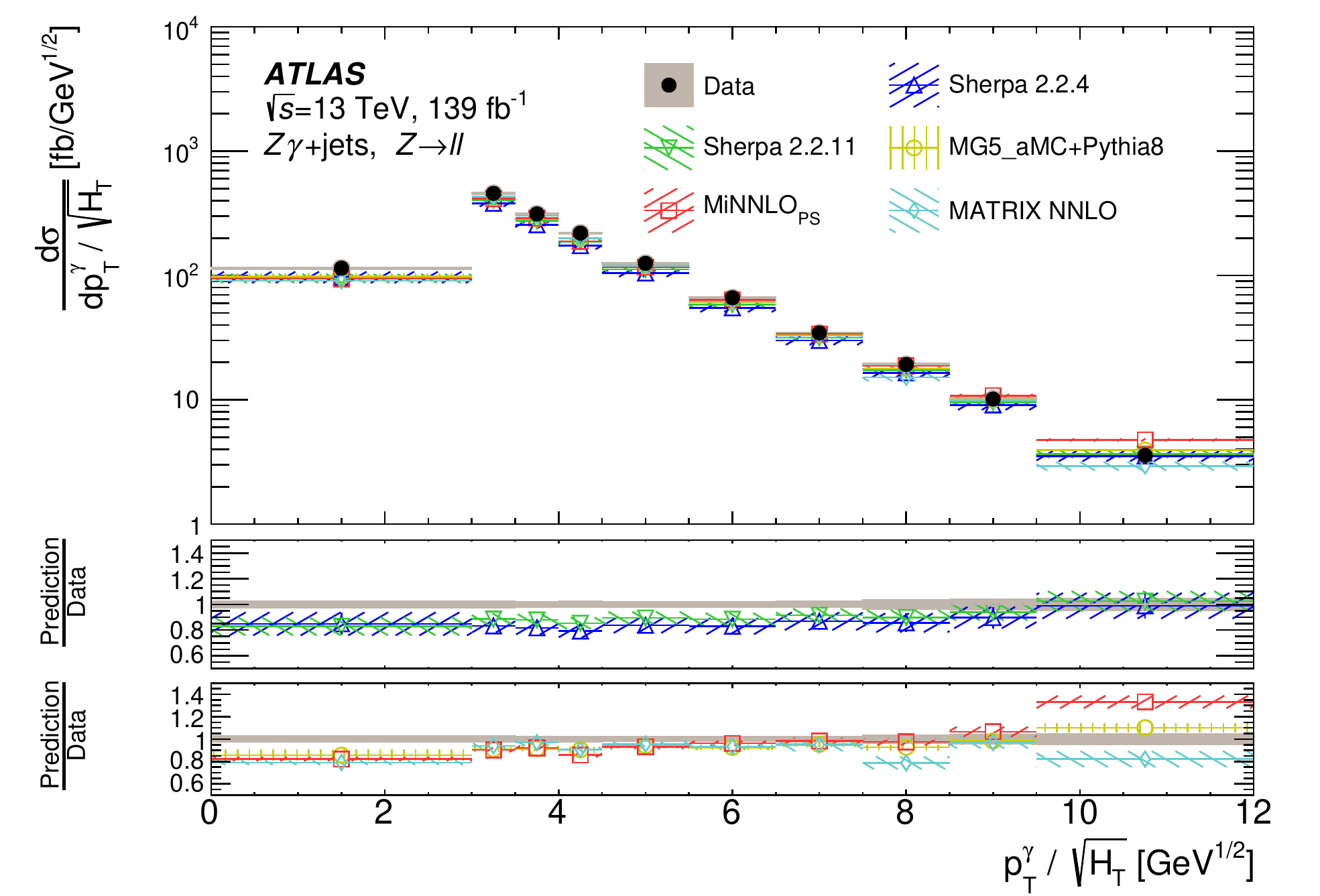}}\\
\subfigure[]{\includegraphics[width=.49\columnwidth]{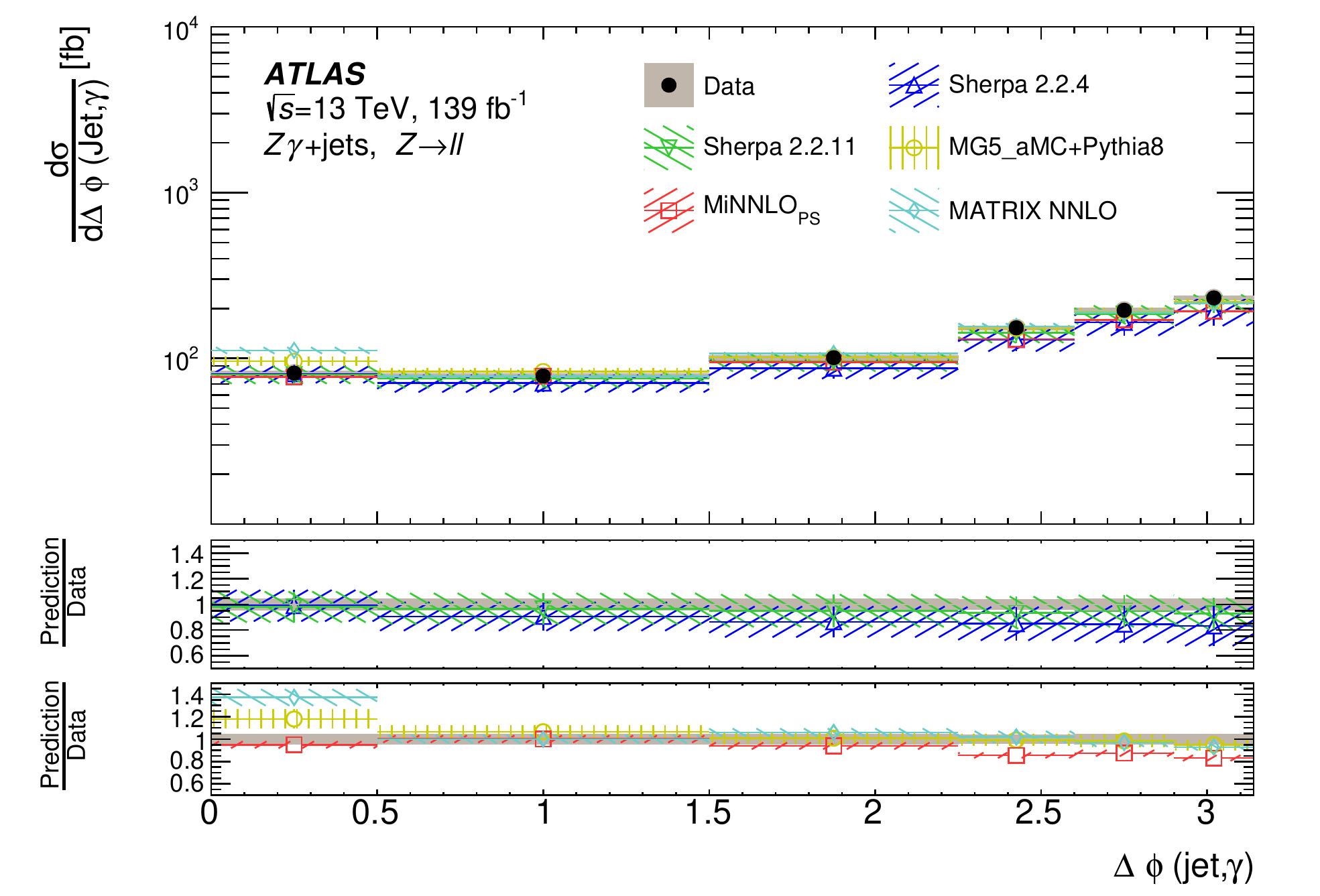}}
\subfigure[]{\includegraphics[width=.49\columnwidth]{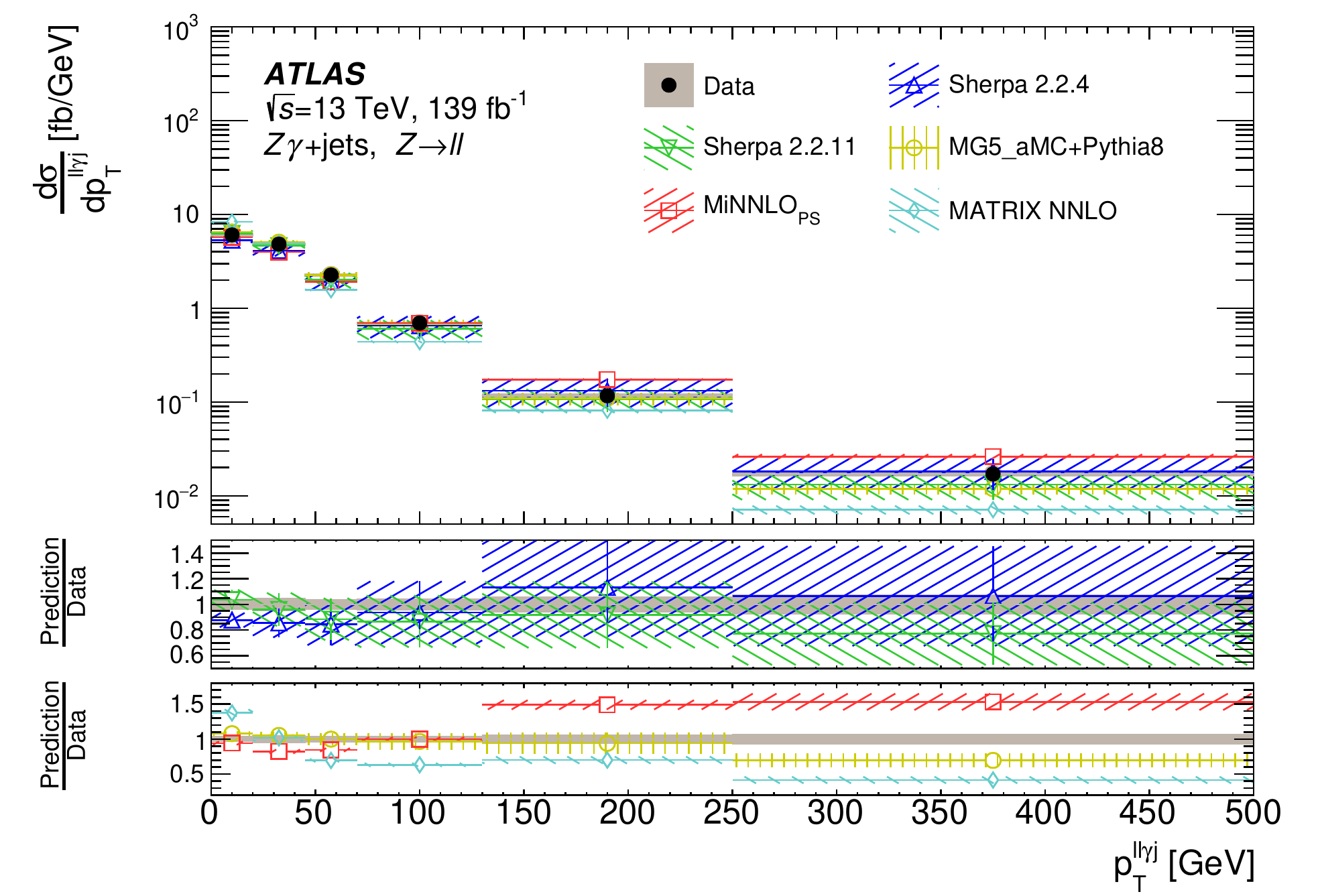}}
\caption{Measured differential cross section (black data points) as a function of the observables
(a) \HT , (b) \ptysqrtHT, (c) \DPhiJetY, and (d) \pTZyj. Error bands on the data points show the statistical uncertainty, while the grey area shows the total uncertainty in the unfolded result. Measured cross sections are compared with SM predictions from event generators at particle level: \Sherpa[2.2.4], \Sherpa[2.2.11], \MGNLOPY[8] (MG5\_aMC+Pythia8 in the legend), and MiNNLO$_\text{PS}$. Fixed-order calculation results from MATRIX NNLO are also shown. Dashed bands represent the statistical uncertainty and theoretical uncertainty (PDF and scale variations). The bottom panels show the ratio of the SM prediction to the measured cross section.}
\label{fig:DiffTheoryPredData4}
\end{figure}

\begin{figure}[h!]
\centering
\subfigure[]{\includegraphics[width=.49\columnwidth]{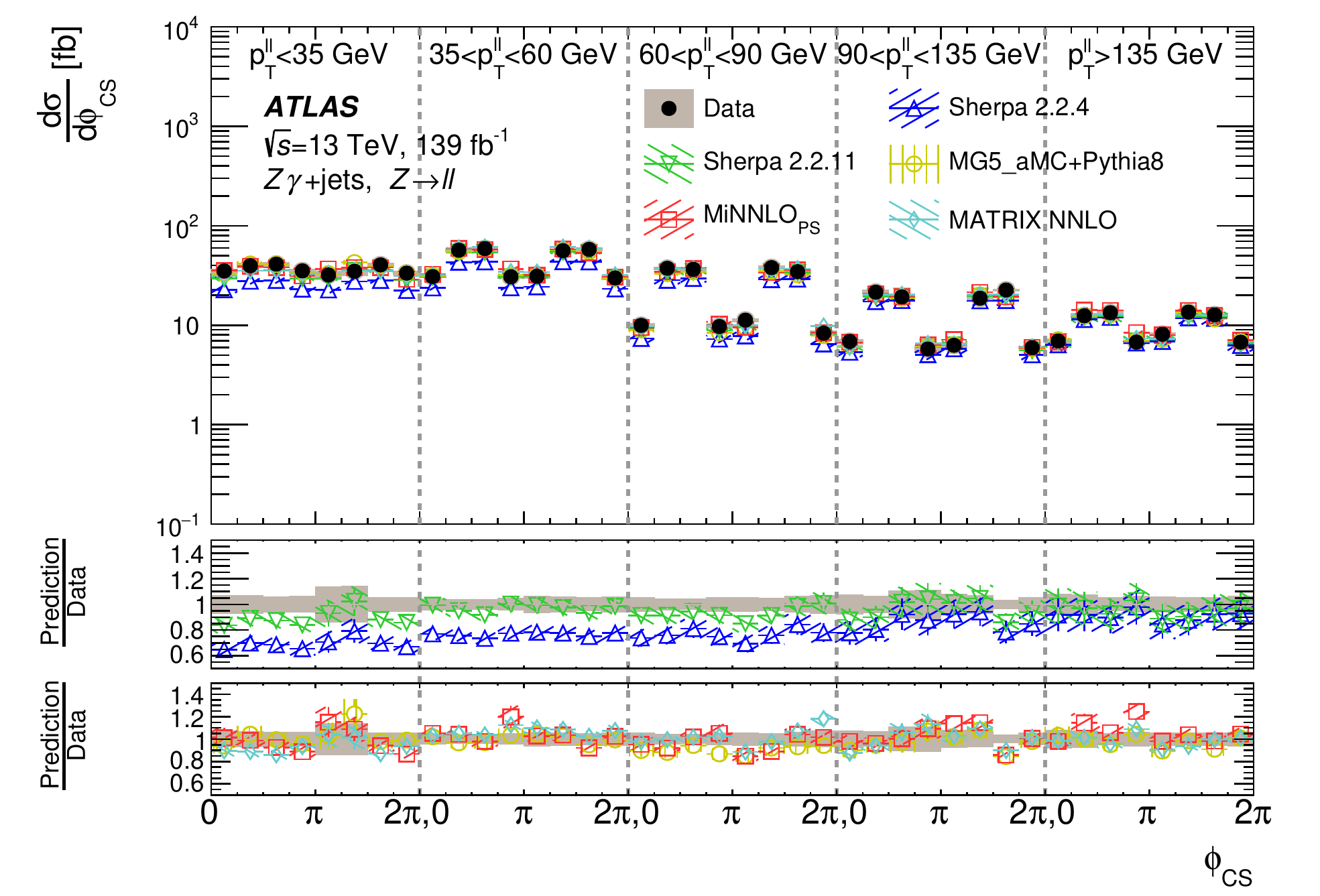}}
\subfigure[]{\includegraphics[width=.49\columnwidth]{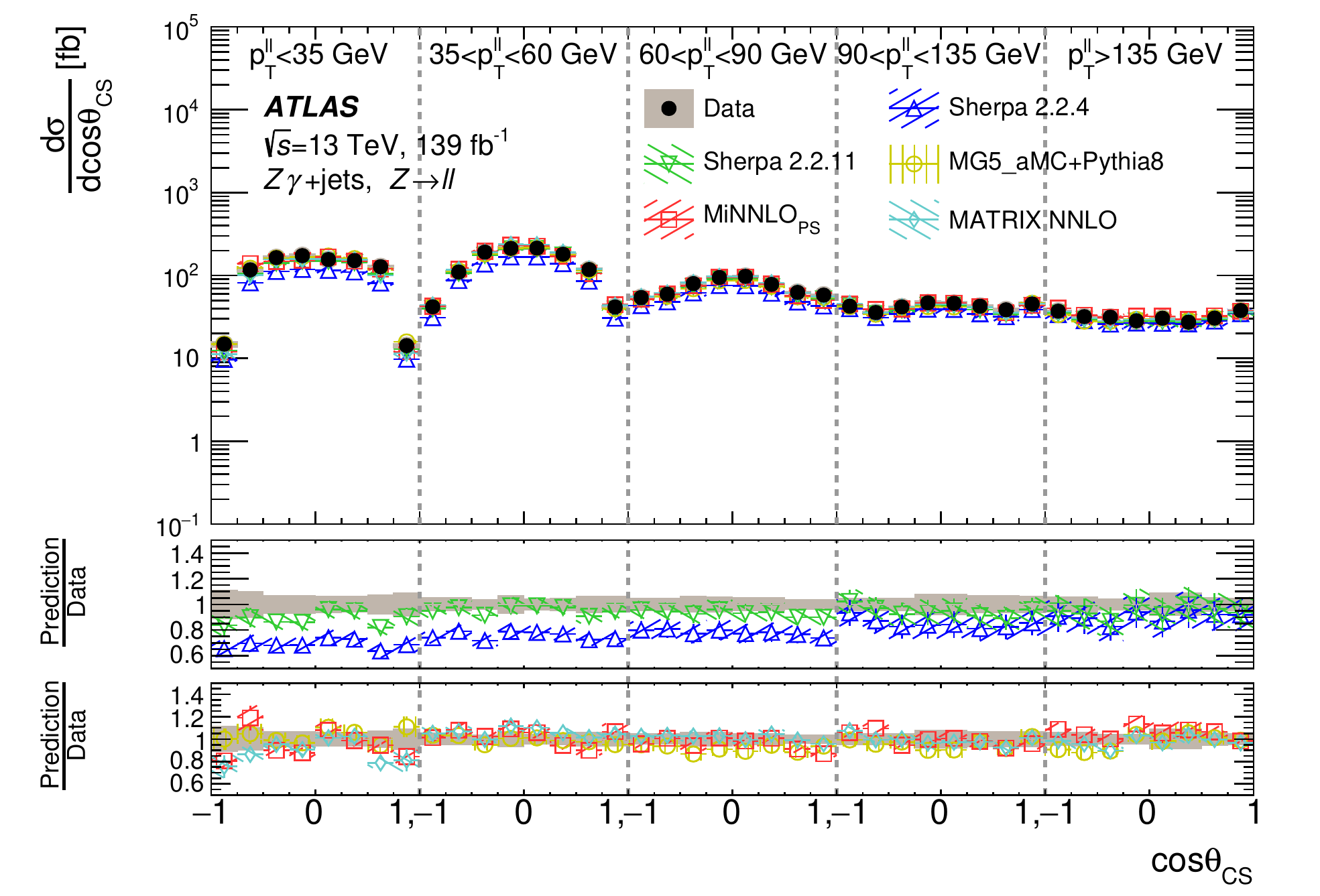}}\\
\caption{Measured differential cross section (black data points) as a function of the observables
(a) $\phi_\text{CS}$ and (b) $\cos\theta_\text{CS}$, in different bins of \pTZ. Error bands on the data points show the statistical uncertainty, while the grey area shows the total uncertainty in the unfolded result. Measured cross sections are compared with SM predictions from event generators at particle level: \Sherpa[2.2.4], \Sherpa[2.2.11], \MGNLOPY[8] (MG5\_aMC+Pythia8 in the legend), and MiNNLO$_\text{PS}$. Fixed-order calculation results from MATRIX NNLO are also shown. Dashed bands represent the statistical uncertainty and theoretical uncertainty (PDF and scale variations). The bottom panels show the ratio of the SM prediction to the measured cross section.}
\label{fig:DiffTheoryPredData5}
\end{figure}

\begin{figure}[h!]
\centering
\subfigure[]{\includegraphics[width=.49\columnwidth]{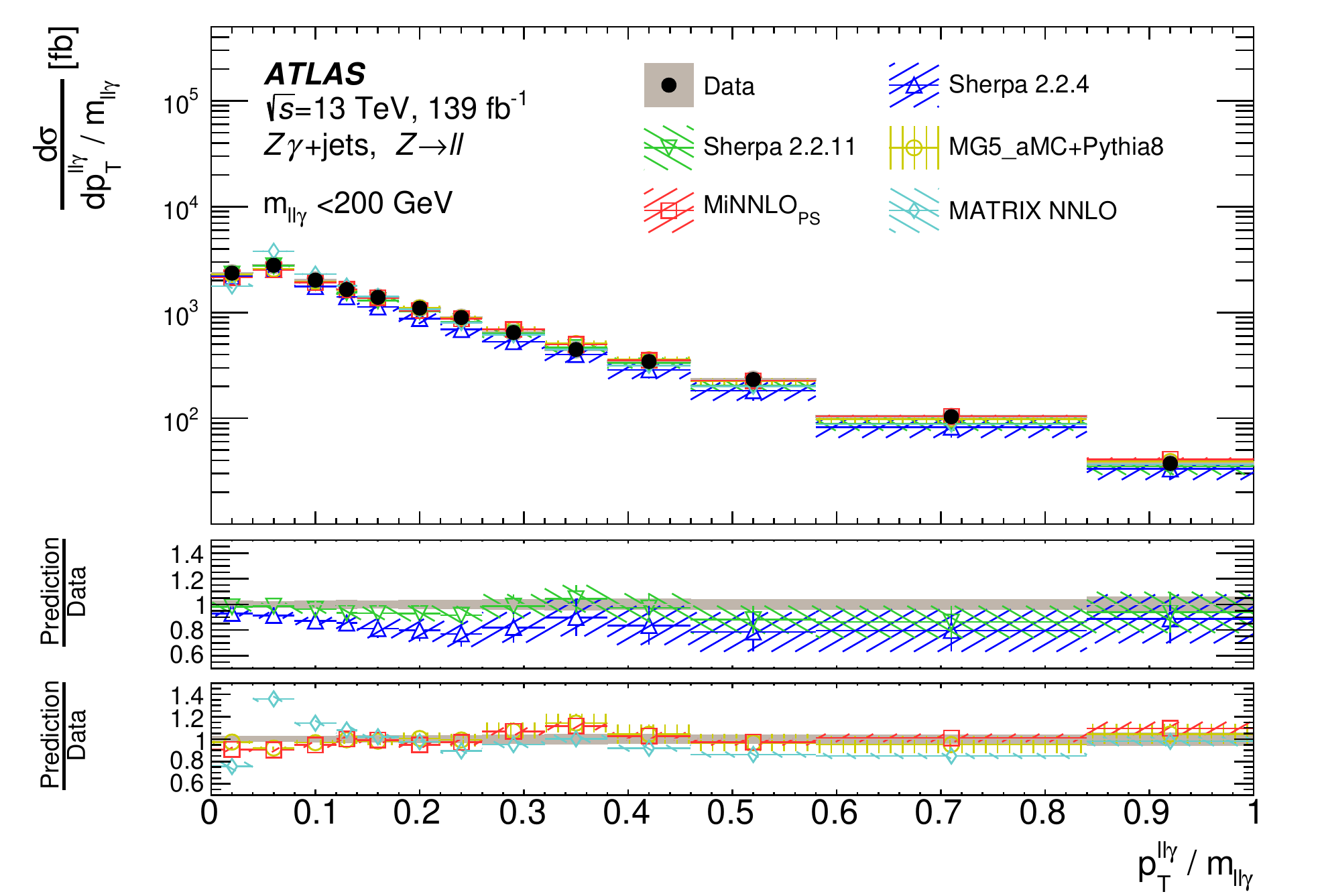}}
\subfigure[]{\includegraphics[width=.49\columnwidth]{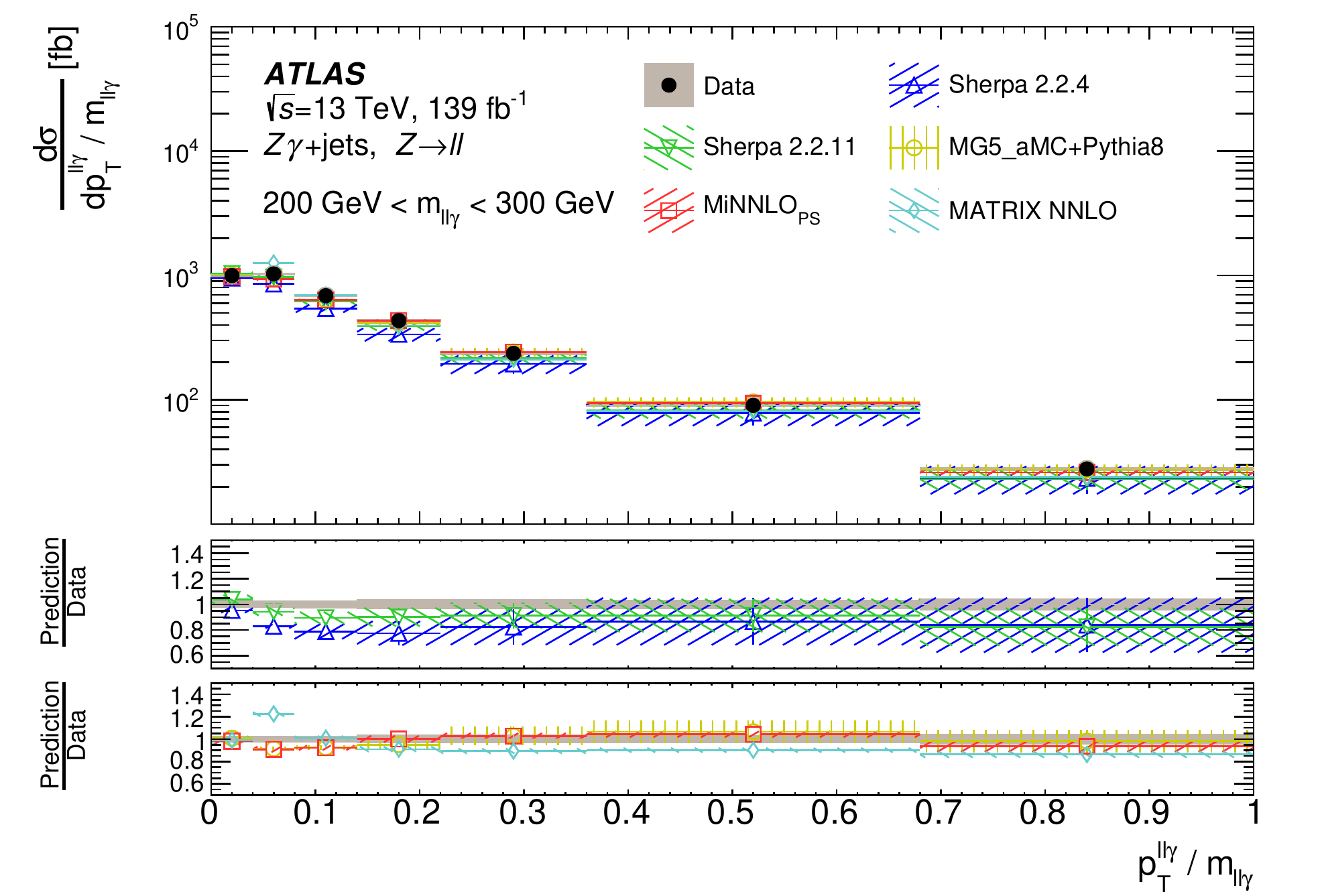}}\\
\subfigure[]{\includegraphics[width=.49\columnwidth]{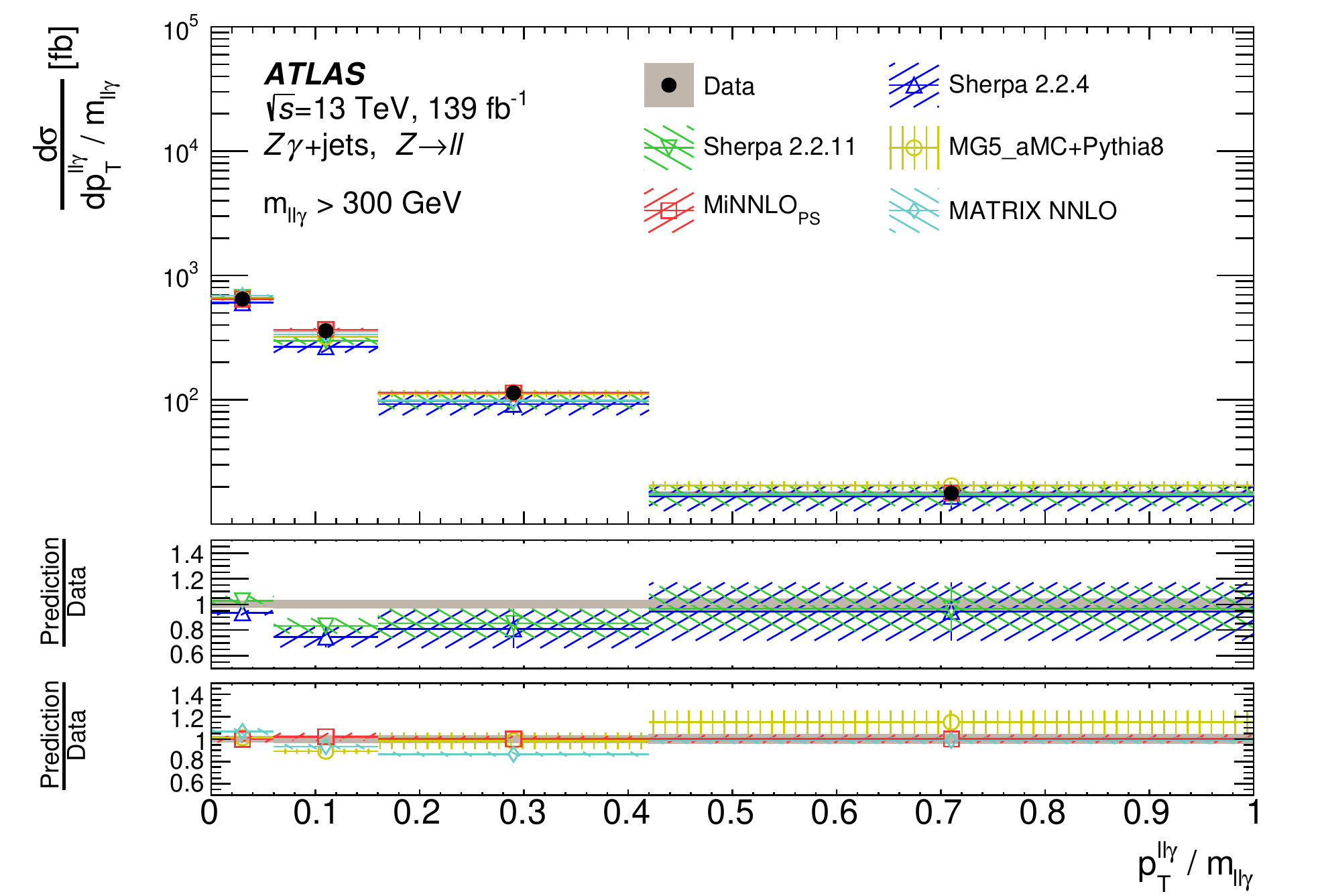}}
\caption{Measured differential cross section (black data points) as a function of \RZy
(a) in $\mZy < 200~\GeV$, (b) in $200~\GeV < \mZy<300~\GeV$, and (c) in $\mZy > 300~\GeV$. Error bands on the data points show the statistical uncertainty, while the grey area shows the total uncertainty in the unfolded result. Measured cross sections are compared with SM predictions from event generators at particle level: \Sherpa[2.2.4], \Sherpa[2.2.11], \MGNLOPY[8] (MG5\_aMC+Pythia8 in the legend), and MiNNLO$_\text{PS}$. Fixed-order calculation results from MATRIX NNLO are also shown. Dashed bands represent the statistical uncertainty and theoretical uncertainty (PDF and scale variations). The bottom panels show the ratio of the SM prediction to the measured cross section.}
\label{fig:DiffTheoryPredData6}
\end{figure}

\begin{figure}[h!]
\centering
\subfigure[]{\includegraphics[width=.49\columnwidth]{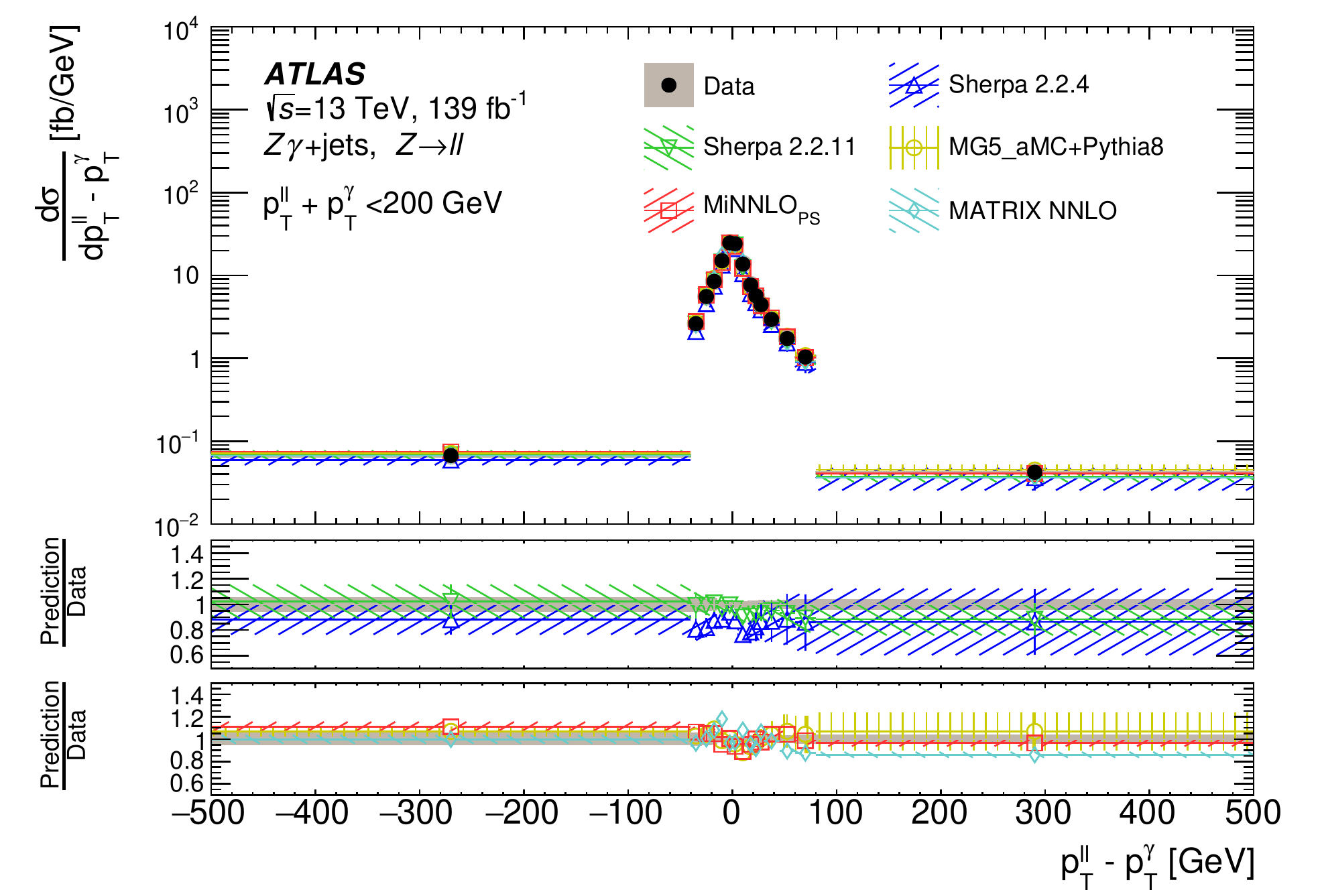}}
\subfigure[]{\includegraphics[width=.49\columnwidth]{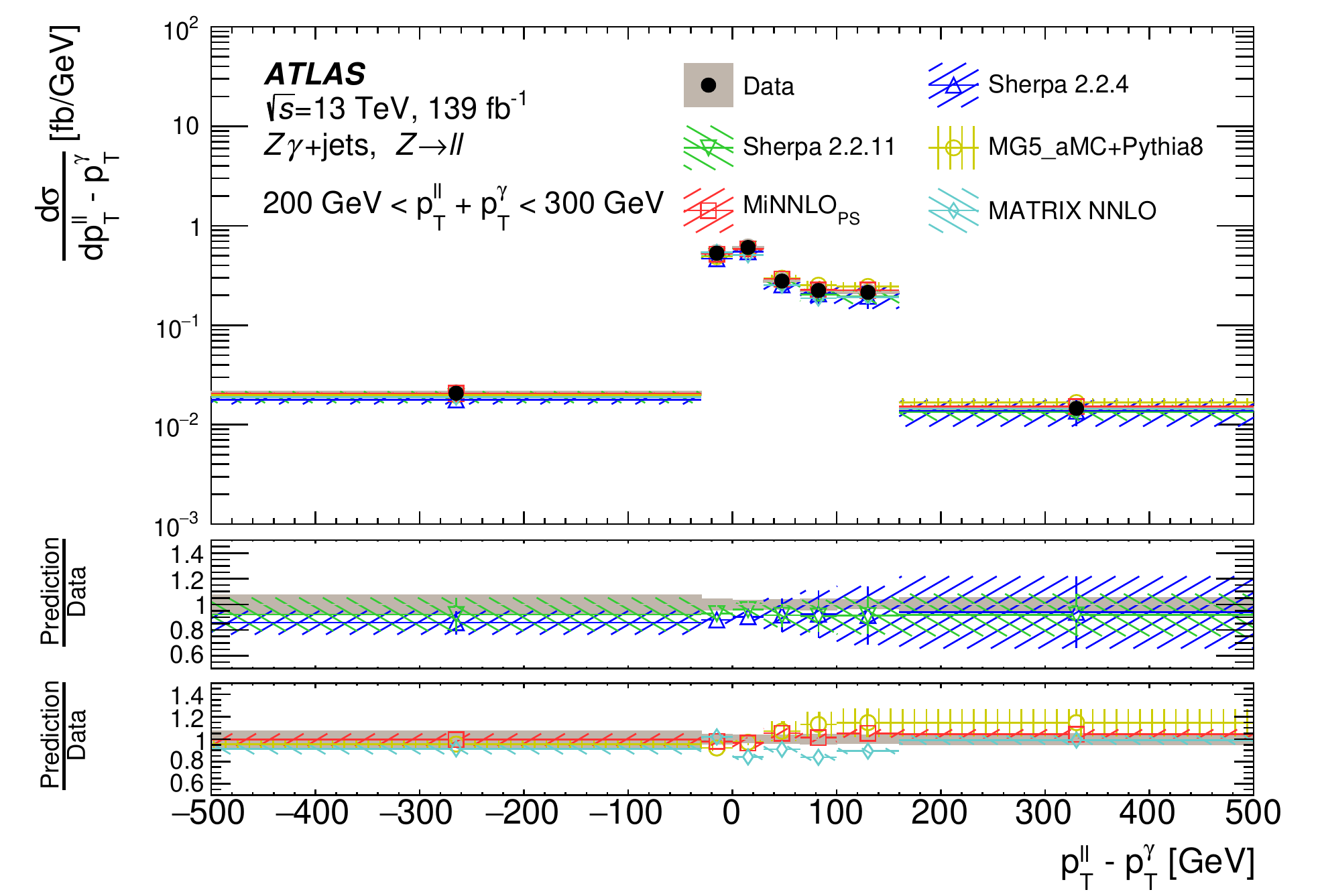}}\\
\subfigure[]{\includegraphics[width=.49\columnwidth]{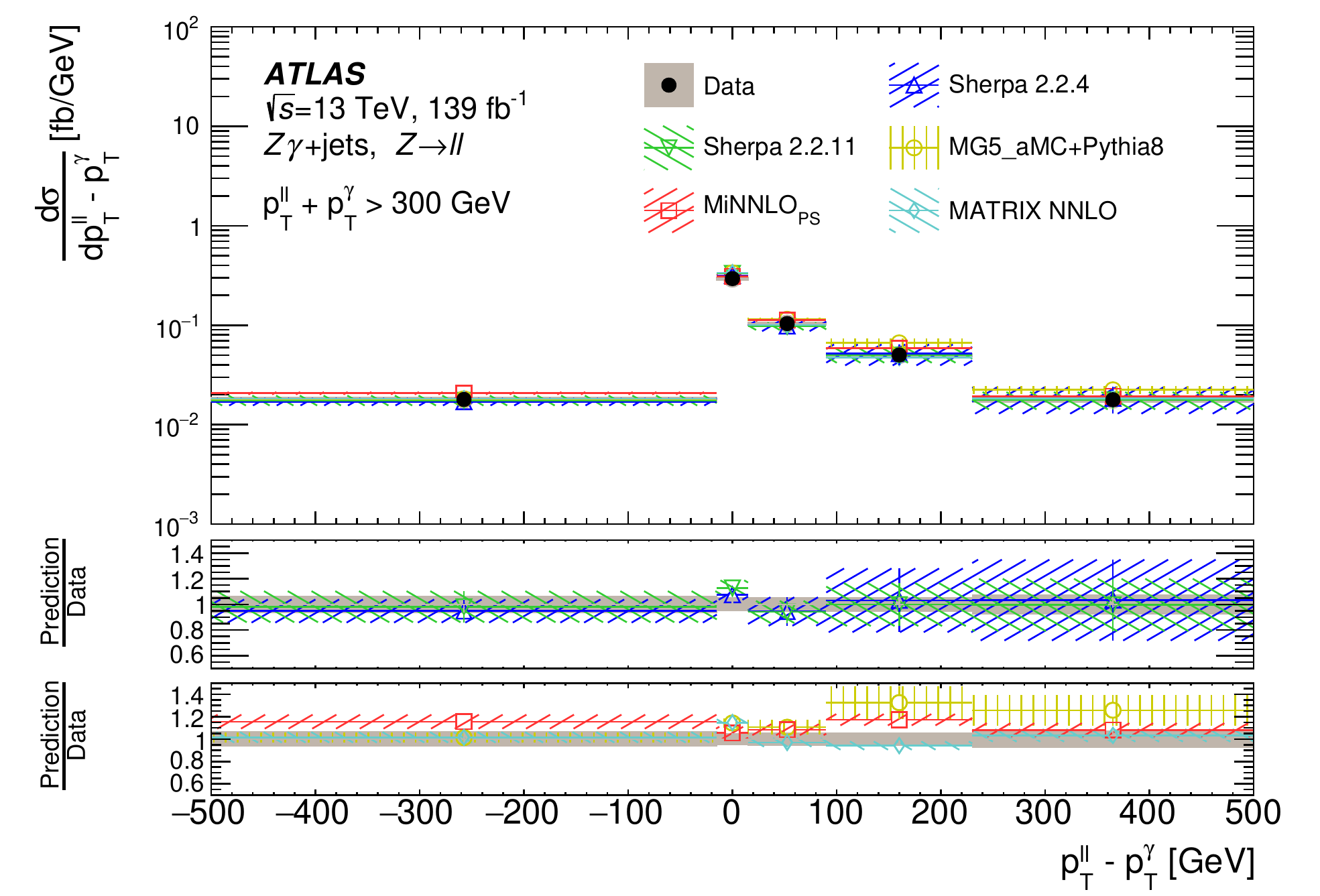}}
\caption{Measured differential cross section (black data points) as a function of \DiffpT
(a) in $\SumpT < 200~\GeV$, (b) in $200~\GeV < \SumpT<300~\GeV$, and (c) in $\SumpT > 300~\GeV$. Error bands on the data points show the statistical uncertainty, while the grey area shows the total uncertainty in the unfolded result. Measured cross sections are compared with SM predictions from event generators at particle level: \Sherpa[2.2.4], \Sherpa[2.2.11], \MGNLOPY[8] (MG5\_aMC+Pythia8 in the legend), and MiNNLO$_\text{PS}$. Fixed-order calculation results using MATRIX NNLO are also shown. Dashed bands represent the statistical uncertainty and theoretical uncertainty (PDF and scale variations). The bottom panels show the ratio of the SM prediction to the measured cross section.}
\label{fig:DiffTheoryPredData7}
\end{figure}

\begin{figure}[h!]
\centering
\subfigure[]{\includegraphics[width=.49\columnwidth]{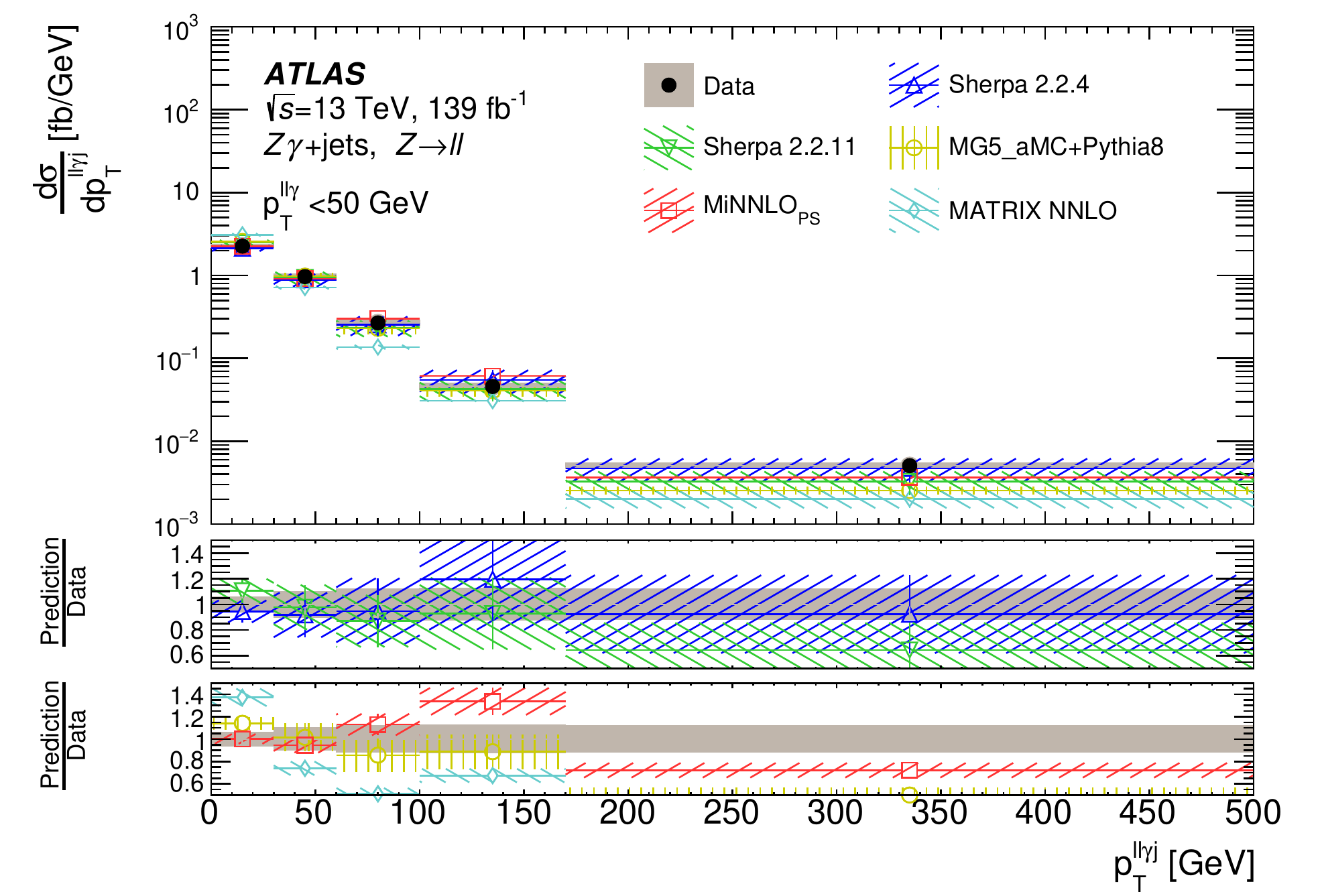}}
\subfigure[]{\includegraphics[width=.49\columnwidth]{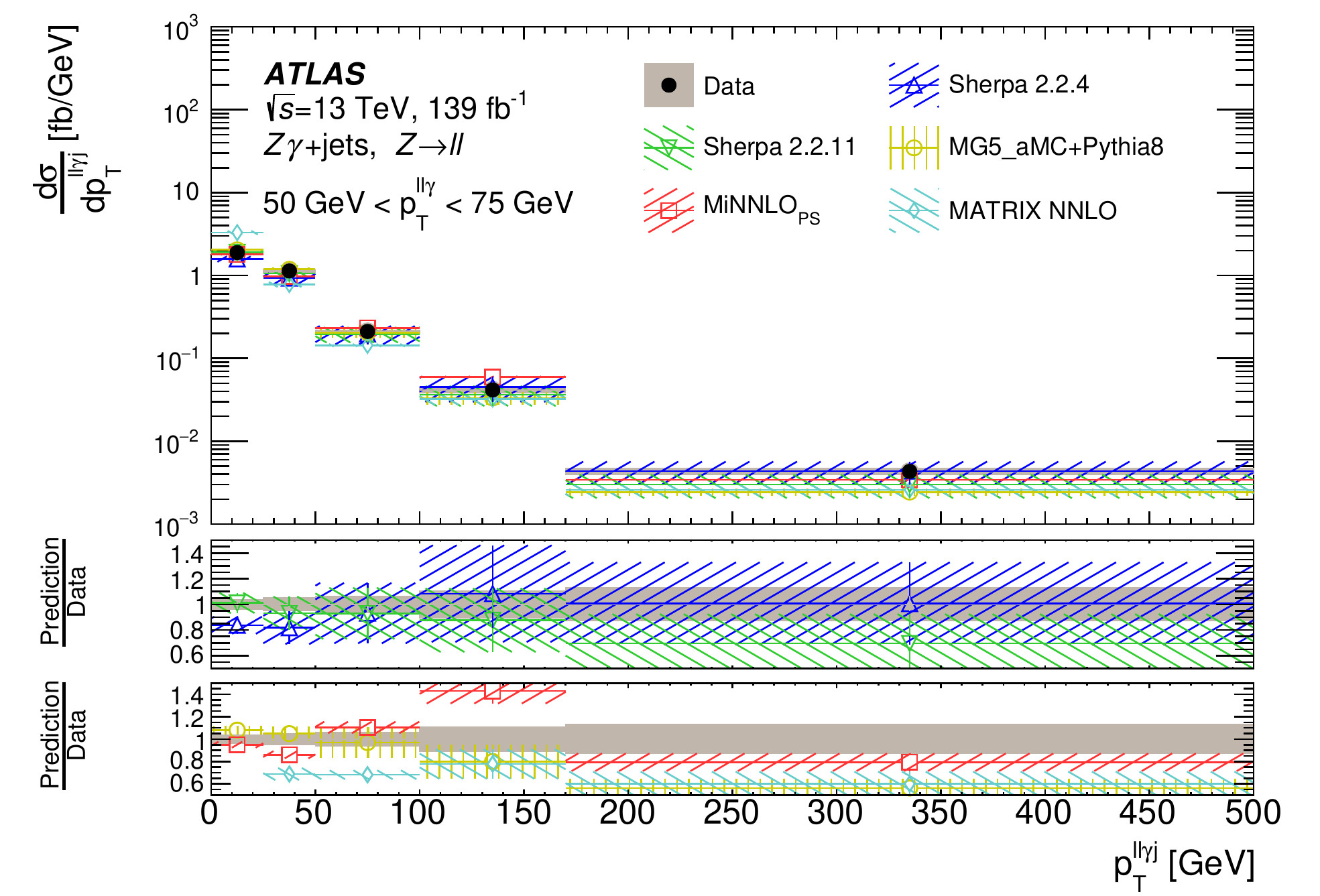}}\\
\subfigure[]{\includegraphics[width=.49\columnwidth]{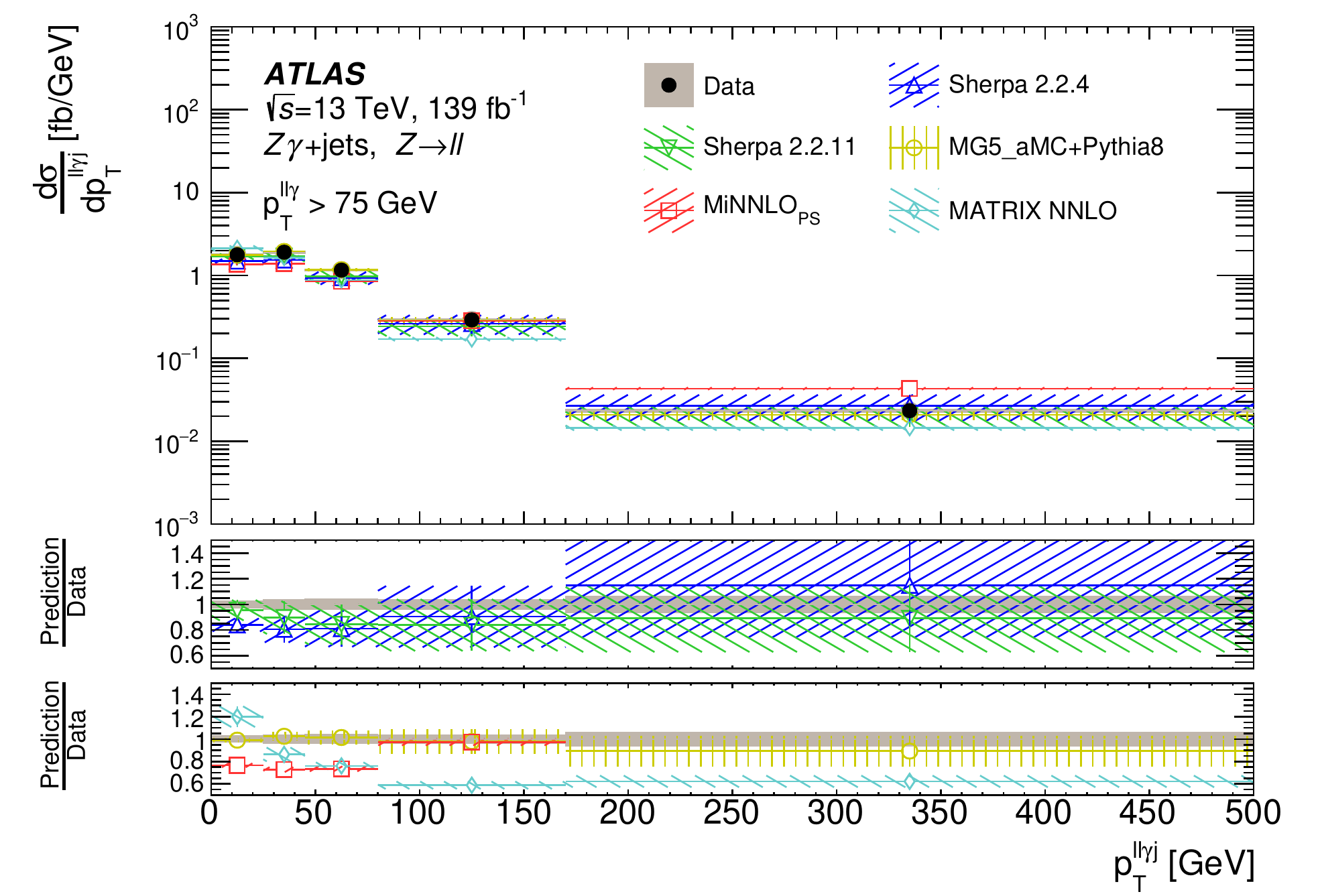}}
\caption{Measured differential cross section (black data points) as a function of \pTZyj
(a) in $\pTZy<50~\GeV$, (b) in $50~\GeV < \pTZy<75~\GeV$, and (c) in $75~\GeV < \pTZy$. Error bands on the data points show the statistical uncertainty, while the grey area shows the total uncertainty in the unfolded result. Measured cross sections are compared with SM predictions from event generators at particle level: \Sherpa[2.2.4], \Sherpa[2.2.11], \MGNLOPY[8] (MG5\_aMC+Pythia8 in the legend), and MiNNLO$_\text{PS}$. Fixed-order calculation results from MATRIX NNLO are also shown. Dashed bands represent the statistical uncertainty and theoretical uncertainty (PDF and scale variations). The bottom panels show the ratio of the SM prediction to the measured cross section.}
\label{fig:DiffTheoryPredData8}
\end{figure}


\FloatBarrier

\section{Conclusion}

Measurements of several differential cross sections for \Zg production in association with jets are presented, in the final state where the $Z$ boson decays into two opposite-sign same-flavour leptons ($e^+e^-$ or $\mu^+\mu^-$). The measurements are performed in a fiducial phase space enhanced in ISR photons, where the sum of the invariant mass of the leptons and the invariant mass of the leptons and the photon is greater than twice the mass of the $Z$ boson. The measurements are performed using data collected by the ATLAS detector from LHC $pp$ collisions at $\sqrt{s}=13~\TeV$, using  a total integrated luminosity of 139~fb$^{-1}$.
 
Differential cross sections are measured as functions of the kinematics of jets, leptons, and photons. Both one-dimensional and two-dimensional distributions are chosen to enhance the separation of hard-scatter effects from soft collinear radiation. A precise measurement of \Zg production in association with jets is obtained, with a total uncertainty between 4\% and 10\% depending on the number of jets. The results are compared with QCD predictions from MC generators involving different precision of multileg merging at LO and NLO, as well as recent predictions at NNLO, including from MiNNLO$_{\textrm{PS}}$, and fixed-order calculations such as with MATRIX. The predictions are in general in good agreement with the measurements within the experimental uncertainties. Jet activity is generally well described, but some trends are observed in the different predictions. Observables sensitive to polarisation effects of the $Z$ boson are well modelled by all predictions. The measurements of \Zg production in association with jets have the potential to constrain the QCD predictions and improve resummation calculations in regions where Sudakov-logarithm terms dominate.


\section*{Acknowledgements}


We thank CERN for the very successful operation of the LHC, as well as the
support staff from our institutions without whom ATLAS could not be
operated efficiently.
 
We acknowledge the support of
ANPCyT, Argentina;
YerPhI, Armenia;
ARC, Australia;
BMWFW and FWF, Austria;
ANAS, Azerbaijan;
CNPq and FAPESP, Brazil;
NSERC, NRC and CFI, Canada;
CERN;
ANID, Chile;
CAS, MOST and NSFC, China;
Minciencias, Colombia;
MEYS CR, Czech Republic;
DNRF and DNSRC, Denmark;
IN2P3-CNRS and CEA-DRF/IRFU, France;
SRNSFG, Georgia;
BMBF, HGF and MPG, Germany;
GSRI, Greece;
RGC and Hong Kong SAR, China;
ISF and Benoziyo Center, Israel;
INFN, Italy;
MEXT and JSPS, Japan;
CNRST, Morocco;
NWO, Netherlands;
RCN, Norway;
MEiN, Poland;
FCT, Portugal;
MNE/IFA, Romania;
MESTD, Serbia;
MSSR, Slovakia;
ARRS and MIZ\v{S}, Slovenia;
DSI/NRF, South Africa;
MICINN, Spain;
SRC and Wallenberg Foundation, Sweden;
SERI, SNSF and Cantons of Bern and Geneva, Switzerland;
MOST, Taiwan;
TENMAK, T\"urkiye;
STFC, United Kingdom;
DOE and NSF, United States of America.
In addition, individual groups and members have received support from
BCKDF, CANARIE, Compute Canada and CRC, Canada;
PRIMUS 21/SCI/017 and UNCE SCI/013, Czech Republic;
COST, ERC, ERDF, Horizon 2020 and Marie Sk{\l}odowska-Curie Actions, European Union;
Investissements d'Avenir Labex, Investissements d'Avenir Idex and ANR, France;
DFG and AvH Foundation, Germany;
Herakleitos, Thales and Aristeia programmes co-financed by EU-ESF and the Greek NSRF, Greece;
BSF-NSF and MINERVA, Israel;
Norwegian Financial Mechanism 2014-2021, Norway;
NCN and NAWA, Poland;
La Caixa Banking Foundation, CERCA Programme Generalitat de Catalunya and PROMETEO and GenT Programmes Generalitat Valenciana, Spain;
G\"{o}ran Gustafssons Stiftelse, Sweden;
The Royal Society and Leverhulme Trust, United Kingdom.
 
The crucial computing support from all WLCG partners is acknowledged gratefully, in particular from CERN, the ATLAS Tier-1 facilities at TRIUMF (Canada), NDGF (Denmark, Norway, Sweden), CC-IN2P3 (France), KIT/GridKA (Germany), INFN-CNAF (Italy), NL-T1 (Netherlands), PIC (Spain), ASGC (Taiwan), RAL (UK) and BNL (USA), the Tier-2 facilities worldwide and large non-WLCG resource providers. Major contributors of computing resources are listed in Ref.~\cite{ATL-SOFT-PUB-2021-003}.


\clearpage

\printbibliography

@Article{as:7:1,
  author =   	"{B. Efron}",
  title =    	"{Bootstrap Methods: Another Look at the Jackknife}",
  journal =  	"Annals Statist.",
  volume =   	"7",
  year =     	"1979",
  pages =    	"1",
	doi        	= "10.1214/aos/1176344552",
	primaryClass   = "hep-ph",
}

@article{Grazzini:2017mhc,
    author = "Grazzini, Massimiliano and Kallweit, Stefan and Wiesemann, Marius",
    title = "{Fully differential NNLO computations with MATRIX}",
    eprint = "1711.06631",
    archivePrefix = "arXiv",
    primaryClass = "hep-ph",
    reportNumber = "ZU-TH-30-17, CERN-TH-2017-232, ZU-TH 30/17",
    doi = "10.1140/epjc/s10052-018-5771-7",
    journal = "Eur. Phys. J. C",
    volume = "78",
    number = "7",
    pages = "537",
    year = "2018"
    
    }

@article{Gehrmann:2011ab,
      author         = "Gehrmann, Thomas and Tancredi, Lorenzo",
      title          = "{Two-loop QCD helicity amplitudes for $q\bar{q} \rightarrow  W^{\pm}\gamma$ and $q\bar{q} \rightarrow  Z^0\gamma$  }",
      journal        = "JHEP",
      volume         = "02",
      year           = "2012",
      pages          = "004",
      doi            = "10.1007/JHEP02(2012)004",
      eprint         = "1112.1531",
      archivePrefix  = "arXiv",
      primaryClass   = "hep-ph",
      reportNumber   = "ZU-TH-26-11",
      SLACcitation   = "%%CITATION = ARXIV:1112.1531;%%"
}

@article{Cascioli:2011va,
      author         = "Cascioli, Fabio and Maierhofer, Philipp and Pozzorini,
                        Stefano",
      title          = "{Scattering Amplitudes with Open Loops}",
      journal        = "Phys. Rev. Lett.",
      volume         = "108",
      year           = "2012",
      pages          = "111601",
      doi            = "10.1103/PhysRevLett.108.111601",
      eprint         = "1111.5206",
      archivePrefix  = "arXiv",
      primaryClass   = "hep-ph",
      reportNumber   = "ZU-TH-23-11, LPN11-66",
      SLACcitation   = "%%CITATION = ARXIV:1111.5206;%%"
}

@article{Denner:2016kdg,
      author         = "Denner, Ansgar and Dittmaier, Stefan and Hofer, Lars",
      title          = "{Collier: a fortran-based Complex One-Loop LIbrary in
                        Extended Regularizations}",
      journal        = "Comput. Phys. Commun.",
      volume         = "212",
      year           = "2017",
      pages          = "220-238",
      doi            = "10.1016/j.cpc.2016.10.013",
      eprint         = "1604.06792",
      archivePrefix  = "arXiv",
      primaryClass   = "hep-ph",
      reportNumber   = "FR-PHENO-2016-003, ICCUB-16-016",
      SLACcitation   = "%%CITATION = ARXIV:1604.06792;%%"
}

@article{Catani:2012qa,
      author         = "Catani, Stefano and Cieri, Leandro and de Florian, Daniel
                        and Ferrera, Giancarlo and Grazzini, Massimiliano",
      title          = "{Vector boson production at hadron colliders:
                        hard-collinear coefficients at the NNLO}",
      journal        = "Eur. Phys. J. C",
      volume         = "72",
      year           = "2012",
      pages          = "2195",
      doi            = "10.1140/epjc/s10052-012-2195-7",
      eprint         = "1209.0158",
      archivePrefix  = "arXiv",
      primaryClass   = "hep-ph",
      reportNumber   = "ZU-TH-16-12",
      SLACcitation   = "%%CITATION = ARXIV:1209.0158;%%"
}

@article{Catani:2007vq,
      author         = "Catani, Stefano and Grazzini, Massimiliano",

      title          = "{Next-to-Next-to-Leading-Order Subtraction Formalism in Hadron Collisions and its Application to Higgs-Boson Production at the Large Hadron Collider}",
      journal        = "Phys. Rev. Lett.",
      volume         = "98",
      year           = "2007",
      pages          = "222002",
      doi            = "10.1103/PhysRevLett.98.222002",
      eprint         = "hep-ph/0703012",
      archivePrefix  = "arXiv",
      primaryClass   = "hep-ph",
      SLACcitation   = "%%CITATION = HEP-PH/0703012;%%"
}

@article{Ebert_2017,

    author = "Ebert, Markus A. and Tackmann, Frank J.",
    title = "{Resummation of Transverse Momentum Distributions in Distribution Space}",
    eprint = "1611.08610",
    archivePrefix = "arXiv",
    primaryClass = "hep-ph",
    reportNumber = "DESY-16-215",
    doi = "10.1007/JHEP02(2017)110",
    journal = "JHEP",
    volume = "02",
    pages = "110",
    year = "2017"

}

@article{Krause_2018,
	doi = {10.1140/epjc/s10052-018-5627-1},
  
    eprint = {1708.06283},
    archivePrefix = {arXiv},
    primaryClass = {hep-ph},

	year = 2018,
	month = {feb},
  
	publisher = {Springer Science and Business Media {LLC}},
  
	volume = {78},
  
	number = {2},
  
	author = {Johannes Krause and Frank Siegert},
  
	title = {{NLO} {QCD} predictions for $Z+\gamma+j$ production with Sherpa},
  
	journal = {Eur. Phys. J. C}
}

@article{Wiesemann_2020,
	doi = {10.1016/j.physletb.2020.135718},
  
  
	year = 2020,
	month = {oct},
  
	publisher = {Elsevier {BV}},
  
	volume = {809},
  
	pages = {135718},
  
	author = {Marius Wiesemann and Luca Rottoli and Paolo Torrielli},
  
	title = {The Z$\upgamma$ transverse-momentum spectrum at NNLO+N3LL},
  
	journal = {Phys. Lett. B},
	eprint = {2006.09338},
    archivePrefix = {arXiv},
    primaryClass = {hep-ph}

}

@article{Stirling_2012,
    author = "Stirling, W. J. and Vryonidou, E.",
    title = "{Electroweak gauge boson polarisation at the LHC}",
    eprint = "1204.6427",
    archivePrefix = "arXiv",
    primaryClass = "hep-ph",
    doi = "10.1007/JHEP07(2012)124",
    journal = "JHEP",
    volume = "07",
    pages = "124",
    year = "2012"
    
}

@article{Brivio_2019,
	doi = {10.1016/j.physrep.2018.11.002},
      eprint = {1706.08945},
    archivePrefix = {arXiv},
    primaryClass = {hep-ph},

	year = 2019,
	month = {feb},
  
	publisher = {Elsevier {BV}},
  
	volume = {793},
  
	pages = {1--98},
  
	author = {Ilaria Brivio and Michael Trott},
  
	title = {The standard model as an effective field theory},
  
	journal = {Physics Reports}
}

@article{Gavela_2020,
	doi = {10.1103/physrevlett.124.051802},
      eprint = {1905.12953},
    archivePrefix = {arXiv},
    primaryClass = {hep-ph},

	year = 2020,
	month = {feb},
  
	publisher = {American Physical Society ({APS})},
  
	volume = {124},
  
	number = {5},
   author = {Gavela, M. B. and No, J. M. and Sanz, V. and de Troc\'oniz, J. F.},
 	 journal = {Phys. Rev. Lett.},
	title = {Nonresonant Searches for Axionlike Particles at the {LHC}},

}

@article{PRESKILL1983127,
    author = "Preskill, John and Wise, Mark B. and Wilczek, Frank",
    title = "{Cosmology of the Invisible Axion}",
    reportNumber = "HUTP-82-A048, NSF-ITP-82-103",
    doi = "10.1016/0370-2693(83)90637-8",
    journal = "Phys. Lett. B",
    volume = "120",
    pages = "127--132",
    year = "1983"
}

@article{Lustermans_2019,
    author = "Lustermans, Gillian and Michel, Johannes K. L. and Tackmann, Frank J. and Waalewijn, Wouter J.",
    title = "{Joint two-dimensional resummation in $q_{T}$ and $0$-jettiness at NNLL}",
    eprint = "1901.03331",
    archivePrefix = "arXiv",
    primaryClass = "hep-ph",
    reportNumber = "DESY-19-004, DESY 19-004, NIKHEF 2018-053",
    doi = "10.1007/JHEP03(2019)124",
    journal = "JHEP",
    volume = "03",
    pages = "124",
    year = "2019"
}

@article{Frixione:1998jh,
      author         = "Frixione, Stefano",
      title          = "{Isolated photons in perturbative QCD}",
      journal        = "Phys. Lett. B",
      volume         = "429",
      year           = "1998",
      pages          = "369-374",
      doi            = "10.1016/S0370-2693(98)00454-7",
      eprint         = "hep-ph/9801442",
      archivePrefix  = "arXiv",
      XprimaryClass   = "hep-ph",
      reportNumber   = "ETH-TH-97-40",
      SLACcitation   = "%%CITATION = HEP-PH/9801442;%%"
}

@article{MatrixZy_2015,
    author = "Grazzini, Massimiliano and Kallweit, Stefan and Rathlev, Dirk",
    title = "{$W\gamma$ and $Z\gamma$ production at the LHC in NNLO QCD}",
    eprint = "1504.01330",
    archivePrefix = "arXiv",
    primaryClass = "hep-ph",
    reportNumber = "ZU-TH-04-15, MITP-15-021",
    doi = "10.1007/JHEP07(2015)085",
    journal = "JHEP",
    volume = "07",
    pages = "085",
    year = "2015"
}

@article{ATLAS:JETM-2018-05,
    author         = "{ATLAS Collaboration}",
    collaboration = "ATLAS",
    title = "{Jet energy scale and resolution measured in proton\textendash{}proton collisions at $\sqrt{s}=13$~TeV with the ATLAS detector}",
    eprint = "2007.02645",
    archivePrefix = "arXiv",
    primaryClass = "hep-ex",
    reportNumber = "CERN-EP-2020-083",
    doi = "10.1140/epjc/s10052-021-09402-3",
    journal = "Eur. Phys. J. C",
    volume = "81",
    number = "8",
    pages = "689",
    year = "2021"
}

@article{PDFNNLO2015,
    author = "Ball, Richard D. and others",
    collaboration = "NNPDF",
    title = "{Parton distributions for the LHC Run II}",
    eprint = "1410.8849",
    archivePrefix = "arXiv",
    primaryClass = "hep-ph",
    reportNumber = "EDINBURGH-2014-15, IFUM-1034-FT, CERN-PH-TH-2013-253, OUTP-14-11P, CAVENDISH-HEP-14-11",
    doi = "10.1007/JHEP04(2015)040",
    journal = "JHEP",
    volume = "04",
    pages = "040",
    year = "2015"
}

@article{TopoClust2017,
   title={Topological cell clustering in the ATLAS calorimeters and its performance in LHC Run 1},
   volume={77},
   ISSN={1434-6052},
   url={http://dx.doi.org/10.1140/epjc/s10052-017-5004-5},
   DOI={10.1140/epjc/s10052-017-5004-5},
   number={7},
   journal={The European Physical Journal C},
   publisher={Springer Science and Business Media LLC},
   author={Aad, G. and Abbott, B. and Abdallah, J. and Abdinov, O. and Aben, R. and Abolins, M. and AbouZeid, O. S. and Abramowicz, H. and Abreu, H. and Abreu, R. and Abulaiti, Y. and Acharya, B. S. and et al.},
   year={2017},
   month={Jul} }

@article{ttyCalc2011,
    author = "Melnikov, Kirill and Schulze, Markus and Scharf, Andreas",
    title = "{QCD corrections to top quark pair production in association with a photon at hadron colliders}",
    eprint = "1102.1967",
    archivePrefix = "arXiv",
    primaryClass = "hep-ph",
    doi = "10.1103/PhysRevD.83.074013",
    journal = "Phys. Rev. D",
    volume = "83",
    pages = "074013",
    year = "2011"
}

@article{D0TeV2009,
%%      author         = "Abazov, V. M. and others",
      author         = "{D0 Collaboration}",
      title          = "{Measurement of the $Z\gamma\to\nu\bar{\nu}\gamma$ Production Cross Section and Limits on Anomalous $ZZ\gamma$ and
                        $Z\gamma\gamma$ Couplings in $p\bar{p}$ Collisions at $\sqrt{s} = 1.96$\,TeV}",
      collaboration  = "D0",
      journal        = "Phys. Rev. Lett.",
      volume         = "102",
      year           = "2009",
      pages          = "201802",
      doi            = "10.1103/PhysRevLett.102.201802",
      eprint         = "0902.2157",
      archivePrefix  = "arXiv",
      primaryClass   = "hep-ex",
      reportNumber   = "FERMILAB-PUB-09-046-E",
      SLACcitation   = "%%CITATION = ARXIV:0902.2157;%%"
}

@article{OPAL2000,
  %%      author         = "Abbiendi, G. and others",
      author         = "{OPAL Collaboration}",
      title          = "{Search for trilinear neutral gauge boson couplings in $Z\gamma$ production at $\sqrt{s} = 189$\,GeV at LEP}",
      collaboration  = "OPAL",
      journal        = "Eur. Phys. J. C",
      volume         = "17",
      year           = "2000",
      pages          = "553-566",
      doi            = "10.1007/s100520000490",
      eprint         = "hep-ex/0007016",
      archivePrefix  = "arXiv",
      primaryClass   = "hep-ex",
      reportNumber   = "CERN-EP-2000-067",
      SLACcitation   = "%%CITATION = HEP-EX/0007016;%%"
}

@article{LEPZZZ2007,
%%      author         = "Abdallah, J. and others",
      author         = "{DELPHI Collaboration}",
      title          = "{Study of triple-gauge-boson couplings $ZZZ$, $ZZ\gamma$ and $Z\gamma\gamma$ at LEP}",
      collaboration  = "DELPHI",
      journal        = "Eur. Phys. J. C",
      volume         = "51",
      year           = "2007",
      pages          = "525-542",
      doi            = "10.1140/epjc/s10052-007-0345-0",
      eprint         = "0706.2741",
      archivePrefix  = "arXiv",
      primaryClass   = "hep-ex",
      reportNumber   = "HEP-PH-EP-2006-008",
      SLACcitation   = "%%CITATION = ARXIV:0706.2741;%%"
}

@article{ZyLEP2004,
%%      author         = "Achard, P. and others",
      author         = "{L3 Collaboration}",
      title          = "{Study of the $e^{+} e^{-} \to Z \gamma$ process at LEP and limits on triple neutral-gauge-boson couplings}",
      collaboration  = "L3",
      journal        = "Phys. Lett. B",
      volume         = "597",
      year           = "2004",
      pages          = "119-130",
      doi            = "10.1016/j.physletb.2004.07.002",
      eprint         = "hep-ex/0407012",
      archivePrefix  = "arXiv",
      primaryClass   = "hep-ex",
      reportNumber   = "CERN-PH-EP-2004-014",
      SLACcitation   = "%%CITATION = HEP-EX/0407012;%%"
}

@article{Abazov:2011qp,
%%      author         = "Abazov, Victor Mukhamedovich and others",
      author         = "{D0 Collaboration}",
      title          = "{$Z\gamma$ production and limits on anomalous $ZZ\gamma$
                        and $Z\gamma\gamma$ couplings in $p\bar{p}$ collisions at
                        $\sqrt{s}=1.96$ TeV}",
      collaboration  = "D0",
      journal        = "Phys. Rev. D",
      volume         = "85",
      year           = "2012",
      pages          = "052001",
      doi            = "10.1103/PhysRevD.85.052001",
      eprint         = "1111.3684",
      archivePrefix  = "arXiv",
      primaryClass   = "hep-ex",
      reportNumber   = "FERMILAB-PUB-11-616-E",
      SLACcitation   = "%%CITATION = ARXIV:1111.3684;%%"
}

@misc{lumi, 
  author="{ATLAS luminosity group}",
  howpublished="\url{https://twiki.cern.ch/twiki/bin/viewauth/Atlas/LuminosityForPhysics/}",
  year={2018}
}

@article{LUCID2,
  author={G. Avoni and others},
  title={The new LUCID-2 detector for luminosity measurement and monitoring in ATLAS},
  journal={JINST},
  volume={13},
  number={07},
  pages={P07017},
  doi="10.1088/1748-0221/13/07/P07017",
  year={2018},
}

@article{Butterworth_2016,
   title={PDF4LHC recommendations for LHC Run II},
   volume={43},
   ISSN={1361-6471},
    eprint = {1510.03865},
    archivePrefix = {arXiv},
    primaryClass = {hep-ph},
   DOI={10.1088/0954-3899/43/2/023001},
   number={2},
   journal={J. Phys. G: Nucl. Part. Phys.},
   publisher={IOP Publishing},
   author={Butterworth, Jon and Carrazza, Stefano and Cooper-Sarkar, Amanda and Roeck, Albert De and Feltesse, Joël and Forte, Stefano and Gao, Jun and Glazov, Sasha and Huston, Joey and Kassabov, Zahari and et al.},
   year={2016},
   month={Jan},
   pages={023001}
}

@article{Ebert_2021,
    author = "Ebert, Markus A. and Michel, Johannes K. L. and Stewart, Iain W. and Tackmann, Frank J.",
    title = "{Drell-Yan $q_{T}$ resummation of fiducial power corrections at N$^{3}$LL}",
    eprint = "2006.11382",
    archivePrefix = "arXiv",
    primaryClass = "hep-ph",
    reportNumber = "DESY-20-016, DESY 20-016, MIT-CTP 5205",
    doi = "10.1007/JHEP04(2021)102",
    journal = "JHEP",
    volume = "04",
    pages = "102",
    year = "2021"
}

@article{GEORGI198673,
title = {Manifesting the invisible axion at low energies},
journal = {Phys. Lett. B},
volume = {169},
number = {1},
pages = {73-78},
year = {1986},
issn = {0370-2693},
doi = {https://doi.org/10.1016/0370-2693(86)90688-X},
author = {Howard Georgi and David {B. Kaplan} and Lisa Randall}
}

@article{Ball2017,
   title={Parton distributions from high-precision collider data},
   volume={77},
   ISSN={1434-6052},
   DOI={10.1140/epjc/s10052-017-5199-5},
   number={10},
   journal={Eur. Phys. J. C},
   eprint = {1706.00428},
    archivePrefix = {arXiv},
    primaryClass = {hep-ph},
   publisher={Springer Science and Business Media LLC},
   author={Ball, Richard D. and Bertone, Valerio and Carrazza, Stefano and Debbio, Luigi Del and Forte, Stefano and Groth-Merrild, Patrick and Guffanti, Alberto and Hartland, Nathan P. and Kassabov, Zahari and Latorre, José I. and et al.},
   year={2017},
   month={Oct}
}

@article{DAGOSTINI1995487,
title = {A multidimensional unfolding method based on Bayes' theorem},
journal = {Nucl. Instrum. Methods. Phys. Res. A},
volume = {362},
number = {2},
pages = {487-498},
year = {1995},
issn = {0168-9002},
doi = "10.1016/0168-9002(95)00274-X",
author = {G. D'Agostini},
}

@article{Hoeche:2011fd,
      author         = "H{\"o}che, Stefan and Krauss, Frank and Sch{\"o}nherr, Marek and
                        Siegert, Frank",
      title          = "{A critical appraisal of NLO+PS matching methods}",
      journal        = "JHEP",
      volume         = "09",
      year           = "2012",
      pages          = "049",
      doi            = "10.1007/JHEP09(2012)049",
      eprint         = "1111.1220",
      archivePrefix  = "arXiv",
      primaryClass   = "hep-ph",
      reportNumber   = "SLAC-PUB-14661, IPPP-11-67, DCPT-11-134, LPN11-58,
                        FR-PHENO-2011-019, MCNET-11-24",
      SLACcitation   = "%%CITATION = ARXIV:1111.1220;%%"
}

@article{Hoeche:2012yf,
      author         = "H{\"o}che, Stefan and Krauss, Frank and Sch{\"o}nherr, Marek and
                        Siegert, Frank",
      title          = "{QCD matrix elements + parton showers. The NLO case}",
      journal        = "JHEP",
      volume         = "04",
      year           = "2013",
      pages          = "027",
      doi            = "10.1007/JHEP04(2013)027",
      eprint         = "1207.5030",
      archivePrefix  = "arXiv",
      primaryClass   = "hep-ph",
      reportNumber   = "SLAC-PUB-15191, IPPP-12-52, DCPT-12-104, LPN12-081,
                        FR-PHENO-2012-017, MCNET-12-09, --FR-PHENO-2012-017",
      SLACcitation   = "%%CITATION = ARXIV:1207.5030;%%"
}

@article{Catani:2001cc,
      author         = "Catani, S. and Krauss, F. and Kuhn, R. and Webber, B. R.",
      title          = "{QCD Matrix Elements + Parton Showers}",
      journal        = "JHEP",
      volume         = "11",
      year           = "2001",
      pages          = "063",
      doi            = "10.1088/1126-6708/2001/11/063",
      eprint         = "hep-ph/0109231",
      archivePrefix  = "arXiv",
      reportNumber   = "CERN-TH-2000-367, CAVENDISH-HEP-00-03",
      SLACcitation   = "%%CITATION = HEP-PH/0109231;%%"
}

@Article{Hoeche:2009rj,
      author         = "H{\"o}che, Stefan and Krauss, Frank and Schumann, Steffen and Siegert, Frank",
      title          = "{QCD matrix elements and truncated showers}",
      journal        = "JHEP",
      volume         = "05",
      pages          = "053",
      doi            = "10.1088/1126-6708/2009/05/053",
      year           = "2009",
      eprint         = "0903.1219",
      archivePrefix  = "arXiv",
      primaryClass   = "hep-ph",
}

@article{Bothmann:2019yzt,
      author = "Bothmann, Enrico and others",
      title = "{Event generation with Sherpa 2.2}",
      journal = "SciPost Phys.",
      volume = "7",
      year = "2019",
      number = "3",
      pages = "034",
      doi = "10.21468/SciPostPhys.7.3.034",
      reportNumber = "FERMILAB-PUB-19-218-T, SLAC-PUB-17433, IPPP/19/42, MCNET-19-11",
      eprint = "1905.09127",
      archivePrefix = "arXiv",
      primaryClass = "hep-ph",
}

@article{Alioli:2008gx,
      author         = "Alioli, Simone and Nason, Paolo and Oleari, Carlo and Re,
                        Emanuele",
      title          = "{NLO vector-boson production matched with shower in
                        POWHEG}",
      journal        = "JHEP",
      volume         = "07",
      year           = "2008",
      pages          = "060",
      doi            = "10.1088/1126-6708/2008/07/060",
      eprint         = "0805.4802",
      archivePrefix  = "arXiv",
      primaryClass   = "hep-ph",
      reportNumber   = "NSF-KITP-08-74",
      SLACcitation   = "%%CITATION = ARXIV:0805.4802;%%"
}

@Article{Sjostrand:2007gs,
	author    = "Sj{\"o}strand, T. and Mrenna, S. and Skands, P.",
	title     = "{A brief introduction to PYTHIA 8.1}",
	journal   = "Comput. Phys. Commun.",
	volume    = "178",
	year      = "2008",
	pages     = "852-867",
	eprint    = "0710.3820",
	archivePrefix = "arXiv",
	primaryClass  =  "hep-ph",
	doi       = "10.1016/j.cpc.2008.01.036",
	SLACcitation  = "%%CITATION = 0710.3820;%%"
}

@article{Lai:2010vv,
      author         = "Lai, H.-L. and others",
      title          = "{New parton distributions for collider physics}",
      journal        = "Phys. Rev. D",
      volume         = "82",
      pages          = "074024",
      doi            = "10.1103/PhysRevD.82.074024",
      year           = "2010",
      eprint         = "1007.2241",
      archivePrefix  = "arXiv",
      primaryClass   = "hep-ph",
      reportNumber   = "MSUHEP-100707, SMU-HEP-10-10",
      SLACcitation   = "%%CITATION = ARXIV:1007.2241;%%",
}

@Article{Pumplin:2002vw,
     author    = "Pumplin, J. and others",
     title     = "{New Generation of Parton Distributions with Uncertainties from Global QCD Analysis}",
     journal   = "JHEP",
     volume    = "07",
     year      = "2002",
     pages     = "012",
     doi       = "10.1088/1126-6708/2002/07/012",
     eprint    = "hep-ph/0201195",
     archivePrefix = "arXiv",
     SLACcitation  = "%%CITATION = HEP-PH/0201195;%%"
}

@article{Sjostrand:2014zea,
      author         = "Sj{\"o}strand, Torbj{\"o}rn and Ask, Stefan and Christiansen,
                        Jesper R. and Corke, Richard and Desai, Nishita and Ilten,
                        Philip and Mrenna, Stephen and Prestel, Stefan and
                        Rasmussen, Christine O. and Skands, Peter Z.",
      title          = "{An introduction to PYTHIA 8.2}",
      journal        = "Comput. Phys. Commun.",
      volume         = "191",
      year           = "2015",
      pages          = "159",
      doi            = "10.1016/j.cpc.2015.01.024",
      eprint         = "1410.3012",
      archivePrefix  = "arXiv",
      primaryClass   = "hep-ph",
      reportNumber   = "LU-TP-14-36, MCNET-14-22, CERN-PH-TH-2014-190,
                        FERMILAB-PUB-14-316-CD, DESY-14-178, SLAC-PUB-16122,
                        --FERMILAB-PUB-14-316-CD",
      SLACcitation   = "%%CITATION = ARXIV:1410.3012;%%"
}

@article{Carra:2021ycg,
    author = "Carra, Sonia and Goumarre, Vincent and Gupta, Ruchi and Heim, Sarah and Heinemann, Beate and Kuechler, Jan and Meloni, Federico and Yap, Yee-Chinn",
    title = "{Constraining off-shell production of axion-like particles with $Z\gamma$ and $WW$ differential cross-section measurements}",
    eprint = "2106.10085",
    archivePrefix = "arXiv",
    primaryClass = "hep-ex",
    reportNumber = "DESY-21-063",
 	doi = "10.1103/PhysRevD.104.092005",
    journal = "Phys. Rev. D",
    volume = "104",
    number = "9",
    pages = "092005",
    year = "2021"
}

@article{Collins:1977iv,
    author = "Collins, John C. and Soper, Davison E.",
    title = "{Angular Distribution of Dileptons in High-Energy Hadron Collisions}",
    reportNumber = "Print-77-0288 (PRINCETON)",
    doi = "10.1103/PhysRevD.16.2219",
    journal = "Phys. Rev. D",
    volume = "16",
    pages = "2219",
    year = "1977"
}

@article{ZyNNLOPS2021,

    author = "Lombardi, Daniele and Wiesemann, Marius and Zanderighi, Giulia",
    title = "{Advancing M\i{}NNLO$_{PS}$ to diboson processes: Z\ensuremath{\gamma} production at NNLO+PS}",
    eprint = "2010.10478",
    archivePrefix = "arXiv",
    primaryClass = "hep-ph",
    reportNumber = "MPP-2020-188",
    doi = "10.1007/JHEP06(2021)095",
    journal = "JHEP",
    volume = "06",
    pages = "095",
    year = "2021"
}

@article{MiNNLOPS2020,
    author = "Monni, Pier Francesco and Nason, Paolo and Re, Emanuele and Wiesemann, Marius and Zanderighi, Giulia",
    title = "{MiNNLO$_{PS}$: a new method to match NNLO QCD to parton showers}",
    eprint = "1908.06987",
    archivePrefix = "arXiv",
    primaryClass = "hep-ph",
    reportNumber = "CERN-TH-2019-117, LAPTH-042/19, MPP-2019-177",
    doi = "10.1007/JHEP05(2020)143",
    journal = "JHEP",
    volume = "05",
    pages = "143",
    year = "2020",
    note = "[Erratum: JHEP 02, 031 (2022)]"


}

@article{PowhegBox2010,
      author         = "Alioli, Simone and Nason, Paolo and Oleari, Carlo and Re,
                        Emanuele",
      title          = "{A general framework for implementing NLO calculations in
                        shower Monte Carlo programs: the POWHEG BOX}",
      journal        = "JHEP",
      volume         = "06",
      pages          = "043",
      doi            = "10.1007/JHEP06(2010)043",
      year           = "2010",
      eprint         = "1002.2581",
      archivePrefix  = "arXiv",
      primaryClass   = "hep-ph",

}

@Article{Alioli:2010xd,
%      author         = "Alioli, Simone and Nason, Paolo and Oleari, Carlo and Re,
%                        Emanuele",
%      title          = "{A general framework for implementing NLO calculations in
%                        shower Monte Carlo programs: the POWHEG BOX}",
%      journal        = "JHEP",
%      volume         = "06",
%      pages          = "043",
%      doi            = "10.1007/JHEP06(2010)043",
%      year           = "2010",
%      eprint         = "1002.2581",
%      archivePrefix  = "arXiv",
%      primaryClass   = "hep-ph",
%}

@Article{Evans:2008zzb,
      author         = "Evans, Lyndon and Bryant, Philip",
      title          = "{LHC Machine}",
      journal        = "JINST",
      volume         = "3",
      pages          = "S08001",
      doi            = "10.1088/1748-0221/3/08/S08001",
      year           = "2008",
      SLACcitation   = "%%CITATION = JINST,3,S08001;%%",
}

@Article{Cacciari:2008gp,
     author    = "Cacciari, Matteo and Salam, Gavin P. and Soyez, Gregory",
     title     = "{The anti-\(k_{t}\) jet clustering algorithm}",
     journal   = "JHEP",
     volume    = "04",
     year      = "2008",
     pages     = "063",
     eprint    = "0802.1189",
     archivePrefix = "arXiv",
     primaryClass  =  "hep-ph",
     doi       = "10.1088/1126-6708/2008/04/063",
     SLACcitation  = "%%CITATION = 0802.1189;%%"
}

@Article{Fastjet,
      author         = "Cacciari, Matteo and Salam, Gavin P. and Soyez, Gregory",
      title          = "{FastJet user manual}",
      journal        = "Eur. Phys. J. C",
      volume         = "72",
      year           = "2012",
      pages          = "1896",
      doi            = "10.1140/epjc/s10052-012-1896-2",
      eprint         = "1111.6097",
      archivePrefix  = "arXiv",
      primaryClass   = "hep-ph",
      reportNumber   = "CERN-PH-TH-2011-297",
      SLACcitation   = "%%CITATION = ARXIV:1111.6097;%%"
}

@article{Dulat:2015mca,
      author         = "Dulat, Sayipjamal and Hou, Tie-Jiun and Gao, Jun and
                        Guzzi, Marco and Huston, Joey and Nadolsky, Pavel and
                        Pumplin, Jon and Schmidt, Carl and Stump, Daniel and Yuan,
                        C. P.",
      title          = "{New parton distribution functions from a global analysis
                        of quantum chromodynamics}",
      journal        = "Phys. Rev. D",
      volume         = "93",
      year           = "2016",
      number         = "3",
      pages          = "033006",
      doi            = "10.1103/PhysRevD.93.033006",
      eprint         = "1506.07443",
      archivePrefix  = "arXiv",
      primaryClass   = "hep-ph",
      SLACcitation   = "%%CITATION = ARXIV:1506.07443;%%"
}

@article{Ball:2014uwa,
      author         = "{The NNPDF Collaboration} and Ball, Richard D. and others",
      title          = "{Parton distributions for the LHC run II}",
      journal        = "JHEP",
      volume         = "04",
      year           = "2015",
      pages          = "040",
      doi            = "10.1007/JHEP04(2015)040",
      eprint         = "1410.8849",
      archivePrefix  = "arXiv",
      primaryClass   = "hep-ph",
      reportNumber   = "EDINBURGH-2014-15, IFUM-1034-FT, CERN-PH-TH-2013-253,
                        OUTP-14-11P, CAVENDISH-HEP-14-11",
      SLACcitation   = "%%CITATION = ARXIV:1410.8849;%%"
}

@Article{Alwall:2014hca,
      author         = "Alwall, J. and Frederix, R. and Frixione, S. and Hirschi,
                        V. and Maltoni, F. and Mattelaer, O. and Shao, H. -S. and
                        Stelzer, T. and Torrielli, P. and Zaro, M.",
      title          = "{The automated computation of tree-level and
                        next-to-leading order differential cross sections, and
                        their matching to parton shower simulations}",
      journal        = "JHEP",
      volume         = "07",
      year           = "2014",
      pages          = "079",
      doi            = "10.1007/JHEP07(2014)079",
      eprint         = "1405.0301",
      archivePrefix  = "arXiv",
      primaryClass   = "hep-ph",
      reportNumber   = "CERN-PH-TH-2014-064, CP3-14-18, LPN14-066, MCNET-14-09,
                        ZU-TH-14-14",
      SLACcitation   = "%%CITATION = ARXIV:1405.0301;%%"
}

@Article{Nason:2004rx,
      author         = "Nason, Paolo",
      title          = "{A new method for combining NLO QCD with shower Monte Carlo algorithms}",
      journal        = "JHEP",
      volume         = "11",
      pages          = "040",
      doi            = "10.1088/1126-6708/2004/11/040",
      year           = "2004",
      eprint         = "hep-ph/0409146",
      archivePrefix  = "arXiv",
}

@Article{Frixione:2007vw,
      author         = "Frixione, Stefano and Nason, Paolo and Oleari, Carlo",
      title          = "{Matching NLO QCD computations with parton shower
                        simulations: the POWHEG method}",
      journal        = "JHEP",
      volume         = "11",
      pages          = "070",
      doi            = "10.1088/1126-6708/2007/11/070",
      year           = "2007",
      eprint         = "0709.2092",
      archivePrefix  = "arXiv",
      primaryClass   = "hep-ph",
}

@article{deFlorian:2016spz,
      author         = "de Florian, D. and others",
      title          = "{Handbook of LHC Higgs Cross Sections: 4. Deciphering the
                        Nature of the Higgs Sector}",
      collaboration  = "LHC Higgs Cross Section Working Group",
      doi            = "10.23731/CYRM-2017-002",
      year           = "2016",
      eprint         = "1610.07922",
      archivePrefix  = "arXiv",
      primaryClass   = "hep-ph",
      reportNumber   = "FERMILAB-FN-1025-T, CERN-2017-002-M",
      SLACcitation   = "%%CITATION = ARXIV:1610.07922;%%"
}

@Article{Gleisberg:2008fv,
      author         = "Gleisberg, Tanju and H{\"o}che, Stefan",
      title          = "{Comix, a new matrix element generator}",
      journal        = "JHEP",
      volume         = "12",
      pages          = "039",
      doi            = "10.1088/1126-6708/2008/12/039",
      year           = "2008",
      eprint         = "0808.3674",
      archivePrefix  = "arXiv",
      primaryClass   = "hep-ph",
}

@Article{Schumann:2007mg,
      author         = "Schumann, Steffen and Krauss, Frank",
      title          = "{A parton shower algorithm based on Catani--Seymour dipole factorisation}",
      journal        = "JHEP",
      volume         = "03",
      pages          = "038",
      doi            = "10.1088/1126-6708/2008/03/038",
      year           = "2008",
      eprint         = "0709.1027",
      archivePrefix  = "arXiv",
      primaryClass   = "hep-ph",
}

@article{Buccioni:2019sur,
    author = "Buccioni, Federico and Lang, Jean-Nicolas and Lindert, Jonas M. and Maierh{\"o}fer, Philipp and Pozzorini, Stefano and Zhang, Hantian and Zoller, Max F.",
    title = "{OpenLoops 2}",
    eprint = "1907.13071",
    archivePrefix = "arXiv",
    primaryClass = "hep-ph",
    reportNumber = "IPPP/19/62, FR-PHENO-2019-12, PSI-PR-19-15, ZU-TH 37/19",
    doi = "10.1140/epjc/s10052-019-7306-2",
    journal = "Eur. Phys. J. C",
    volume = "79",
    number = "10",
    pages = "866",
    year = "2019"
}

@article{Agostinelli:2002hh,
      author         = "Agostinelli, S. and others",
      title          = "{\textsc{Geant4} -- a simulation toolkit}",
      journal        = "Nucl. Instrum. Meth. A",
      volume         = "506",
      year           = "2003",
      pages          = "250",
      doi            = "10.1016/S0168-9002(03)01368-8",
      reportNumber   = "SLAC-PUB-9350, FERMILAB-PUB-03-339",
      SLACcitation   = "%%CITATION = NUIMA,A506,250;%%"
}

@Article{PERF-2007-01,
    author         = "{ATLAS Collaboration}",
    title          = "{The ATLAS Experiment at the CERN Large Hadron Collider}",
    journal        = "JINST",
    volume         = "3",
    year           = "2008",
    pages          = "S08003",
    doi            = "10.1088/1748-0221/3/08/S08003",
    primaryClass   = "hep-ex",
}

@Article{SOFT-2010-01,
    author         = "{ATLAS Collaboration}",
    title          = "{The ATLAS Simulation Infrastructure}",
    journal        = "Eur. Phys. J. C",
    volume         = "70",
    year           = "2010",
    pages          = "823",
    doi            = "10.1140/epjc/s10052-010-1429-9",
    eprint         = "1005.4568",
    archivePrefix  = "arXiv",
    primaryClass   = "physics.ins-det",
}

@Article{STDM-2010-08,
    author         = "{ATLAS Collaboration}",
    title          = "{Measurement of the inclusive isolated prompt photon cross section in \(pp\) collisions at \(\sqrt{s} = 7\,\text{TeV}\) with the ATLAS detector}",
    journal        = "Phys. Rev. D",
    volume         = "83",
    year           = "2011",
    pages          = "052005",
    doi            = "10.1103/PhysRevD.83.052005",
    reportNumber   = "CERN-PH-EP-2010-068",
    eprint         = "1012.4389",
    archivePrefix  = "arXiv",
    primaryClass   = "hep-ex",
}

@Article{STDM-2012-23,
    author         = "{ATLAS Collaboration}",
    title          = "{Measurement of the \(Z/\gamma^*\) boson transverse momentum distribution in \(pp\) collisions at \(\sqrt{s} = 7\,\text{TeV}\) with the ATLAS detector}",
    journal        = "JHEP",
    volume         = "09",
    year           = "2014",
    pages          = "145",
    doi            = "10.1007/JHEP09(2014)145",
    reportNumber   = "CERN-PH-EP-2014-075",
    eprint         = "1406.3660",
    archivePrefix  = "arXiv",
    primaryClass   = "hep-ex",
}

@Article{PERF-2014-03,
    author         = "{ATLAS Collaboration}",
    title          = "{Performance of pile-up mitigation techniques for jets in \(pp\) collisions at \(\sqrt{s} = 8\,\text{TeV}\) using the ATLAS detector}",
    journal        = "Eur. Phys. J. C",
    volume         = "76",
    year           = "2016",
    pages          = "581",
    doi            = "10.1140/epjc/s10052-016-4395-z",
    reportNumber   = "CERN-PH-EP-2015-206",
    eprint         = "1510.03823",
    archivePrefix  = "arXiv",
    primaryClass   = "hep-ex",
}

@Article{STDM-2014-01,
    author         = "{ATLAS Collaboration}",
    title          = "{Measurements of \(Z\gamma\) and \(Z\gamma\gamma\) production in \(pp\) collisions at \(\sqrt{s} = 8\,\text{TeV}\) with the ATLAS detector}",
    journal        = "Phys. Rev. D",
    volume         = "93",
    year           = "2016",
    pages          = "112002",
    doi            = "10.1103/PhysRevD.93.112002",
    reportNumber   = "CERN-EP-2016-049",
    eprint         = "1604.05232",
    archivePrefix  = "arXiv",
    primaryClass   = "hep-ex",
}

@Article{STDM-2014-09,
    author         = "{ATLAS Collaboration}",
    title          = "{Measurement of the inclusive isolated prompt photon cross section in \(pp\) collisions at \(\sqrt{s} = 8\,\text{TeV}\) with the ATLAS detector}",
    journal        = "JHEP",
    volume         = "08",
    year           = "2016",
    pages          = "005",
    doi            = "10.1007/JHEP08(2016)005",
    reportNumber   = "CERN-EP-2016-035",
    eprint         = "1605.03495",
    archivePrefix  = "arXiv",
    primaryClass   = "hep-ex",
}

@Article{PERF-2015-09,
    author         = "{ATLAS Collaboration}",
    title          = "{Jet reconstruction and performance using particle flow with the ATLAS Detector}",
    journal        = "Eur. Phys. J. C",
    volume         = "77",
    year           = "2017",
    pages          = "466",
    doi            = "10.1140/epjc/s10052-017-5031-2",
    reportNumber   = "CERN-EP-2017-024",
    eprint         = "1703.10485",
    archivePrefix  = "arXiv",
    primaryClass   = "hep-ex",
}

@Article{STDM-2015-21,
    author         = "{ATLAS Collaboration}",
    title          = "{Studies of \(Z\gamma\) production in association with a high-mass dijet system in \(pp\) collisions at \(\sqrt{s} = 8\,\text{TeV}\) with the ATLAS detector}",
    journal        = "JHEP",
    volume         = "07",
    year           = "2017",
    pages          = "107",
    doi            = "10.1007/JHEP07(2017)107",
    reportNumber   = "CERN-EP-2017-046",
    eprint         = "1705.01966",
    archivePrefix  = "arXiv",
    primaryClass   = "hep-ex",
}

@Article{PERF-2017-02,
    author         = "{ATLAS Collaboration}",
    title          = "{Measurement of the photon identification efficiencies with the ATLAS detector using LHC Run~2 data collected in 2015 and 2016}",
    journal        = "Eur. Phys. J. C",
    volume         = "79",
    year           = "2019",
    pages          = "205",
    doi            = "10.1140/epjc/s10052-019-6650-6",
    reportNumber   = "CERN-EP-2018-216",
    eprint         = "1810.05087",
    archivePrefix  = "arXiv",
    primaryClass   = "hep-ex",
}

@Article{TOPQ-2017-14,
    author         = "{ATLAS Collaboration}",
    title          = "{Measurements of inclusive and differential fiducial cross-sections of \(t\bar{t}\gamma\) production in leptonic final states at \(\sqrt{s} =  13\,\text{TeV}\) in ATLAS}",
    journal        = "Eur. Phys. J. C",
    volume         = "79",
    year           = "2019",
    pages          = "382",
    doi            = "10.1140/epjc/s10052-019-6849-6",
    reportNumber   = "CERN-EP-2018-302",
    eprint         = "1812.01697",
    archivePrefix  = "arXiv",
    primaryClass   = "hep-ex",
}

@Article{DAPR-2018-01,
    author         = "{ATLAS Collaboration}",
    title          = "{ATLAS data quality operations and performance for 2015--2018 data-taking}",
    journal        = "JINST",
    volume         = "15",
    year           = "2020",
    pages          = "P04003",
    doi            = "10.1088/1748-0221/15/04/P04003",
    reportNumber   = "CERN-EP-2019-207",
    eprint         = "1911.04632",
    archivePrefix  = "arXiv",
    primaryClass   = "physics.ins-det",
}

@Article{EGAM-2018-01,
    author         = "{ATLAS Collaboration}",
    title          = "{Electron and photon performance measurements with the ATLAS detector using the 2015--2017 LHC proton--proton collision data}",
    journal        = "JINST",
    volume         = "14",
    year           = "2019",
    pages          = "P12006",
    doi            = "10.1088/1748-0221/14/12/P12006",
    reportNumber   = "CERN-EP-2019-145",
    eprint         = "1908.00005",
    archivePrefix  = "arXiv",
    primaryClass   = "hep-ex",
}

@Article{JETM-2018-05,
    author         = "{ATLAS Collaboration}",
    title          = "{Jet energy scale and resolution measured in proton--proton collisions at \(\sqrt{s} = 13\,\text{TeV}\) with the ATLAS detector}",
    journal        = "Eur. Phys. J. C",
    volume         = "81",
    year           = "2020",
    pages          = "689",
    doi            = "10.1140/epjc/s10052-021-09402-3",
    reportNumber   = "CERN-EP-2020-083",
    eprint         = "2007.02645",
    archivePrefix  = "arXiv",
    primaryClass   = "hep-ex",
}

@Article{MUON-2018-03,
    author         = "{ATLAS Collaboration}",
    title          = "{Muon reconstruction and identification efficiency in ATLAS using the full Run~2 \(pp\) collision data set at \(\sqrt{s} = 13\,\text{TeV}\)}",
    journal        = "Eur. Phys. J. C",
    volume         = "81",
    year           = "2021",
    pages          = "578",
    doi            = "10.1140/epjc/s10052-021-09233-2",
    reportNumber   = "CERN-EP-2020-199",
    eprint         = "2012.00578",
    archivePrefix  = "arXiv",
    primaryClass   = "hep-ex",
}

@Article{STDM-2018-03,
    author         = "{ATLAS Collaboration}",
    title          = "{Measurement of \(W^{\pm}Z\) production cross sections and gauge boson polarisation in \(pp\) collisions at \(\sqrt{s} = 13\,\text{TeV}\) with the ATLAS detector}",
    journal        = "Eur. Phys. J. C",
    volume         = "79",
    year           = "2019",
    pages          = "535",
    doi            = "10.1140/epjc/s10052-019-7027-6",
    reportNumber   = "CERN-EP-2018-327",
    eprint         = "1902.05759",
    archivePrefix  = "arXiv",
    primaryClass   = "hep-ex",
}

@Article{STDM-2018-04,
    author         = "{ATLAS Collaboration}",
    title          = "{Measurement of the \(Z(\to\ell^+\ell^-)\gamma\) production cross-section in \(pp\) collisions at \(\sqrt{s} = 13\,\text{TeV}\) with the ATLAS detector}",
    journal        = "JHEP",
    volume         = "03",
    year           = "2020",
    pages          = "054",
    doi            = "10.1007/JHEP03(2020)054",
    reportNumber   = "CERN-EP-2019-228",
    eprint         = "1911.04813",
    archivePrefix  = "arXiv",
    primaryClass   = "hep-ex",
}

@Article{STDM-2018-34,
    author         = "{ATLAS Collaboration}",
    title          = "{Measurements of \(W^{+}W^{-} + \ge 1\) jet production cross-sections in \(pp\) collisions at \(\sqrt{s} = 13\,\text{TeV}\) with the ATLAS detector}",
    journal        = "JHEP",
    volume         = "06",
    year           = "2021",
    pages          = "003",
    doi            = "10.1007/JHEP06(2021)003",
    reportNumber   = "CERN-EP-2021-030",
    eprint         = "2103.10319",
    archivePrefix  = "arXiv",
    primaryClass   = "hep-ex",
}

@Article{TRIG-2018-01,
    author         = "{ATLAS Collaboration}",
    title          = "{Performance of the ATLAS muon triggers in Run~2}",
    journal        = "JINST",
    volume         = "15",
    year           = "2020",
    pages          = "P09015",
    doi            = "10.1088/1748-0221/15/09/p09015",
    reportNumber   = "CERN-EP-2020-031",
    eprint         = "2004.13447",
    archivePrefix  = "arXiv",
    primaryClass   = "hep-ex",
}

@Article{TRIG-2018-05,
    author         = "{ATLAS Collaboration}",
    title          = "{Performance of electron and photon triggers in ATLAS during LHC Run~2}",
    journal        = "Eur. Phys. J. C",
    volume         = "80",
    year           = "2020",
    pages          = "47",
    doi            = "10.1140/epjc/s10052-019-7500-2",
    reportNumber   = "CERN-EP-2019-169",
    eprint         = "1909.00761",
    archivePrefix  = "arXiv",
    primaryClass   = "hep-ex",
}

@Article{STDM-2020-14,
    author         = "{ATLAS Collaboration}",
    title          = "{Measurements of \(Z\gamma+\)jets differential cross sections in \(pp\) collisions at \(\sqrt{s} = 13\,\text{TeV}\) with the ATLAS detector}",
    year           = "2022",
    reportNumber   = "CERN-EP-2022-240",
    eprint         = "2212.07184",
    archivePrefix  = "arXiv",
    primaryClass   = "hep-ex",
}

@Booklet{ATL-SOFT-PUB-2021-001,
    author         = "{ATLAS Collaboration}",
    title          = "{The ATLAS Collaboration Software and Firmware}",
    howpublished   = "{ATL-SOFT-PUB-2021-001}",
    url            = "https://cds.cern.ch/record/2767187",
    year           = "2021",
}

@Booklet{ATL-SOFT-PUB-2021-003,
    author         = "{ATLAS Collaboration}",
    title          = "{ATLAS Computing Acknowledgements}",
    howpublished   = "{ATL-SOFT-PUB-2021-003}",
    url            = "https://cds.cern.ch/record/2776662",
    year           = "2021",
}

@Article{CMS-SMP-13-010,
    author         = "{CMS Collaboration}",
    title          = "{Angular coefficients of \(Z\) bosons produced in \(pp\) collisions at \(\sqrt{s} = 8\,\text{TeV}\) and decaying to \(\mu^+ \mu^-\) as a function of transverse momentum and rapidity}",
    journal        = "Phys. Lett. B",
    volume         = "750",
    year           = "2015",
    pages          = "154",
    doi            = "10.1016/j.physletb.2015.08.061",
    reportNumber   = "CERN-PH-EP-2015-046",
    eprint         = "1504.03512",
    archivePrefix  = "arXiv",
    primaryClass   = "hep-ex",
}

@Article{CMS-SMP-13-014,
    author         = "{CMS Collaboration}",
    title          = "{Measurement of the \(Z\gamma\) production cross section in \(pp\) collisions at \(8\,\text{TeV}\) and search for anomalous triple gauge boson couplings}",
    journal        = "JHEP",
    volume         = "04",
    year           = "2015",
    pages          = "164",
    doi            = "10.1007/JHEP04(2015)164",
    reportNumber   = "CERN-PH-EP-2014-303",
    eprint         = "1502.05664",
    archivePrefix  = "arXiv",
    primaryClass   = "hep-ex",
}

@Booklet{ATLAS-CONF-2019-021,
    author         = "{ATLAS Collaboration}",
    title          = "{Luminosity determination in \(pp\) collisions at \(\sqrt{s} = 13\,\text{TeV}\) using the ATLAS detector at the LHC}",
    howpublished   = "{ATLAS-CONF-2019-021}",
    url            = "https://cds.cern.ch/record/2677054",
    year           = "2019",
}

@Booklet{ATL-PHYS-PUB-2014-021,
    author         = "{ATLAS Collaboration}",
    title          = "{ATLAS Pythia~8 tunes to \(7~\text{TeV}\) data}",
    howpublished   = "{ATL-PHYS-PUB-2014-021}",
    url            = "https://cds.cern.ch/record/1966419",
    year           = "2014",
}

@Booklet{ATL-PHYS-PUB-2016-017,
    author         = "{ATLAS Collaboration}",
    title          = "{The Pythia~8 A3 tune description of ATLAS minimum bias and inelastic measurements incorporating the Donnachie--Landshoff diffractive model}",
    howpublished   = "{ATL-PHYS-PUB-2016-017}",
    url            = "https://cds.cern.ch/record/2206965",
    year           = "2016",
}

\clearpage
 
\begin{flushleft}
\hypersetup{urlcolor=black}
{\Large The ATLAS Collaboration}

\bigskip

\AtlasOrcid[0000-0002-6665-4934]{G.~Aad}$^\textrm{\scriptsize 101}$,
\AtlasOrcid[0000-0002-5888-2734]{B.~Abbott}$^\textrm{\scriptsize 119}$,
\AtlasOrcid[0000-0002-7248-3203]{D.C.~Abbott}$^\textrm{\scriptsize 102}$,
\AtlasOrcid[0000-0002-1002-1652]{K.~Abeling}$^\textrm{\scriptsize 55}$,
\AtlasOrcid[0000-0002-8496-9294]{S.H.~Abidi}$^\textrm{\scriptsize 29}$,
\AtlasOrcid[0000-0002-9987-2292]{A.~Aboulhorma}$^\textrm{\scriptsize 35e}$,
\AtlasOrcid[0000-0001-5329-6640]{H.~Abramowicz}$^\textrm{\scriptsize 150}$,
\AtlasOrcid[0000-0002-1599-2896]{H.~Abreu}$^\textrm{\scriptsize 149}$,
\AtlasOrcid[0000-0003-0403-3697]{Y.~Abulaiti}$^\textrm{\scriptsize 116}$,
\AtlasOrcid[0000-0003-0762-7204]{A.C.~Abusleme~Hoffman}$^\textrm{\scriptsize 136a}$
% ...
\bigskip
\\

% ...
$^{29}$Physics Department, Brookhaven National Laboratory, Upton NY; United States of America.\\
% ...
$^{35}$$^{(a)}$Facult\'e des Sciences Ain Chock, R\'eseau Universitaire de Physique des Hautes Energies - Universit\'e Hassan II, Casablanca;$^{(b)}$Facult\'{e} des Sciences, Universit\'{e} Ibn-Tofail, K\'{e}nitra;$^{(c)}$Facult\'e des Sciences Semlalia, Universit\'e Cadi Ayyad, LPHEA-Marrakech;$^{(d)}$LPMR, Facult\'e des Sciences, Universit\'e Mohamed Premier, Oujda;$^{(e)}$Facult\'e des sciences, Universit\'e Mohammed V, Rabat;$^{(f)}$Institute of Applied Physics, Mohammed VI Polytechnic University, Ben Guerir; Morocco.\\
% ...
$^{55}$II. Physikalisches Institut, Georg-August-Universit\"{a}t G\"ottingen, G\"ottingen; Germany.\\
% ...
$^{101}$CPPM, Aix-Marseille Universit\'e, CNRS/IN2P3, Marseille; France.\\
$^{102}$Department of Physics, University of Massachusetts, Amherst MA; United States of America.\\
% ...
$^{116}$Department of Physics, New York University, New York NY; United States of America.\\
% ...
$^{119}$Homer L. Dodge Department of Physics and Astronomy, University of Oklahoma, Norman OK; United States of America.\\
% ...
$^{136}$$^{(a)}$Departamento de F\'isica, Pontificia Universidad Cat\'olica de Chile, Santiago;$^{(b)}$Millennium Institute for Subatomic physics at high energy frontier (SAPHIR), Santiago;$^{(c)}$Instituto de Investigaci\'on Multidisciplinario en Ciencia y Tecnolog\'ia, y Departamento de F\'isica, Universidad de La Serena;$^{(d)}$Universidad Andres Bello, Department of Physics, Santiago;$^{(e)}$Instituto de Alta Investigaci\'on, Universidad de Tarapac\'a, Arica;$^{(f)}$Departamento de F\'isica, Universidad T\'ecnica Federico Santa Mar\'ia, Valpara\'iso; Chile.\\
% ...
$^{149}$Department of Physics, Technion, Israel Institute of Technology, Haifa; Israel.\\
$^{150}$Raymond and Beverly Sackler School of Physics and Astronomy, Tel Aviv University, Tel Aviv; Israel.\\
% ...
$^{a}$ Also Affiliated with an institute covered by a cooperation agreement with CERN.\\
$^{b}$ Also at An-Najah National University, Nablus; Palestine.\\
$^{c}$ Also at Borough of Manhattan Community College, City University of New York, New York NY; United States of America.\\
$^{d}$ Also at Bruno Kessler Foundation, Trento; Italy.\\
$^{e}$ Also at Center for High Energy Physics, Peking University; China.\\
$^{f}$ Also at Center for Interdisciplinary Research and Innovation (CIRI-AUTH), Thessaloniki ; Greece.\\
% ...

\end{flushleft}


\end{document}